\documentclass[twocolumn]{aastex62}

\usepackage{latexsym,amssymb}

\usepackage{amsmath}

\usepackage{natbib}
\usepackage{rotating}

\newcommand{\du}{\mathrm{d}}

\newcommand{\Sim}{\mathord{\sim}}

\usepackage{graphicx,xcolor}

\shorttitle{Gyroresonant Streaming Instabilities}
\shortauthors{Holcomb \& Spitkovsky}

\begin{document}

\title{On the Growth and Saturation of the Gyroresonant Streaming Instabilities}

\correspondingauthor{Cole Holcomb}
\email{cole.holcomb@gmail.com}

\author[0000-0002-2467-8962]{Cole Holcomb}
\author{Anatoly Spitkovsky}
\affil{Department of Astrophysical Sciences, Princeton University, \\ 4 Ivy Ln., Princeton, NJ 08540, USA}

\begin{abstract}
The self-regulation of cosmic ray (CR) transport in the interstellar and intracluster media has long been viewed through the lenses of linear and quasilinear kinetic plasma physics. Such theories are believed to capture the essence of CR behavior in the presence of self-generated turbulence, but cannot describe potentially critical details arising from the nonlinearities of the problem. We utilize the particle-in-cell numerical method to study the time-dependent nonlinear behavior of the gyroresonant streaming instabilities, self-consistently following the combined evolution of particle distributions and self-generated wave spectra in one-dimensional periodic simulations. We demonstrate that the early growth of instability conforms to the predictions from linear physics, but that the late-time behavior can vary depending on the properties of the initial CR distribution. We emphasize that the nonlinear stages of instability depend strongly on the initial anisotropy of CRs -- highly anisotropic CR distributions do not efficiently reduce to Alfv\'enic drift velocities, owing to reduced production of left-handed resonant modes. We derive estimates for the wave amplitudes at saturation and the time scales for nonlinear relaxation of the CR distribution, then demonstrate the applicability of these estimates to our simulations. Bulk flows of the background plasma due to the presence of resonant waves are observed in our simulations, confirming the microphysical basis of CR-driven winds.
\end{abstract}

  
\section{Introduction}\label{sec:int} 

Cosmic rays (CRs) are an energetically significant component of interstellar and intracluster media (see \citealt{2013PhPl...20e5501Z} for a review). For the Milky Way galaxy (and similar galaxies), CRs are roughly in equipartition with the magnetic, thermal, and turbulent components of the galactic energy density \citep{2011piim.book.....D}. Despite the relatively small number densities of CRs, the plentiful kinetic energy associated with their relativistic velocities implies the potential to have outsized dynamical influence on their surroundings. 

In the interstellar medium (ISM) this influence is thought to play a crucial role in stellar feedback, the processes by which stars redistribute momentum and energy and thus regulate galactic-scale thermodynamics and subsequent star formation. The forces that couple CRs to the interstellar plasma can drive bulk motions in the latter -- the so-called  CR-driven galactic winds \citep{1975ApJ...196..107I,1991A&A...245...79B,2008ApJ...674..258E,2008ApJ...687..202S}. Much attention has been given to this mechanism in recent years, both in large-scale fluid simulation studies \citep{2016ApJ...816L..19G,2017MNRAS.467..906W,2017ApJ...834..208R,2018MNRAS.tmp.1588G,2018ApJ...854....5J,2019MNRAS.tmp..260T}, and in analytical fluid studies \citep{2016MNRAS.462.4227R,2018ApJ...854...89M}.

The streaming of CRs through intracluster media (ICM) is also thought to be an important component of the feedback processes by which active galactic nuclei inject heat into their surrounding media, thus preventing the ``catastrophic" formation of massive galaxies at the cluster core with star-formation rates beyond what is observed \citep{1991ApJ...377..392L,2008MNRAS.384..251G}. The recent fluid simulations of \cite{2017ApJ...844...13R} demonstrated the viability of CR streaming as a mechanism for maintaining the thermal equilibrium of ICM. Additionally, CRs have been utilized in models to explain the observed bimodality of galaxy-cluster radio halos \citep{2013MNRAS.434.2209W,2018MNRAS.473.3095W}.

The cross-section for Coulomb collisions becomes sufficiently small at energies above $\Sim 1$ GeV that CRs are not effectively confined to the galactic disk by the interstellar gas. Instead, the predominant mechanism for CR interactions is via collisionless scattering on magnetic fluctuations. The present-day canon is based on the self-confinement paradigm \citep{1969ApJ...156..445K,1969ApJ...156..303W,1971ApJ...170..265S,2005ppfa.book.....K}, whereby CRs generate the magnetic fluctuations that they subsequently scatter on (n.b., \ turbulent confinement is an increasingly viable alternative; \citealt{2002PhRvL..89B1102Y}).  CRs are believed to induce waves in the galactic magnetic field via the gyroresonant streaming instability  \citep{1967ApJ...147..689L,1969ApJ...156..445K}. These waves would allow CRs to indirectly couple to the ISM plasma, thus facilitating the transfer of momentum and energy.

The separation of scales between typical CR gyroradii ($\Sim$au) and galactic structures ($\Sim$kpc) necessitates a two-pronged approach to understanding the physical influence of CRs. One approach utilizes fluid approximations to study the influence of CRs on large scales. In this scheme, the physics of CRs is parameterized within the conservation equations of the fluid framework (e.g.,\ \citealt{2017PhPl...24e5402Z,2018ApJ...854....5J}). The other approach, which we adopt here, analyzes the behavior of CRs on kinetic scales by fully resolving the physics of wave-particle interactions. We employ the particle-in-cell method (PIC), which has previously been used to study the nonresonant branch of the CR streaming instability (a.k.a.\ Bell or CR current-driven instability) in the context of magnetic-field amplification around shock waves \citep{2008ApJ...684.1174N,2009ApJ...694..626R,2009ApJ...706...38S} and the gyroresonant instability of high-density ion beams \citep{2019ApJ...873...57W}. Additionally, the MHD-PIC numerical scheme \citep{2015ApJ...809...55B} has been used to study the closely related pressure anisotropy-driven gyroresonance instability \citep{2018MNRAS.476.2779L}.

  In this work we study the behavior of the gyroresonant streaming instability using a range of CR distributions. Numerical simulation allows us to follow the instability through the nonlinear stages of evolution and observe the feedback between particle distributions and wave spectra. The CR distributions are chosen to emphasize the role of initial CR anisotropy in determining the qualitative evolution of instability, particularly in the nonlinear instability phase. We demonstrate that CR distributions with large degrees of anisotropy excite right-handed Alfv\'enic waves, while those with small anisotropy produce linearly polarized modes. The properties of these waves then determine the temporal evolution of CR streaming in the nonlinear phase. In section \ref{sec:physics}, we summarize the physics of the gyroresonant streaming instability and derive scaling relations that quantitatively predict its behaviors. In section \ref{sec:num}, we describe our numerical methods and simulation setups. We present the results of our simulations in section \ref{sec:res}. In section \ref{sec:disc}, we discuss the implications and applications of our results. Finally, we summarize in section \ref{sec:conc}.

\section{Physics of CR Streaming Instabilities}\label{sec:physics}

Energetic charged particles streaming along a large-scale magnetic field embedded in a background plasma can interact with transverse electromagnetic fluctuations, absorbing or exciting waves that travel parallel (or antiparallel) to the bulk particle motion. There are two well known and distinct mechanisms for streaming CRs to grow waves in the ISM/ICM. The first is known as the (gyroresonant) CR streaming instability \citep{1967ApJ...147..689L,1969ApJ...156..445K,1969ApJ...156..303W}, which occurs when CRs impart their momentum to Alfv\'en waves by adjusting their pitch angle. The second is the current-driven nonresonant instability (a.k.a.\ Bell instability; \citealt{2004MNRAS.353..550B,2005MNRAS.358..181B}), where a large CR current $J_{\rm cr}$ drives the growth of slowly propagating waves via the $\boldsymbol{J}\times \boldsymbol{B}$ force (however, see \citealt{2019ApJ...872...48W} for an alternative interpretation).

\cite{2009MNRAS.392.1591A} demonstrated that these two behaviors can be jointly derived in the framework of linearized kinetic theory. The nonresonant branch is expected to be important only near the sources of CRs, where the associated current is large, e.g.,\ supernova remnants (SNRs; \citealt{2010ApJ...717.1054R}). Here we focus on the resonant branch of the instability, which is believed to be the predominant CR-driven instability for the majority of the ISM/ICM. The kinetic dispersion relation predicts that the current-driven instability is subdominant so long as the relation  $U_{\rm cr}/U_{B}\lesssim c/v_{\rm dr}$ holds, where $v_{\rm dr}$ is the bulk drift velocity of CRs and $U_{\rm cr}$ and $U_B$ are the energy densities of CRs and the background magnetic field, respectively \citep{2010ApJ...709.1412Z}.

\subsection{Resonant Scattering}

\begin{figure}[t]
\centering\includegraphics[width=\linewidth,clip=true]{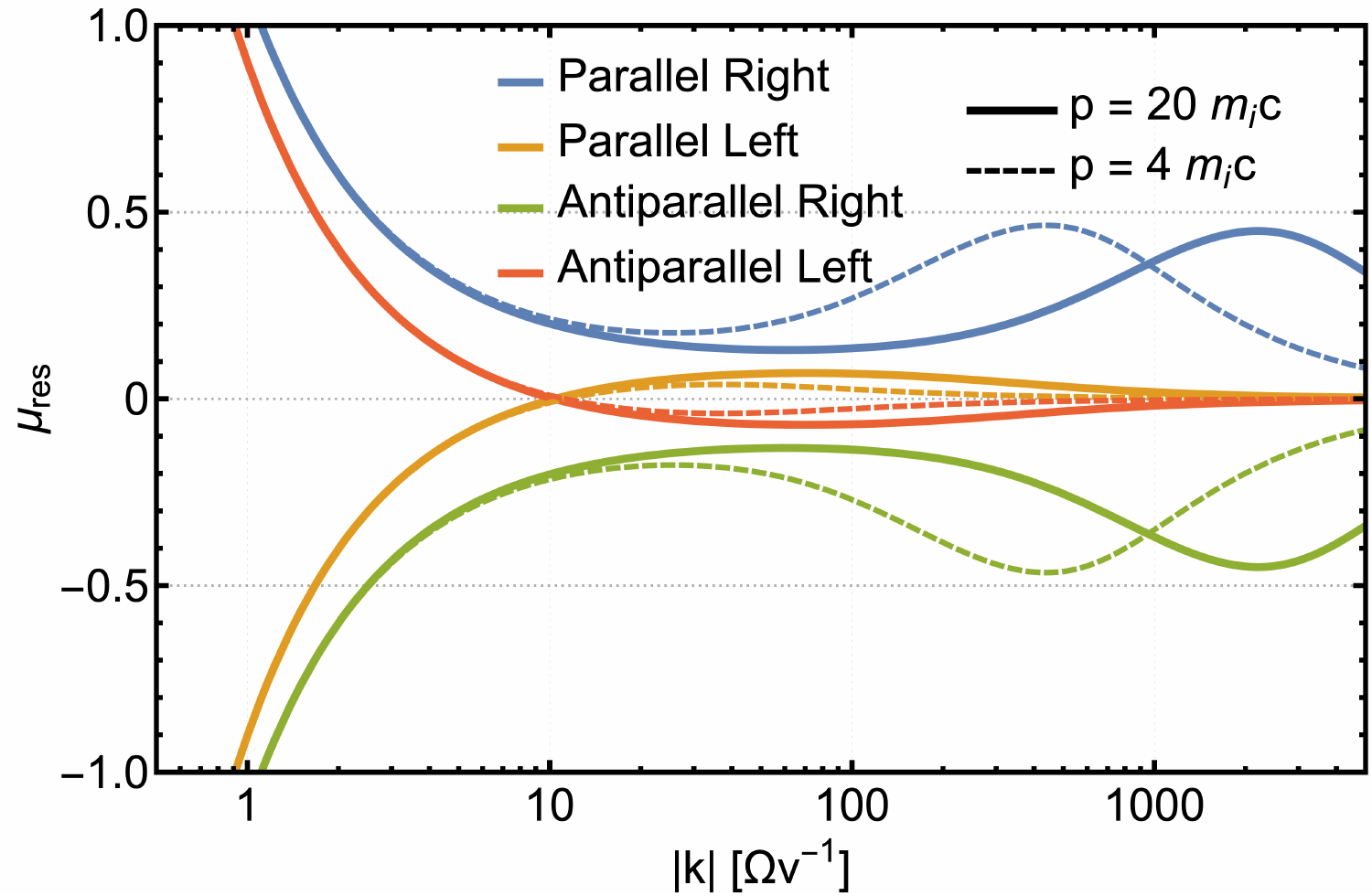}
\caption{Resonant pitch-angle cosine $\mu_{\rm res}$ as a function of wavenumber $k$ for a given particle momentum $p$. The resonance condition (Eq.\ \eqref{eqn:mures}) splits into four separate conditions depending on the propagation direction (parallel/antiparallel) and polarization handedness (right/left) of the wave. We show the resonance conditions for positively charged CRs with momentum $p= 20\ m_{\rm i} c$ (solid lines) and $p=4\ m_{\rm i} c$ (dashed lines), and the wavenumber $k$, presented in units of $\Omega(p)/v(p)$, is scaled separately for both values of momentum. The momentum $p$ and pitch-angle cosine $\mu$ determine which types of waves a particle may resonate with. For most configurations, a single particle may simultaneously resonate with more than one type of wave and at more than one wavenumber -- typically a parallel-propagating wave and an antiparallel-propagating wave of opposite polarization. }
\label{fig:mures}
\end{figure}

The foundation of the streaming instability, and of the subsequent pitch-angle diffusion, is the mechanism by which CRs scatter in electromagnetic fluctuations. Charged particles propagating in a magnetized plasma execute gyromotion when moving with perpendicular velocity to the background magnetic field ($\boldsymbol{B}_0 = B_0 \hat{x}$ throughout this work). This behavior allows particles to strongly interact with transverse waves, such as Alfv\'en waves, under certain conditions. Wave-particle pairs are said to be in resonance when they satisfy the gyroresonance condition
\begin{align}\label{eqn:res}
\omega - v_x k &= \Omega,
\end{align}
where $\omega$ and $k$ are the frequency and parallel wavenumber of the wave, and $v_x$ and $\Omega$ are the velocity parallel to the background magnetic field and the relativistic gyrofrequency. This is a special case of a more general set of conditions (e.g., \citealt{2013PhPl...20e5501Z}), where we have assumed transverse electromagnetic waves propagating parallel or antiparallel to the background magnetic field. 

We can rearrange Eq.\  \eqref{eqn:res} to produce the resonant pitch-angle cosine, given particle total momentum $p$ and wavenumber $k$,
\begin{align}\label{eqn:mures}
\mu_{\rm res}(k,p) &=\frac{v_{\rm ph}(k)}{v(p)} - \frac{\Omega_0}{k\gamma(p)v(p)},
\end{align}
where $\gamma(p) = \sqrt{1+ (p/mc)^2}$ is the particle Lorentz factor, $v(p) = p/(m\gamma(p))$ is the total particle velocity, $\Omega_0$ is the nonrelativistic gyrofrequency, $\mu = p_x/p$, and $v_{\rm ph} = \omega/k$ is the wave phase velocity. In Figure \ref{fig:mures} we show $\mu_{\rm res}$ as a function of the absolute value of the wavenumber with ions of fixed particle momentum $p= 20\ m_{\rm i} c$ (solid lines) and $p=4\ m_{\rm i} c$ (dashed lines), and Alfv\'en waves of all four combinations of polarization and propagation direction (the polarization convention is described in appendix \ref{sec:pol}). The phase velocity of waves is obtained by solving the kinetic plasma dispersion relation for parallel/antiparallel-propagating waves in the cold and low-frequency limits (including the ion-cyclotron and whistler regimes around $k \gtrsim 100\ \Omega/v$). The spread of the resonance curves around $\mu = v_{\rm A}/v$ is exaggerated by the choice of the parameters $m_{\rm i}/m_{\rm e} = 100$ (the ion-to-electron mass ratio) and $v_{\rm A} = 0.1 c$ (the Alfv\'en speed), which we adopt in our simulations for reasons described in section \ref{sec:num}. From this representation of the resonance condition (Eq.\ \eqref{eqn:mures}) we note two important features of this wave-particle interaction. The first is that there exists a minimal wavenumber for a given particle momentum $k_{\rm min}(p)$ such that $\mu_{\rm res} = \pm 1$; therefore, waves of longer wavelength cannot satisfy the resonance condition. The second, and more important in the context of this article, is that a given CR is capable of achieving gyroresonance with waves of particular combinations of polarization and propagation direction for limited ranges of the pitch angle.

The transverse electric fields of Alfv\'en waves vanish in the reference frame moving with the wave phase velocity $v_{\rm ph}$. Consequently, CRs elastically scatter on magnetostatic perturbations in the wave frame, changing their propagation directions without exchanging energy (pure pitch-angle scattering). Thus, the action of a monochromatic wave packet is to scatter resonant particles along the trajectory described by 
\begin{align}\label{eqn:constvel}
\gamma_{\rm ph}^2 (p_{\parallel} (t) - \gamma m v_{\rm ph} )^2 + p_{\perp}^2 (t) &= \text{constant,}
\end{align}
where $\gamma_{\rm ph} = (1-v_{\rm ph}^2/c^2)^{-1/2}$ is the Lorentz factor associated with a boost into the wave frame traveling with phase velocity $v_{\rm ph}$. This semiellipse in momentum space describes a constant energy surface when viewed in the wave frame. As scattering ensues in the laboratory frame, waves grow or damp predominantly in exchange for the free momentum associated with the streaming of CRs, while the particle energies remain nearly constant as long as $v \gg v_{\rm ph}$ \citep{2005ppfa.book.....K}. 

\subsection{Distribution Functions \& Instability}\label{sec:resdisp}

To elicit the behavior of the CR streaming instabilities, we utilize two classes of CR distribution functions. The first is the gyrotropic ring distribution, 
\begin{align}\label{eqn:ring}
f_{\rm ring}(p, \mu) &= \frac{n_{\rm cr}}{2\pi p^2} \delta(p - p_{\rm 0}) \delta(\mu - \mu_0),
\end{align}
where the input parameters $p_0$ and $\mu_0$ define the unique total momentum and pitch-angle cosine that all CRs share. The ring distribution has been used to study the growth of nonlinear waves in the solar wind \citep{1997JGR...10222365G,2002JGRA..107.1367S}. Here, we use it as a simple model to study the effects of particle scattering in self-generated quasimonochromatic spectra of small and large-amplitude waves.

By the definition of the ring distribution, the perpendicular momenta of all CRs are randomly oriented in gyrophase but equal in magnitude. These properties ensure that all CRs are initially distributed within the same resonant bands, determined by Eq.\  \eqref{eqn:res}. In this work we study ring distributions with super-Alfv\'enic drift $\mu_0 v(p_0) > v_{\rm A}$, which excite parallel right-handed and antiparallel left-handed waves. We derive the dispersion relation for waves in the presence of the ring distribution in Appendix \ref{sec:disp}.

The second distribution we will utilize is the more familiar power-law distribution with an additional bulk drift along the background magnetic field ($\hat{x}$ axis). In the frame in which the CRs appear isotropic, the distribution takes the form
\begin{align}\label{eqn:pow}
f''_{\rm PL} (p'', \mu'')  &= \frac{n_{\rm cr}}{4\pi}\frac{(\alpha - 3)\Theta_{\rm min} \Theta_{\rm max}}{(p''_{\rm min})^{3-\alpha} - (p''_{\rm max})^{3-\alpha}}(p'')^{-\alpha},
\end{align}
where doubly primed quantities refer to measurements made in the isotropic CR frame, $\alpha$ is the power law index, and $p''_{\rm min}$ and $p''_{\rm max}$ are input parameters specifying the minimum and maximum CR momenta, respectively. The latter are encoded by the Heaviside step functions $\Theta_{\rm min}\equiv \Theta(p'' - p''_{\rm min})$ and $\Theta_{\rm max}\equiv \Theta(p''_{\rm max} - p'')$. Since the distribution function is Lorentz invariant, the lab-frame distribution is obtained using the Lorentz transformation of momentum $p'' = \sqrt{\gamma_{\rm dr}^2(\mu p - m v_{\rm dr} \gamma/c)^2 + (1-\mu^2)p^2}$, where $v_{\rm dr}$ and $\gamma_{\rm dr}$ are the CR drift velocity and associated Lorentz factor.

 The power-law distribution produces a substantially smaller growth rate than the ring distribution with a comparable CR flux $n_{\rm cr}v_{\rm dr}$. As long as the CR density is sufficiently small ($n_{\rm cr} \ll n_{\rm i}$), the real part of the dispersion relation is approximately unaffected and one can reduce the problem of finding the linear growth rate to the solution of the resonant integral 
\begin{align}
\begin{split}\label{eqn:zwei}
\Gamma_{\rm cr}(k) &= \pi^3 q^2\frac{v_A^2}{c^2}\iint \delta(\omega(k) - \Omega(p) - k \mu v(p))  \\
			& \ \ \ \ \ \ \ \ \ \ \ \ \ \ \ \times A[f_{\rm PL}] v(p) p^2 (1-\mu^2)dp d\mu,
\end{split}\\
A[f] &\equiv \frac{\partial f}{\partial p} + \bigg(\frac{kv(p)}{\omega}- \mu\bigg)\frac{1}{p}\frac{\partial f}{\partial \mu}, \label{eqn:a}
\end{align}
where the Dirac delta function encodes the resonance condition \citep{1969ApJ...156..445K,2017PhPl...24e5402Z}. The quantity $(1-\mu^2)vp^2$ in the integrand is strictly nonnegative; therefore, the sign of $A[f_{\rm PL}](k)$ determines whether waves grow or damp via resonant interactions with a given power-law distribution function.

If we assume that CRs form an isotropic power-law distribution in a frame moving (also called ``drifting" or ``streaming") with positive velocity $v_{\rm dr}$ along the background magnetic field, then only the right- and left-handed parallel-propagating modes will have positive growth rates.  If one makes the additional simplifying assumptions that $\omega \ll \Omega(p)$, $v_{A} \lesssim v_{\rm dr} \ll c$, and $p''_{\rm max}\rightarrow\infty$, then the unstable growth rate can be derived as\footnote{The factor of 1/2 arises because we consider the growth of right- and left-handed modes separately, whereas typically the combined growth rate is quoted in the literature.}
\begin{align}\label{eqn:kuls}
\begin{split}
\Gamma_{\rm cr}^{\rm lin}(k) &= \frac{1}{2}\frac{\pi}{4} \frac{\alpha -3}{\alpha -2} \frac{n_{\rm cr} }{n_{\rm i}}  \Omega_0  \bigg( \frac{v _{\rm dr}}{v_{\rm A}} - 1 \bigg) \\
&\times \begin{cases} 
      \bigg(\frac{p_{\rm k}(k)}{p_{\rm min}}\bigg)^{3-\alpha} & p_{\rm k}(k) > p_{\rm min} \\
       \bigg(\frac{p_{\rm k}(k)}{p_{\rm min}}\bigg) & p_{\rm k}(k) \le p_{\rm min}
\end{cases},
\end{split}
\end{align}
where $p_k = m_{\rm i}\Omega_0/|k|$, $\Omega_0$ is the nonrelativistic ion gyrofrequency, and $p_{\rm min}\approx p''_{\rm min}$. Under these assumptions the growth rates for parallel-propagating left/right-circularly polarized waves become degenerate, $\Gamma_{\rm cr}^{\rm PR} = \Gamma_{\rm cr}^{\rm PL} = \Gamma_{\rm cr}^{\rm lin}$, and the resulting combined wave is linearly polarized with total growth rate $2\Gamma_{\rm cr}^{\rm lin}$. It was noted by \cite{1971ApL.....8..189K} that Eq.\  \eqref{eqn:kuls} should be multiplied by $\gamma_{\rm dr}$ and $k$ should be replaced by $\gamma_{\rm dr} k$ to obtain a better approximation when $v_{\rm dr} \sim c$. However, this correction to the growth rate does not capture the dissolution of the left/right-handed degeneracy that occurs when large drift velocities are considered, which we now discuss.

\begin{figure}[t]
\centering\includegraphics[width=\linewidth,clip=true]{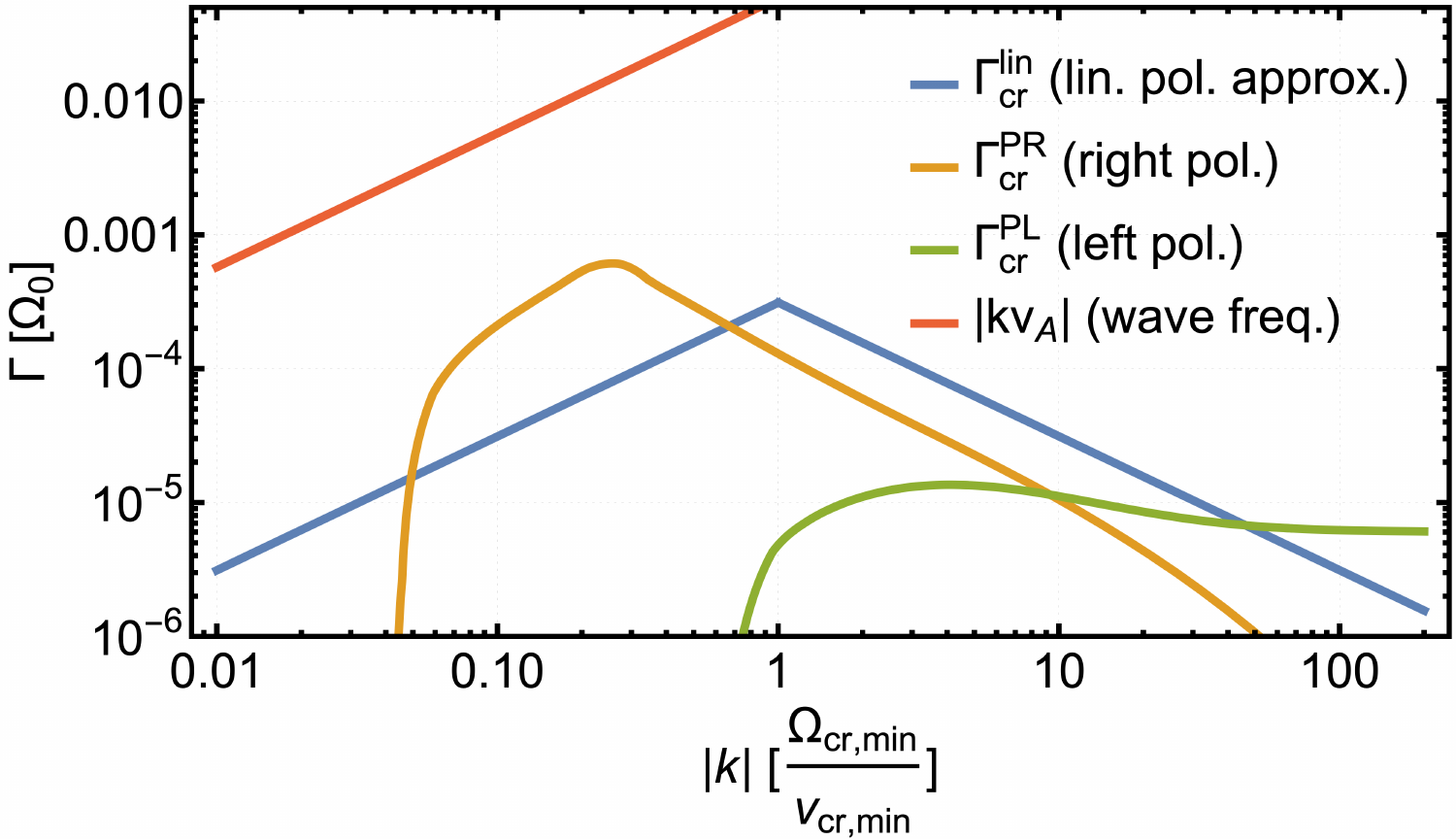}
\caption{Linear growth rates of the gyroresonant instability for a power-law CR distribution with large drift velocity. We compare the growth rate of Eq.\ \eqref{eqn:zwei} (orange and green) against the simplified growth rate of Eq.\ \eqref{eqn:kuls} (blue) for the highly anisotropic parameters of simulation Hi1 (see Table \ref{tab:resultspl}). For large CR anisotropy, growth of right-handed modes ($\Gamma_{\rm cr}^{\rm PR}$) dominates over growth of left-handed modes ($\Gamma_{\rm cr}^{\rm PL}$) . The fastest-growing wavenumber for $\Gamma_{\rm cr}^{\rm lin}$ is $\Omega_{\rm cr,min}/v_{\rm cr,min}$, where ``min" refers to the lowest energy CRs in the distribution. We show the real part of the Alfv\'en dispersion relation $\omega = kv_{\rm A}$ for comparison (red).}
\label{fig:anagrowth}
\end{figure}

Numerical integration must be used to solve Eq.\  \eqref{eqn:zwei} under less stringent assumptions. In Figure \ref{fig:anagrowth} we show the growth rate as a function of wavenumber $k$ using Eqs.\ \eqref{eqn:zwei} and \eqref{eqn:kuls}, selecting parameters suitable for the particle-in-cell simulations of this study. We use a large CR drift velocity to illustrate the degree to which the full integral Eq.\ \eqref{eqn:zwei} can diverge from the approximate formula Eq.\ \eqref{eqn:kuls}. As the initial anisotropy is increased, a greater number of CRs resonate with right-handed waves (orange line) and fewer with left-handed waves (green line). A larger ratio of right-hand resonant to left-hand resonant particles enhances the growth rates of the right-handed modes at the expense of the left-handed mode rates, while the deformation of the CR distribution function shifts the fastest-growing modes of the right- and left-handed polarizations to smaller and larger wavenumbers, respectively. These effects break the growth-rate degeneracy between the right- and left-handed modes, permitting the growth of circularly polarized waves, rather than linearly polarized waves.

\subsection{The Quasilinear Theory \& The 90 degree problem}

The quasilinear theory (QLT; \citealt{1966ApJ...146..480J}) predicts that the pitch-angle diffusion coefficient $D_{\mu'\mu'}$ vanishes as $\mu' \rightarrow 0$, where $\mu'$ is the particle pitch-angle cosine measured in the wave frame, owing to the increasingly small-amplitude of waves as $k\rightarrow \infty$. These \emph{resonance gaps} are the basis of the so-called ``90 degree problem" that threatens to prevent CRs from reaching isotropy ($v_{\rm dr} \approx v_{\rm A}$). Theoretical efforts to alleviate the 90 degree problem typically seek either to append nonresonant scattering effects to the QLT or attempt to capture the essence of nonlinear effects (i.e.,\ the deviations of particles away from their unperturbed gyromotions) in some way. 

The adiabatic mirroring mechanism \citep{1978PhFl...21..347J,1981A&A....98..161A} is an example of a nonresonant effect. A gradient in the total magnetic field produces a ``mirror force" on particles $\boldsymbol{F}_{M} = -M \nabla B$, where $M \equiv \gamma^2 m v_\perp^2/2B$ is the relativistic magnetic moment of the particle. Particles become trapped in the magnetic mirror if the wave-frame pitch-angle cosine becomes smaller than
\begin{align}\label{eqn:mumirr}
\mu'_M &\approx \frac{1}{\sqrt{2}}\bigg(\frac{\delta B}{B_0}\bigg)_M,
\end{align}
where $(\delta B/B_0)_M$ is the peak amplitude of the transverse fluctuation that forms the magnetic mirror \citep{2001ApJ...553..198F}. If resonant scattering can reduce the pitch-angle cosine of a particle to $\mu' \le \mu'_M$, then said particle can reverse direction with respect to the wave, assuming the magnetic moment $M$ is approximately conserved. Such a process would allow CRs to bypass $\mu'\sim 0$ ($\mu\sim v_{\rm ph}/v$ in the laboratory frame) without the need for quasilinear scattering, thereby solving the 90 degree problem. 

If we assume that the mirroring fluctuation is sinusoidal (as was done by \citealt{2001ApJ...553..198F} to obtain Eq.\ \eqref{eqn:mumirr}), then we can derive a time scale for the mirroring process. The effective potential due to the mirror force $\boldsymbol{F}_M$ causes quasiharmonic motion for trapped particles with oscillation frequency
\begin{align}\label{eqn:bounce}
\Omega_{M} &\sim \sqrt{\frac{\gamma v_\perp^2 k_M^2}{2} \bigg(\frac{\delta B}{B_0}\bigg)_M^2},
\end{align}
where $k_M$ is the wavenumber of the mirroring fluctuation. We take $k_M \approx k_{\rm max}$ because the interaction of the fastest-growing unstable mode $k_{\rm max}$ with the randomly phased adjacent modes will typically produce the largest fluctuations in the field envelope, $B = \sqrt{B_0^2 + B_y^2 + B_z^2}$, and thus dominate the mirror force $\boldsymbol{F}_M$.

Nonlinear theories attempt to modify the resonance function away from the sharp delta function form taken by Eq.\ \eqref{eqn:res}. Accounting for the drift of particles away from simple gyromotion about the background magnetic field results in resonance broadening \citep{1966PhFl....9.1773D,1981A&A....98..161A,Shalchi:2009vg}. These effects become increasingly strong as the wave amplitude $\delta B$ becomes larger, allowing particles to interact with a band of modes of finite width or, equivalently, allowing waves to interact with a range of pitch-angle cosines. Extending the influence of wave-particle interactions can facilitate diffusion across $\mu' = 0$, since particles can essentially skip over the region of the wave spectrum where the power vanishes.

Finally, although the aforementioned solutions are relevant, the 90 degree problem can be ameliorated within QLT itself by merely relaxing the magnetostatic approximation that is typically employed ($\omega/k \sim 0$). In particular, \citet{1989ApJ...336..243S} demonstrated that the resonance gaps associated with the vanishing of $D_{\mu\mu}$ are replaced by nonzero values when both parallel \emph{and} antiparallel waves are present with adequate power. This is apparent from Figure \ref{fig:mures} -- if a broad spectrum of all four propagation-polarization combinations are present, then particles can access the entire range of $-1 \le \mu \le 1$ by quasilinear pitch-angle scattering alone.

\subsection{Saturation Amplitudes \& Relaxation Time Scales}\label{sec:physsat}

The expression for the linear growth rate (Eq.\  \eqref{eqn:zwei}) dictates that the magnetic field perturbations will stabilize under the condition that the resonant integral over $A[f]$ vanishes. The excitation of waves is fueled by the momentum extracted from the cosmic ray distribution as gradients in momentum and pitch angle are flattened, reducing $A[f]$ to zero everywhere along the path of the resonant integral. Saturation of a given wave mode occurs locally in the Fourier $k$ space -- an unstable wave will cease to grow when particles are scattered into local isotropy around the associated resonant band. It follows that we can distinguish between saturation of the linear growth phase and the \emph{total saturation} of instability. The former occurs when the fastest-growing mode can no longer grow at the predicted linear rate, while the latter corresponds to the total depletion of free momentum in the CR distribution ($v_{\rm dr} = v_{\rm ph}$).

The preceding discussion suggests that we can estimate the amplitude of the fastest-growing mode at any point in the lifetime of instability by equating the change in the CR momentum density $\Delta p_{\rm cr} = m_{\rm i} n_{\rm cr} \gamma_{\rm cr} (v_{\rm dr,0} - v_{\rm dr}(t))$ to the change in the wave momentum density $\delta B^2 / 4\pi v_{\rm ph}$ (neglecting the contributions of initial seed waves):
\begin{align}\label{eqn:sat}
\bigg(\frac{\delta B(t)}{B_0}\bigg)^2  &\approx \gamma_{\rm cr}\frac{n_{\rm cr}}{n_{\rm i}}\frac{v_{\rm ph}}{v_{\rm A}}\frac{\big(v_{\rm dr,0} - v_{\rm dr}(t)\big)}{v_{\rm A}},
\end{align}
where $\gamma_{\rm cr}$ is a characteristic Lorentz factor of the CR (e.g., the mean value along the resonant contour), $v_{\rm dr,0}$ is the initial CR drift velocity, and $v_{\rm dr}(t)$ is the drift velocity at time $t$. If the linear phase is allowed to progress for a duration of several inverse growth rates $\Gamma_{\rm cr}^{-1}$, then the fastest-growing mode $k_{\rm max}$ will dominate the power spectrum of waves and the resulting dynamical evolution of the particles. Under such conditions, Eq.\  \eqref{eqn:sat} will approximately describe the time dependence of the fastest-growing mode amplitude if $v_{\rm dr}(t)$ is known (or vice versa). The nonlinearities of the coupled evolution equations for the CR distribution function and wave spectrum resist analytical time-dependent solutions. However, reasonable estimates of the linear phase saturation amplitudes and the associated drift velocities can be made if the predominant physics of wave-particle interactions is known. The appropriate physical mechanisms differ between the CR distributions considered here. 

The essence of CR dynamics in the spectrum excited by a ring distribution (Eq.\  \eqref{eqn:ring}) is captured by the interaction of particles with a single transverse wave. The forces generated by the periodic electromagnetic fields of a circularly polarized wave form an effective potential well in which resonant particles can become trapped, oscillating with frequency
\begin{align}\label{eqn:trapfreq}
\Omega_{\rm trap} &= \sqrt{\frac{\delta B}{B_0}kv_\perp \Omega},
\end{align}
where $v_\perp$ is the particle velocity in the direction perpendicular to the background magnetic field \citep{1971JGR....76.4463S}. The trapping frequency sets a limiting time scale over which the instability can grow waves because after $t \sim \Omega_{\rm trap}^{-1}$ particles will isotropize with respect to the effective potential well. Thus, we can derive the approximate saturation amplitude of the fastest-growing mode by setting the trapping frequency equal to the instability growth rate $\Omega_{\rm trap} \approx \Gamma_{\rm ring}$. This procedure predicts the linear phase saturated wave amplitude for trapping to be
\begin{align}\label{eqn:beamsat}
\bigg(\frac{\delta B}{B_0}\bigg)_{\rm trap} &\approx \bigg( \frac{n_{\rm cr}}{n_{\rm i}}\bigg)^{2/3} \bigg( \frac{v_\perp(v_{\rm dr,0} - v_{\rm ph})}{v_{\rm ph}^2}\bigg)^{1/3},
\end{align}
where we have used the growth rate formula for the ring distribution derived by \cite{2002JGRA..107.1367S} and have neglected constants of order unity. 

Pitch-angle diffusion sets in only when resonant particles are able to stochastically interact with many wave modes of relevant amplitudes \citep{2017ApJ...834..161S}. The deflection of particle motion caused by one mode has the effect of untrapping it from the others. For this reason $\Omega_{\rm trap}^{-1}$ is no longer the relevant saturation time scale for instabilities arising from power-law CR distributions. Instead, the physical mechanism that drives saturation is the transfer of CRs from one resonance band into another via resonant scattering. For simplicity, we adopt the resonant scattering rate of QLT, given by
\begin{align}\label{eqn:qltscat}
\nu_{\rm QLT}(k) &= \frac{\pi}{2} \bigg(\frac{\delta B_k}{B_0}\bigg)^2 \Omega,
\end{align}
where $\delta B_k$ is the amplitude of the fluctuations with wavenumber $k$ \citep{1966ApJ...146..480J,1969ApJ...156..445K,1975MNRAS.172..557S}. Making the assumption that $\delta B_{k_{\rm max}} \approx \delta B$, we set the growth rate (Eq.\  \eqref{eqn:kuls} with relativistic correction) equal to the appropriate frequency, $\nu_{\rm QLT} \approx \Gamma_{\rm cr}^{\rm lin}$, to arrive at
\begin{align}\label{eqn:plsat}
\bigg(\frac{\delta B}{B_0}\bigg)_{\rm diff} &\approx \sqrt{\frac{1}{8}\gamma_{\rm dr}\gamma_{\rm cr} \frac{n_{\rm cr}}{n_{\rm i}} \bigg(\frac{v_{\rm dr}}{v_{\rm A}} - 1\bigg)},
\end{align}
assuming the power-law index $\alpha =4$. Here we used the relativistic correction to the growth rate suggested by \cite{1971ApL.....8..189K} because it is appropriate for the simulations in this work (as demonstrated in appendix \ref{sec:disp}). In general, care should be taken in selecting the appropriate form of the growth rate.

The conservation of total momentum (Eq.\ \eqref{eqn:sat}) links the growth of the wave amplitude to the decline of the CR drift velocity. The CR drift velocity that corresponds to a wave amplitude of $(\delta B/B_0)_{\rm diff}$ can be estimated by inserting Eq.\ \eqref{eqn:plsat} into Eq.\  \eqref{eqn:sat}. If we define $t_{\rm sat}^{\rm lin}$ as the time at which $(\delta B/B_0) = (\delta B/B_0)_{\rm diff}$, then the linear saturation velocity of the power-law distribution instability, $v_{\rm diff} \equiv v_{\rm dr} (t_{\rm sat}^{\rm lin})$, reduces to
\begin{align}\label{eqn:plsatvel}
v_{\rm diff} &\approx \frac{\gamma_{\rm dr}v_{\rm A} + v_{\rm dr,0}(8 - \gamma_{\rm dr})}{8},
\end{align}
where we have used $v_{\rm ph} = v_{\rm A}$. An analogous velocity $v_{\rm trap} \equiv v_{\rm dr} (t_{\rm sat}^{\rm lin})$ can be obtained for ring distributed CRs. A precise determination (to within the approximations made herein) of the saturation time $t_{\rm sat}^{\rm lin}$ requires knowledge of the initial wave amplitude. These seed waves can arise from, for example, thermal fluctuations of the background plasma \citep{2014PhPl...21c2109Y,2015PhPl...22g2108S} or the turbulent cascade. In general, a reasonable expectation is that $t_{\rm sat}^{\rm lin}$ is a few times the inverse growth rate $\Gamma_{\rm cr}^{-1}$.

Disruption of the CR pitch-angle distribution by the fastest-growing mode precipitates the end of the linear growth phase. Unstable growth will continue in other modes at slower rates as the CR distribution function adjusts to the presence of resonant waves. The remaining particle momentum will ultimately be absorbed by the magnetic field as parallel-propagating CRs preferentially scatter toward the $\mu = -1$ direction, resulting in the total saturation of the instability across a spectrum of waves. An approximate upper bound on the transverse magnetic-field energy $(\delta B/B_0)_{\rm tot}$ corresponds to realizing complete wave-frame isotropy of CRs, i.e., setting $v_{\rm dr} = v_{\rm A}$ in Eq.\  \eqref{eqn:sat}.

We have stated that the relevant physics that give rise to the evolution of the CR distribution are the resonant scattering interaction at $\mu >  v_{\rm A}/v + \mu'_{M}$, the nonresonant mirror interaction at $ v_{\rm A}/v - \mu'_{M} \le \mu \le  v_{\rm A}/v + \mu'_M$, then the resonant scattering interaction again at $\mu <  v_{\rm A}/v - \mu'_{M}$. If we assume a power-law CR distribution with small initial anisotropy then the excited wave spectrum will consist of parallel-propagating waves with nearly equal power in the right- and left-hand circularly polarized components. In such a spectrum, the time scale for the relaxation of the CR distribution and total saturation of instability is set by the longer of the resonant scattering and mirroring time scales, $t_{\rm sat}^{\rm tot} \sim \max ({t_\mu, t_M}$). We now estimate the resonant scattering time scale $t_\mu$ and the mirroring time scale $t_M$.

The stochastic process of pitch-angle scattering suggests the construction of a mean free time that describes the typical time scale for CRs to resonantly scatter from some fiducial pitch angle down to the mirroring region (Appendix \ref{sec:relax}),
\begin{align}
t_{\mu} &= \frac{3}{8}\int_{\mu_{\rm M}}^{\mu_0}d\mu \frac{(1-\mu^2)^2}{D_{\rm \mu \mu}} \\
		  &\approx \frac{3}{4\pi} \bigg(\frac{\delta B}{B_0} \bigg)_{\rm diff}^{-2} \Omega^{-1} C_\mu, \label{eqn:tmu}
\end{align}
where $D_{\mu \mu} = (1-\mu^2) \nu_{\rm QLT}/2$ is the quasilinear diffusion coefficient for pitch-angle scattering, $\mu_0 \equiv \mu_{\rm res}(k_{\rm max}, p_{\rm cr})$ is the resonant pitch-angle cosine (Eq.\  \eqref{eqn:mures}) of the fastest-growing mode $k_{\rm max}$ for a CR with typical momentum $p_{\rm cr}$, $\mu_{M}$ is the pitch-angle cosine at which magnetic mirroring becomes the dominant process (Eq.\ \eqref{eqn:mumirr} represented in the laboratory frame), and the factor $C_\mu$ (Eq.\  \eqref{eqn:cmu}) depends on the shape of wave spectrum. For the simulations presented in this work, we have $C_\mu \sim 10$.

The time scale for magnetic mirroring is simply $t_M = \pi \Omega_M^{-1}$, where $\Omega_M$ is given by Eq.\  \eqref{eqn:bounce}. It is the duration over which a trapped particle reverses its direction in the magnetic mirror. Taking the ratio of $t_M$ and $t_\mu$, we have
\begin{align}
\frac{t_M}{t_\mu} &= \frac{4\pi^2}{3} \frac{1}{\sqrt{\frac{\gamma_{\rm cr}}{2}\big(1 - \frac{1}{2}(\frac{\delta B}{B_0})^2\big)}}\bigg(\frac{\delta B}{B_0}\bigg) \frac{1}{C_\mu} \\
 			   &\approx \frac{4\pi^2}{3\sqrt{2}} \frac{1}{(\mu_0 - \frac{1}{\sqrt{2}}\frac{\delta B}{B_0})\sqrt{1 - \frac{1}{2}(\frac{\delta B}{B_0})^2}}\bigg(\frac{\delta B}{B_0}\bigg)^2 ,
\end{align}
where we have included only the dominant term of $C_\mu$ (Appendix \ref{sec:relax}) and assumed that $\mu_0 v_{\rm cr} \gg v_{\rm A}$ and $\gamma_{\rm cr}=2$ in the second line. This ratio suggests that the CR distribution relaxation process will be dominated by the resonant-scattering time scale unless the wave amplitude becomes sufficiently large that the pitch angle required for mirroring becomes comparable to the typical pitch angle of resonant particles, $\mu_M \approx \mu_0$. Assuming that $\mu_0 = 0.5$ and $v_{\rm A} = 10^{-4} c$, the time scales become equal around $\delta B/B_0 \approx 0.2$ for $\gamma_{\rm cr} =2$, while larger mirror amplitudes are required for CRs with greater energy content ($\gamma_{\rm cr} > 2$). 

If we now change focus to CR distributions with large anisotropy then the relaxation process is complicated further by the lack of left-handed modes in the wave spectrum. Under conditions of extreme anisotropy, mirroring alone is not sufficient for providing efficient passage of CRs beyond the $\mu \sim v_{\rm A}/v$ resonance gap. Although the right-handed part of the wave spectrum will be able to scatter CRs down to small $\mu$ and mirroring in the gradient of the total magnetic field will bring CRs down to $\mu \approx v_{\rm A}/v - \mu'_M$, there will be little, if any, power in the left-handed waves needed to scatter CRs to $\mu \sim -1$. The result is a flat CR distribution in the $\mu \gtrsim 0$ region of momentum space, with a bulk drift velocity $v_{\rm dr} \approx 0.5 c$. Eventually the buildup of CRs  at small $\mu$ should generate the waves required to achieve total isotropy, but it is not clear \emph{a priori} what the associated time scale would be.

\section{Numerical Methods}\label{sec:num}

We utilize the relativistic electromagnetic particle-in-cell (PIC) code Tristan-MP \citep{2005AIPC..801..345S}. This code has been extensively used to simulate particle acceleration in collisionless shocks \citep{2008ApJ...673L..39S,2009ApJ...698.1523S,2009ApJ...707L..92S,2015PhRvL.114h5003P}, the CR current-driven instabilities in SNRs \citep{2009ApJ...694..626R,2010ApJ...717.1054R}, the generation of pulsar magnetospheres \citep{2015ApJ...801L..19P}, and more. The PIC method allows us to resolve the plasma physics down to electron kinetic scales.

We perform one-dimensional (1D3V) simulations with periodic boundary conditions to explore the linear and nonlinear phases of the CR streaming instability in the initial rest frame of the background plasma. The plasma consists of equal-temperature Maxwellian-distributed ions and electrons with reduced mass ratio $m_{\rm i}/m_{\rm e} = 100$ and ion thermal velocity $v_{\rm th,i} \equiv \sqrt{k_{\rm B} T/m_{\rm i} } =10^{-2} c$. The electric and magnetic fields are initialized with zero amplitude with the exception of a uniform background magnetic field $\boldsymbol{B}_0 = B_0 \hat{x}$, such that the Alfv\'en speed is $v_{\rm A} = 0.1c$. The speed of light $c$ is set to $0.45 $ cells per timestep and the electron skin depth $c \omega_{\rm pe}^{-1} = 10$ cells, where $\omega_{\rm pe}$ is the electron plasma frequency.

Reduction of the ion-to-electron mass ratio affects both the time scales and the length scales of the problem at hand. Each of the relevant time scales (i.e., $\Gamma_{\rm cr}^{-1}$, $\nu_{\rm QLT}^{-1}$, $t_{\mu}$, and $t_{\rm M}$) and the length scale $k_{\rm max}^{-1}$ grow in inverse proportion to the ion gyrofrequency. Therefore a reduced ion mass allows for shorter computation times and smaller simulation domains when $m_{\rm e}$ is fixed. In systems with order-unity mass ratios, CRs of positive and negative charge will grow waves at comparable wavelengths. This would potentially have a nontrivial effect on the evolution of the system, since individual CRs would scatter on waves generated by both species. Our order-of-magnitude reduction of the mass ratio ($m_{\rm i}/m_{\rm e} = 100$) grants us an order-of-magnitude reduction of the computation time of our simulations without qualitatively impacting the CR dynamics, since we maintain a sufficiently large separation between electron and ion length scales.

Cosmic rays are initialized depending on the chosen distribution function. In the case of the ring distribution, CRs are assigned momenta according to Eq.\  \eqref{eqn:ring}. An additional population of cosmic ray electrons (CRe) with zero perpendicular momentum is created to neutralize the charge and current of the CR ions. In the case of the power-law distribution, we initialize CR ions and electrons with a $p^{-4}$ power-law in the background plasma rest frame with Lorentz factors in the range $\gamma = [2,10]$. We then Lorentz boost each individual CR and CRe by $\gamma_{\rm dr}$ in the $\hat{x}$ direction. Note that this procedure does not maintain the Lorentz invariance of the distribution function, i.e., $\gamma_{\rm dr} \neq (1 -v_{\rm dr}^2/c^2)^{-1/2}$, and thus does not produce an accurate representation of Eq.\  \eqref{eqn:pow} in general \citep{2013A&A...558A.133M,2015PhPl...22d2116Z}. The linear growth rates are modified, owing to a factor of $\gamma''/\gamma$ that is appended to the rest-frame CR distribution function $f_{\rm PL}''$ (Eq.\  \eqref{eqn:pow}) by the momentum transformation of individual particles. The anisotropy of our power-law distributions is reduced compared to Lorentz-invariant distributions of the same parameters, and it follows that the right/left-handed unstable modes grow at slower/faster rates than would be expected  (see Appendix \ref{sec:disp}). This effect does not qualitatively change the results of the simulations presented here, and the modification to the growth rate is accounted for in all subsequent calculations.

The number densities of CR ions and electrons are equal to their background plasma counterparts, $n_{\rm cr} = n_{\rm cre} = n_{\rm i} = n_{\rm e} = n/4$, where $n$ is the total number of particles per cell in the simulation. In order to achieve a low \emph{effective} value of the relative density $(n_{\rm cr}/n_{\rm i})_{\rm eff}$, we reduce the mass $m_{\rm cr}$ of CR ions while holding the charge-to-mass ratio $q_{\rm cr}/m_{\rm cr}$ fixed so that $(n_{\rm cr}/n_{\rm i})_{\rm eff} = m_{\rm cr}/m_{\rm i}$ (and similarly for CR electrons). This procedure preserves the electromagnetic dynamics of CRs and the evolution of instability while enhancing the statistical quality of the CR distribution. 

\begin{table}
\centering
\begin{tabular}{|c|c|cc|}
\hline
Simulation & CR Distribution 	& $L_x$ [cell]  	& $n$ [cell$^{-1}$] \\
\hline
\hline
Gy1		& Gyrotropic Ring 	& 196036		& 1000	\\
Gy2		& Gyrotropic Ring 	& 98018		& 500		\\
Gy3		& Gyrotropic Ring 	& 98018		& 50		\\
Gy4		& Gyrotropic Ring 	& 98018 		& 50		\\	
Gy5		& Gyrotropic Ring 	& 98018 		& 50		\\
\hline
Lo		& Power-Law		&1450000 	& 200 \\
\hline
Med	& Power-Law		&1450000 	& 200 \\
\hline
Hi1		& Power-Law		& $328340$	& 250 \\
Hi2		& Power-Law		& $1450000$	& 100 \\
Hi3		& Power-Law		& $1450000$	& 100 \\

\hline

\end{tabular}
\caption{Simulation-specific properties of the domain and plasma.}\label{tab:prop}
\end{table}

In Table \ref{tab:prop} we list the CR distribution, computational domain size, and particle density $n$ chosen for each simulation. The simulation size is chosen to capture roughly ten wavelengths of the fastest-growing mode at minimum. We choose a variety of particle densities with the goal of minimizing noise while maintaining feasible limits on computational expenses. In Table \ref{tab:resultsgy} we show the CR properties for the gyrotropic ring distribution simulations. These five simulations are numbered by increasing CR density from Gy1 to Gy5, with the other parameters fixed. Table \ref{tab:resultspl} summarizes the properties of the power-law distributed CR simulations. The prefixes Lo, Med, and Hi signify the relative overall CR anisotropy in relation to each other, from low anisotropy to high anisotropy, respectively. Here, ``anisotropy" is controlled by the parameter $\gamma_{\rm dr}$ (and the corresponding $v_{\rm dr}$), with larger $\gamma_{\rm dr}$ corresponding to greater numbers of CRs with $\mu > v_{\rm A}/v$ (we discuss the notion of anisotropy further in section \ref{sec:disc}). Finally, the Hi1-3 simulations are numbered by increasing CR density (increasing growth rate), while the CR densities in simulations Lo and Med are chosen such that their maximal growth rates roughly match that of Hi3.

\begin{figure}[t]
\centering\includegraphics[width=\linewidth,clip=true]{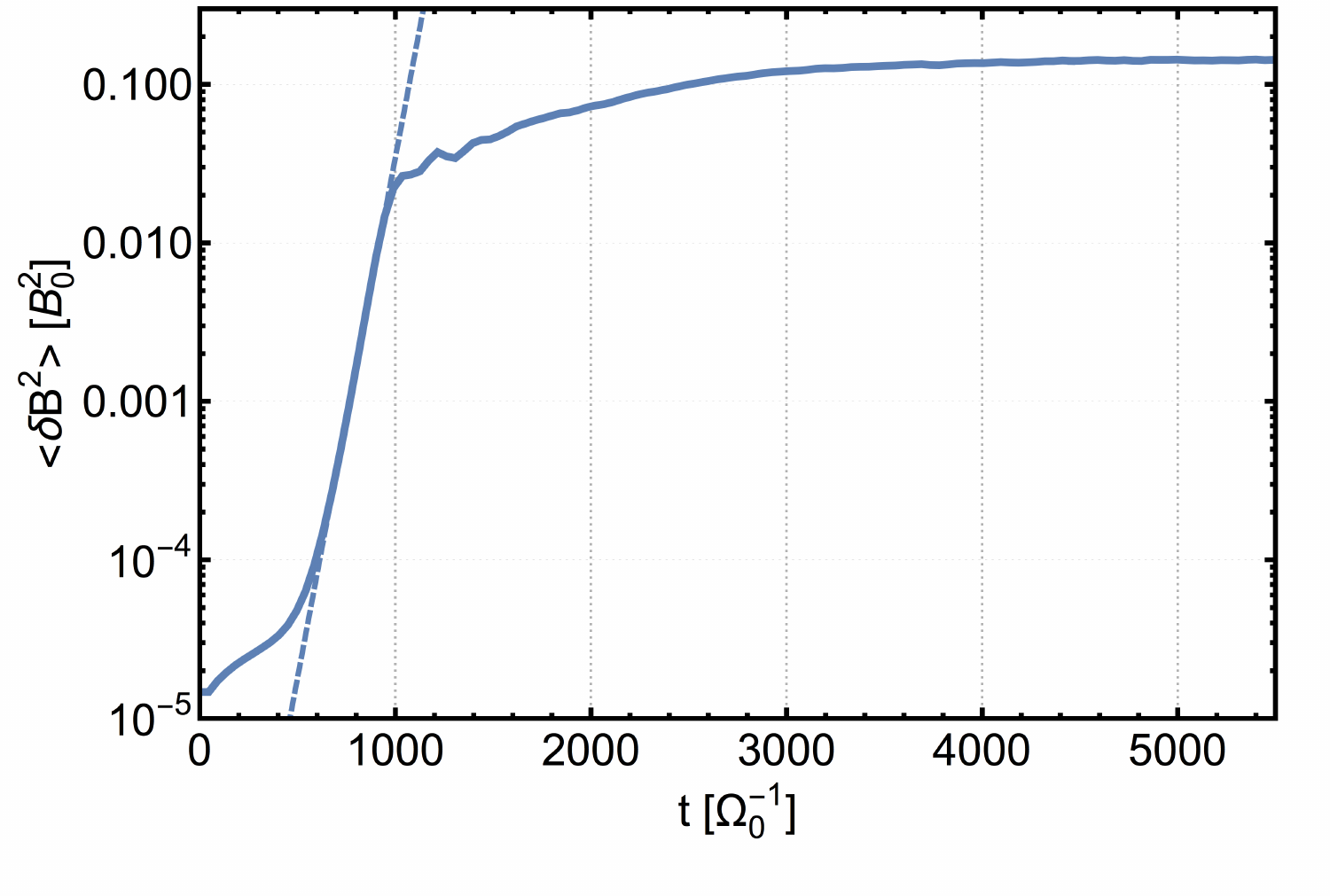}
\caption{Phases of instability, as demonstrated by the growth of the transverse magnetic-field energy in high-density high-anisotropy power-law simulation Hi3. After an initial phase of equilibration, exponential growth begins ($t \sim 500 \ \Omega_0^{-1}$), demarcating the onset of the linear phase of instability. The dotted line shows the exponential development of waves with the growth rate $\Gamma = 7.55\times 10^{-3} \ \Omega_0$ (measured on the total transverse field evolution). By $t \sim 1000 \ \Omega_0^{-1}$ the system transitions to the nonlinear phase of instability as the fastest-growing mode saturates. Growth continues at slower rates in other wave bands, but ultimately these modes cease to grow and the total saturation of instability is obtained ($t \sim 5000 \ \Omega_0^{-1}$).}
\label{fig:gcr}
\end{figure}


\begin{table*}
\centering
\begin{tabular}{|c|ccc|ccccc|}
\hline
Simulation		& $(n_{\rm cr}/n_{\rm i})_{\rm eff}$ 	& $v_{\rm dr}$ [$v_{\rm A}$]	& $\gamma_{0}$ 		& $\Gamma_{\rm meas}$ $[\Omega_0]$	& $\Gamma_{\rm cr}^{\rm PR}$  $[\Omega_0]$ 	& & $k_{\rm meas}$ $[\omega_{\rm pi}c^{-1}]$	& $k_{\rm max}^{\rm PR}$ $[\omega_{\rm pi}c^{-1}]$ \\
\hline
\hline
Gy1		 	& $10^{-4}$				& 5 						& 1.5					& $4.2\times 10^{-2}$				& $4.2\times 10^{-2}$				& & $-1.8\times 10^{-1}$ 					& $-1.7\times 10^{-1}$ \\
Gy2		 	& $10^{-3}$				& 5 						& 1.5					& $9.5\times 10^{-2}$				& $9.5\times 10^{-2}$ 				& & $-1.8\times 10^{-1}$					& $-1.7\times 10^{-1}$ \\
Gy3			& $10^{-2}$				& 5						& 1.5					& $2.1\times 10^{-1}$				& $2.1\times 10^{-1}$				& & $-2.0\times 10^{-1}$ 					& $-2.0\times 10^{-1}$ \\
Gy4		 	& $2\times 10^{-2}$			& 5 						& 1.5					& $2.6\times 10^{-1}$				& $2.6\times 10^{-1}$				& & $-2.5\times 10^{-1}$					& $-2.2\times 10^{-1}$  \\
Gy5			& $5\times 10^{-2}$			& 5						& 1.5 				& $3.6\times 10^{-1}$				& $3.6\times 10^{-1}$				& & $-2.5\times 10^{-1}$					& $-2.8\times 10^{-1}$\\
\hline
\end{tabular}
\caption{CR Parameters \& Results for the Ring Distribution Simulations}\label{tab:resultsgy}
\end{table*}

\begin{table*}
\centering
\begin{tabular}{|c|ccc|ccccc|}
\hline
Simulation& $(n_{\rm cr}/n_{\rm i})_{\rm eff}$ 	& $v_{\rm dr}$ [$v_{\rm A}$] 	& $\gamma_{\rm dr}$ 	& $\Gamma_{\rm meas}$ 	$[\Omega_0]$	& $\Gamma_{\rm cr}$ $[\Omega_0]$	& & $k_{\rm meas}$ $[\omega_{\rm pi}c^{-1}]$ 	& $k_{\rm max}^{\rm PR}$ $[\omega_{\rm pi}c^{-1}]$  \\
\hline
\hline
Lo		& $2\times 10^{-2}$			& 1.4						& 1.021				& $9.6 \times 10^{-3}$				& $4.8\times 10^{-3}$			& & $-5.3\times 10^{-2}$					& $-5.2\times 10^{-2}$ \\
\hline
Med	& $7\times 10^{-3}$			& 2.9						& 1.091				& $8.7 \times 10^{-3}$				& $5.8\times 10^{-3}$			& & $-4.3\times 10^{-2}$					& $-4.1\times 10^{-2}$ \\
\hline
Hi1		& $2\times 10^{-4}$	 		& 7.9 					& 2.25				& $1.5\times 10^{-3}$    				& $6.8\times 10^{-4}$ 			& &	$-1.4\times10^{-2}$					& $-1.4\times10^{-2}$	 \\
Hi2		& $7\times 10^{-4}$			& 7.9						& 2.25				& $3.8\times 10^{-3}$				& $2.4\times 10^{-3}$ 			& &	$-1.4\times10^{-2}$					& $-1.4\times10^{-2}$ \\
Hi3 	 	& $2\times 10^{-3}$	  		& 7.9						& 2.25				& $8.2\times 10^{-3}$ 				& $6.8\times 10^{-3}$			& &	$-1.4\times10^{-2}$					& $-1.4\times10^{-2}$	 \\
\hline

\end{tabular}
\caption{CR Parameters \& Results for the Power-Law Distribution Simulations}\label{tab:resultspl}
\end{table*}

\section{Results}\label{sec:res}

We observe that the evolution of the transverse magnetic energy $\delta B^2/ 8\pi$ generally progresses through four (semi)distinct phases (as embodied by power-law simulation Hi3 in Figure \ref{fig:gcr}): 1.\ Initial Equilibration Phase, 2.\ Linear Instability Phase, 3.\ Nonlinear Instability Phase, and 4.\ Total Saturation Phase. The early stage of the simulation is characterized by a period of thermalization as electromagnetic fluctuations equilibrate with particles, setting the ``noise floor" of the electromagnetic spectrum. The initial production of fluctuations is the physical result of thermal currents produced by the random initialization of the simulated particles \citep{2014PhPl...21c2109Y,2015PhPl...22g2108S}. The amplitude of the early-time noise floor is determined by the input parameters (e.g.,  particles per Debye volume). On longer timescales, the particle and electromagnetic energies grow from numerical heating because of interpolation effects of the PIC simulation grid \citep{1991ppcs.book.....B,2013A&A...558A.133M}. The input parameters for our simulations are chosen such that unstable waves reach significant amplitudes compared to the noise floor. 

The thermal perturbations serve as seed waves for the subsequent linear phase of streaming instability. Exponential growth becomes evident at around $t\sim 500 \ \Omega_0^{-1}$ in Figure \ref{fig:gcr}. After a duration of a few $\Gamma_{\rm cr}^{-1}$, the fastest-growing mode saturates ($t\sim 1000 \ \Omega_0^{-1}$ in Figure \ref{fig:gcr}). The fastest-growing mode reaches an amplitude such that the CR distribution is significantly disrupted from its initial state. CRs are scattered into resonance with other modes, particularly those with $k > k_{\rm max}$, allowing growth to continue outside of the $k_{\rm max}$ band at slower rates compared to $\Gamma_{\rm cr}$. Ultimately, these slower growing modes cease to extract bulk momentum from the CRs, leading to the total saturation of the streaming instability ($t \sim 5000 \ \Omega_0^{-1}$ in Figure \ref{fig:gcr}).

In the following subsections we study the evolution of the simulations beyond the initial equilibration phase by examining the growth rates of the transverse magnetic energy in the linear instability phase, following the change in the wave spectra and CR distribution functions into the nonlinear instability phase, and looking into the final state of the system at total saturation. We highlight the differing behaviors of systems with ring and power-law CRs.

\subsection{Growth Rates}

If the linear growth phase is able to persist for more than a few $\Gamma_{\rm cr}^{-1}$, then the fastest-growing mode will dominate the wave power spectrum. Simple linear regression on the logarithm of the total transverse magnetic-field energy can then be used to measure the growth rate of the fastest-growing mode (e.g.,\ the dashed line in Figure \ref{fig:gcr}). However, this method can be inaccurate because nearby modes with slower growth rates are mixed into the regression. A more direct measurement comes from the spectral representation of the transverse magnetic energy in each mode. 

In Table \ref{tab:resultsgy} we compare the spectrally measured growth rates $\Gamma_{\rm meas}$ for ring distributed CRs against the predicted maximal growth rates $\Gamma_{\rm cr}^{\rm PR}$, where the superscript ${\rm PR}$ refers to the parallel right-handed branch (as opposed to the antiparallel left-handed branch; Appendix \ref{sec:disp}). These measured values come from the temporal profiles of the Fourier-transformed transverse magnetic field. The theoretical maximal growth rates are drawn from the numerical solution to the ring-distribution dispersion relation (Eq.\  \eqref{eqn:ringdisp}). All growth rates observed here are in accordance with the predicted values. 

In Table \ref{tab:resultspl} we compare the spectrally measured maximal growth rates $\Gamma_{\rm meas}$ against the predicted fastest-growing modes $\Gamma_{\rm cr}$ (equation \eqref{eqn:zwei}) for the simulations with power-law distributed CRs. The high-anisotropy simulations (Hi1-3) and intermediate anisotropy simulation Med are dominated by right-handed modes, while the low-anisotropy simulation Lo has significant power in both left- and right-handed modes. As a result, the overall growth rates for the Hi1-3 runs follow $\Gamma_{\rm cr} = \Gamma_{\rm cr}^{\rm PR}$, while the expected growth rate for the Med and Lo runs is $\Gamma_{\rm cr} = \Gamma_{\rm cr}^{\rm PR} +\Gamma_{\rm cr}^{\rm PL}$, where the superscripts ${\rm PR}$ and ${\rm PL}$ refer to parallel-propagating right- and left-hand polarized modes, respectively. Here, the results are generally within a factor of a few of the predicted rates, but  are less satisfactory compared to the ring distribution simulations. The relatively slower growth rates make the power-law simulations more susceptible to the deficiencies of PIC, such as numerical heating. A larger number of simulated particles per cell may improve the quality of the growth rate measurements by increasing the signal-to-noise ratio for a fixed unstable growth rate. 

\subsection{Wave Spectra}

For ion rings with super-Alfv\'enic drift velocities, the parallel right-handed and antiparallel left-handed waves (both negative helicity) have the greatest growth rates in the wavenumber range of interest (Appendix \ref{sec:disp}). Rearranging Eq.\  \eqref{eqn:res} and inserting the ring distribution input parameters $p_0$ and $\mu_0$, we have
\begin{align}\label{eqn:kres}
k_{\rm res}(p_0,\mu_0) &= \frac{-\Omega(p_0)}{\mu_0 v(p_0) - v_{\rm ph}}.
\end{align}
For the parameters of simulation Gy1, this results in $k_{\rm res}^{\rm PR} = -0.17 \ \omega_{\rm pi}c^{-1}$ for $v_{\rm ph} > 0$ and $k_{\rm res}^{\rm AL}= -0.11 \ \omega_{\rm pi} c^{-1}$ for $v_{\rm ph} < 0$, where the superscripts ${\rm PR}$ and ${\rm AL}$ refer to parallel-propagating right-handed and antiparallel-propagating left-handed modes, respectively, and $v_{\rm ph}$ is determined by the appropriate dispersion relation. The frequencies and wavenumbers corresponding to parallel-propagating left-handed and antiparallel-propagating right-handed waves do not satisfy the gyroresonance condition under the constraint of super-Alfv\'enic drift velocity.

\begin{figure}[t]
\centering\includegraphics[width=\linewidth,clip=true]{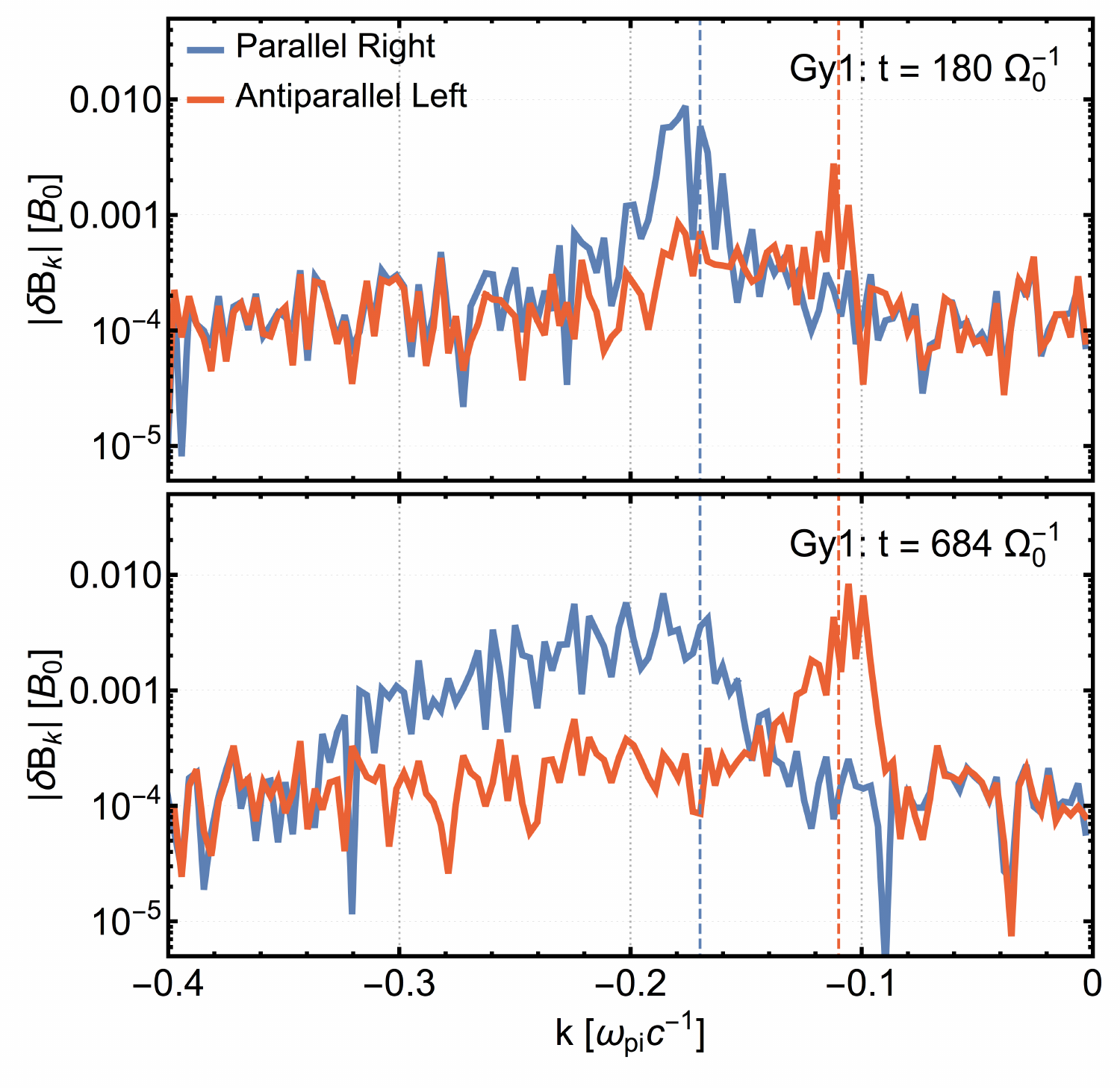}
\caption{Propagation/polarization-decomposed wave spectra for the low CR density simulation Gy1. Near the saturation of the linear phase (top panel, $t=180 \ \Omega_0^{-1}$), the peaks roughly correspond to the predicted unstable wavenumbers $k_{\rm res}^{\rm PR/AL}$ (dashed lines).  The spectrum extends to shorter wavelengths in the later nonlinear stage of instability (bottom panel, $t=684 \ \Omega_0^{-1}$). }
\label{fig:Gy1spec}
\end{figure}

For this application it is useful to decompose the spectrum into parallel/antiparallel-propagating right/left-hand polarized components. Following our conventions in Appendix \ref{sec:pol}, we Fourier-transform the quantities $(v_y + i v_z)/v_{\rm A} \mp (B_y + iB_z)/B_0$, where the upper sign gives parallel left-handed modes ($k>0$) and parallel right-handed modes ($k<0$), and the lower sign gives antiparallel right-handed modes ($k>0$) and antiparallel left-handed modes ($k<0$). We show the spectra of the low CR density simulation Gy1 decomposed in this way in Figure \ref{fig:Gy1spec}. The strongest wave growth in the linear phase coincides with the predicted wavenumbers,  $k_{\rm res}^{\rm PR} = k_{\rm max}^{\rm PR}$ and $k_{\rm res}^{\rm AL} = k_{\rm max}^{\rm AL}$. In Table \ref{tab:resultsgy}, we record the observed largest amplitude mode in the linear phase $k_{\rm meas}$ of the simulations with ring-distributed CRs. These measurements are in good agreement with the predicted values, with some small discrepancies owing to the noisy amplitudes of seed waves and the shallow decline of the growth rates around $k_{\rm max}^{\rm PR}$ when the CR density becomes large.

\begin{figure}[t]
\centering\includegraphics[width=\linewidth,clip=true]{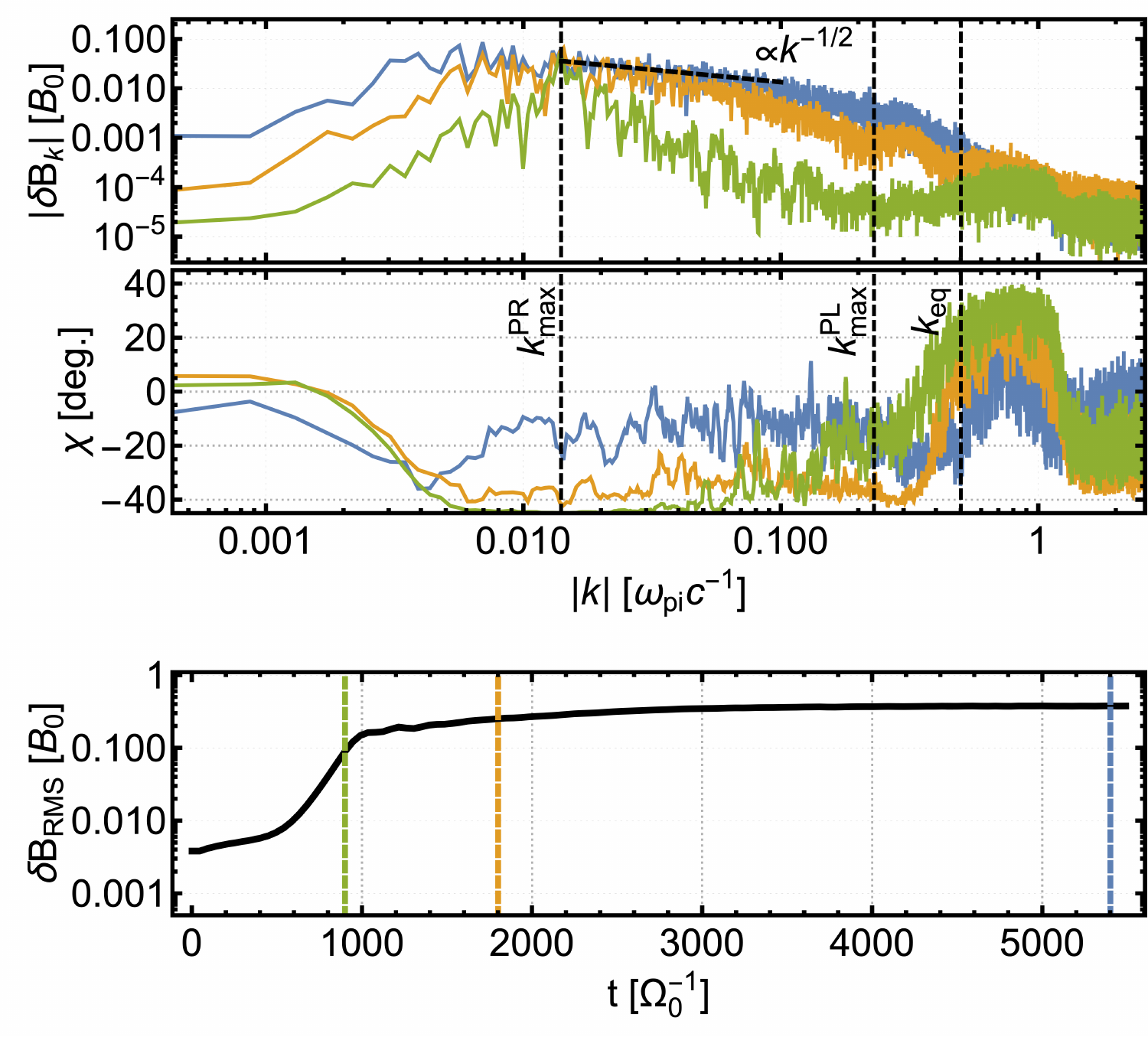}
\caption{Wave-amplitude spectra (top), helicity spectra (middle), and temporal evolution of the root-mean-square transverse field (bottom) in the high-density high-anisotropy power-law simulation Hi3. Three different epochs are shown, as indicated by vertical dashed lines in the bottom panel. Near the end of the linear growth phase ($t = 900 \  \Omega_0^{-1}$, green), the spectrum peaks at $k_{\rm max}^{\rm PR}$ with an exponential falloff on either side. The linear-phase spectrum is predominantly right-hand polarized with left-hand polarization coming in weakly around $k_{\rm max}^{\rm PL}$ and more substantially around the wavenumber at which left- and right-handed growth rates are expected to become equal, $k_{\rm eq}$. In the early nonlinear stage ($t = 1800 \  \Omega_0^{-1}$, orange) the spectrum flattens out as slower growing modes begin to saturate, and the faster growing right-handed modes continue to drown out the left-handed modes near $k_{\rm max}^{\rm PL}$. In the late nonlinear stage near total saturation ($t = 5400 \ \Omega_0^{-1}$, blue), the spectrum flattens further to a $\Sim k^{-1/2}$ scaling. Left-handed modes become significant throughout the majority of the spectrum as CRs are scattered into $\mu < 0$ resonances, resulting in elliptically polarized waves that isotropize the CR distribution.}
\label{fig:d1spec}
\end{figure}

While a cold CR distribution (e.g.,\ the ring distribution) excites a relatively narrow spectrum of waves, a hotter distribution such as an extended power-law will produce a broader spectrum by virtue of having a range of CR momenta and pitch angles. In the top panel of Figure \ref{fig:d1spec} we show the wave-amplitude spectrum of power-law distribution simulation Hi3 at three epochs. The spectrum grows most strongly at the predicted wavenumber $k_{\rm max}^{\rm PR} = -0.014 \ \omega_{\rm pi} c^{-1}$ throughout the linear instability phase. The surrounding modes continue to grow after the saturation of the $k_{\rm max}^{\rm PR}$ mode, resulting in an increasingly shallow spectrum with a rapid drop at small k owing to the high momentum cutoff in the CR distribution function. The spectrum scales roughly in accordance with Eq.\  \eqref{eqn:plsat}, $\delta B_k \propto \Gamma_{\rm cr}^{1/2}(k) \propto |k|^{-1/2}$, at very late times near the total saturation of instability.

The middle panel of Figure \ref{fig:d1spec} shows the helicity angle $\chi$, where $\chi = \pm 45^\circ$ corresponds to positive/negative helicity waves. The angle $\chi$ is typically defined in terms of the Stokes parameters; we obtain it as a function of the wavenumber $k$ using the $y$ and $z$-components of the  Fourier-transformed magnetic fields,
\begin{align*}
\chi \equiv \frac{1}{2}\arcsin \bigg( \frac{2\operatorname{Im} (\hat{B}_y \hat{B}_z^*)}{|\hat{B}_y|^2 + |\hat{B}_z|^2} \bigg),
\end{align*}
where the circumflex indicates Fourier transformation and the asterisk indicates complex conjugatation. We have smoothed each k bin with a seven-bin moving-average filter. The large anisotropy of the CR distribution results in right-handed (negative helicity) waves that emerge around $k_{\rm max}^{\rm PR}$. The fastest-growing left-handed mode $k_{\rm max}^{\rm PL}$ is dominated by the growth of the right-handed mode at the same wavenumber. At larger values of $k$, the right-handed growth rates decrease and become subdominant compared to the left-handed growth rates, resulting in a reversal of wave helicity around $k_{\rm eq} \approx 0.5  \ \omega_{\rm pi} c^{-1}$. In the post-linear phases of instability, additional power is injected in the left-handed modes by the CRs as they scatter towards $\mu \sim -1$.

\begin{figure}[t]
\centering\includegraphics[width=\linewidth,clip=true]{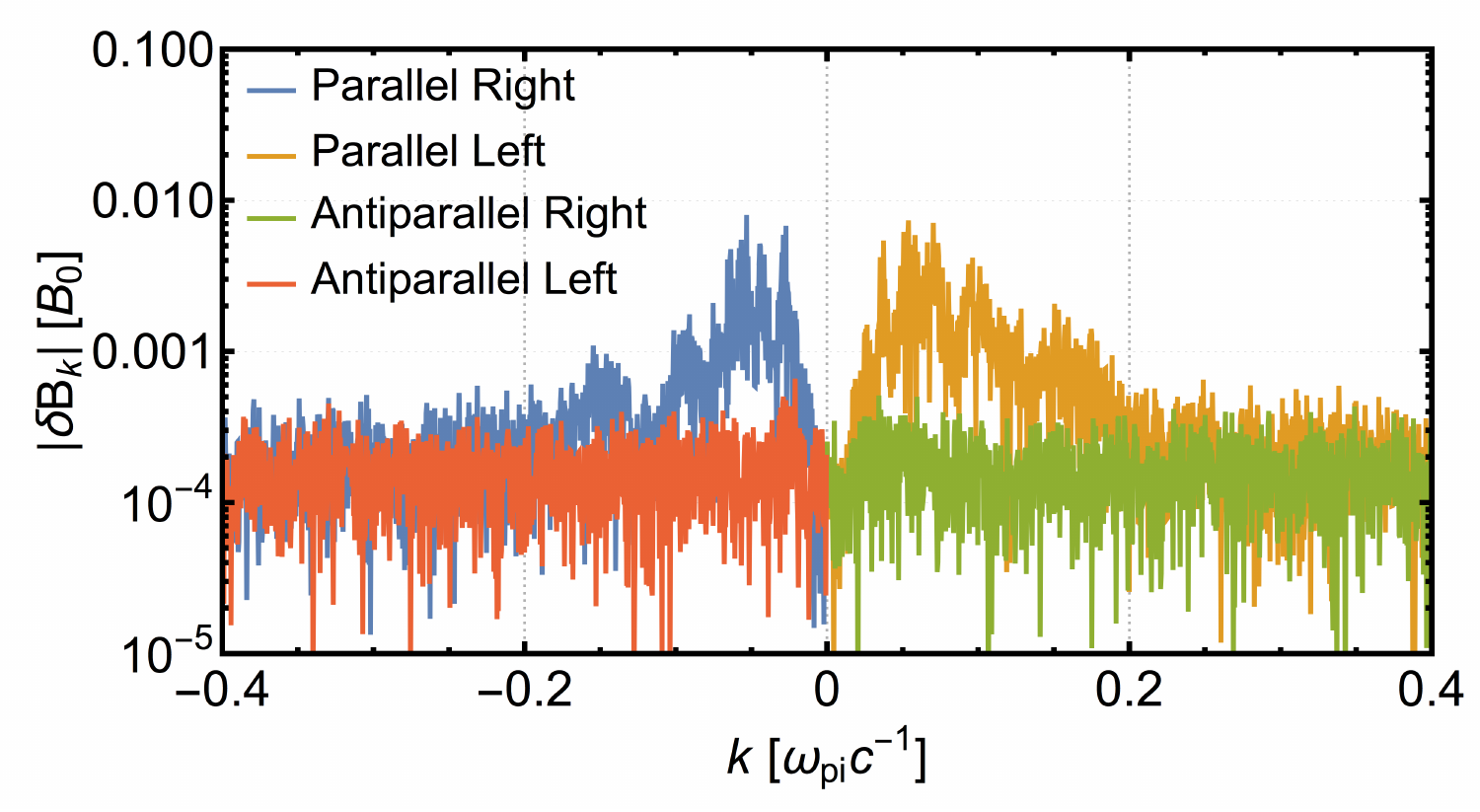}
\caption{Propagation/polarization-decomposed wave spectra for low-anisotropy simulation Lo in the linear phase ($t = 540 \ \Omega_0^{-1}$). The relatively small initial drift velocity ($v_{\rm dr} = 1.4 v_{\rm A}$) results in comparable amplitudes of parallel-propagating left- and right-handed modes. These modes combine into a predominantly linearly polarized total spectrum.}
\label{fig:d6spec}
\end{figure}

Larger proportions of CRs resonate with the parallel-propagating left-handed waves as the initial drift velocity of power-law distributed CRs is reduced. When $v_{\rm dr} \gtrsim v_{\rm A}$, the growth rates of the parallel-propagating left- and right-handed modes become degenerate, and Eq.\  \eqref{eqn:kuls} becomes a good approximation for both. Left- and right-handed modes superimposed on one another combine to produce linear polarization if the component amplitudes are equal. In Figure \ref{fig:d6spec} we show the propagation/polarization-decomposed wave spectrum in the linear instability phase of simulation Lo. The marginally super-Alfv\'enic drift velocity of this simulation, $v_{\rm dr} = 1.4 v_{\rm A}$, allows CRs to excite parallel-propagating left- and right-handed modes in nearly equal measure. The combined spectrum is approximately linearly polarized around the largest amplitude modes.

\subsection{Particle Distributions}\label{sec:dist}

\begin{figure*}[t]
\centering\includegraphics[width=\linewidth,clip=true]{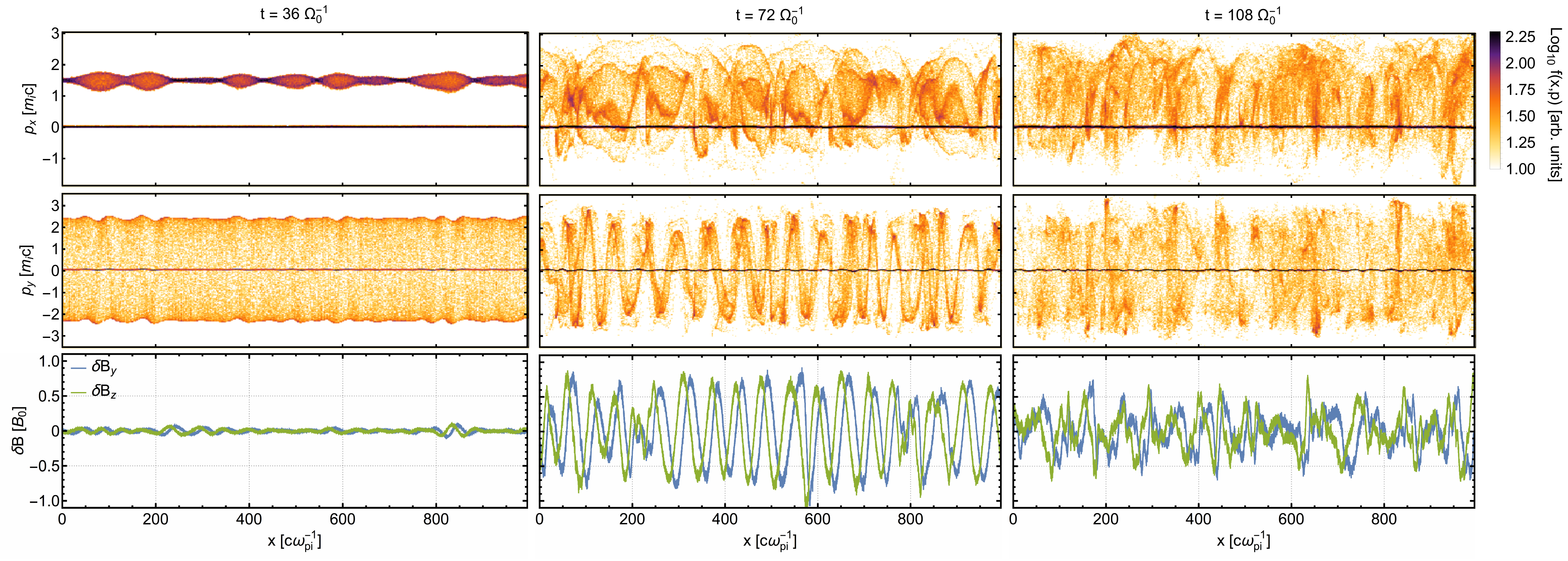}
\caption{Instability development in simulation Gy4. The CR and background ion momentum densities $p_x$ (top row) and $p_y$ (middle row) are shown along with the transverse magnetic-field components $\delta B_{y,z}$ (bottom row) at early ($t=36 \ \Omega_0^{-1}$, left column), intermediate ($t=72 \ \Omega_0^{-1}$, middle column), and late ($t= 108 \ \Omega_0^{-1}$, right column) stages of development. During the linear growth phase, a resonant mode quickly emerges from thermal noise (left column). The CR distribution is not substantially disturbed until the wave amplitude reaches $\delta B/B \gtrsim 0.1$, after which large angle scatters occur, rapidly reducing the average parallel CR momentum and establishing visible oscillations in the transverse momenta (middle column). Disruption of the CR distribution brings linear growth to halt, while CRs continue to excite the modes that ultimately lead to isotropy (right column).}
\label{fig:momdens}
\end{figure*}

The spread of the CR distribution function offers a complementary view of the phases of instability. Figure \ref{fig:momdens} depicts snapshots of the CR and background ion momentum phase space ($p_x(x)$ top row, $p_y(x)$ middle row) and transverse magnetic-field amplitude (bottom row) at various stages of the high CR density simulation Gy4. Initially CRs are scattered by small angles as they resonate with small-amplitude waves ($t=36 \ \Omega_0^{-1}$, left column), and gyroresonant structure can be observed in the transverse motion on the scale of $2\pi/k_{\rm max}^{\rm PR}$. This structure becomes increasingly apparent as waves saturate at large amplitude ($t=72 \ \Omega_0^{-1}$, middle column) and the parallel motion of CRs is substantially disrupted. Eventually the scattering on large-amplitude waves causes the CR distribution to approach isotropy ($t= 108\ \Omega_0^{-1}$, right column), and the waves decay to somewhat smaller amplitudes.

\begin{figure*}[t]
\centering\includegraphics[width=\linewidth,clip=true]{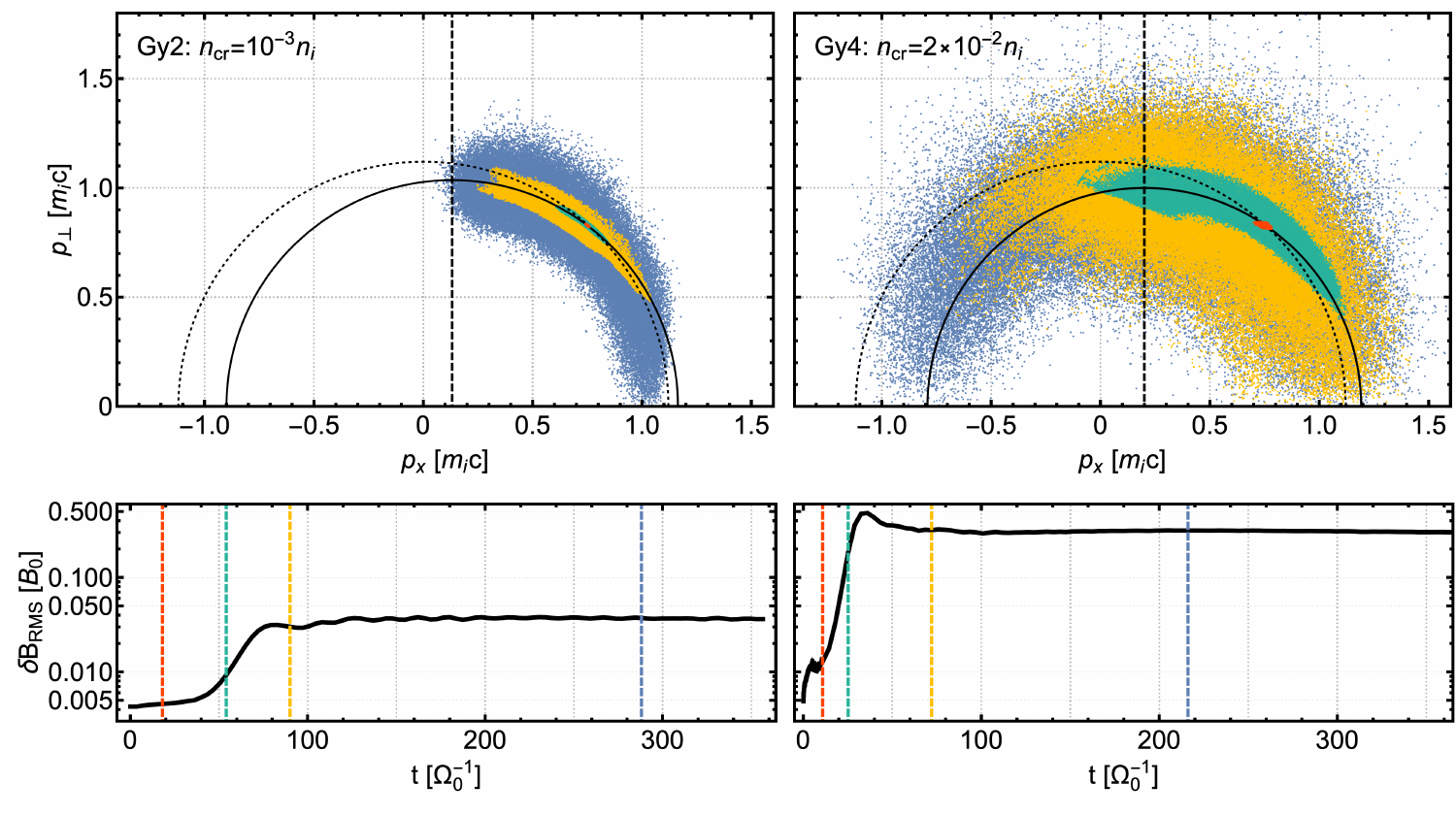}
\caption{Resonant scattering surfaces (top row) and the associated root-mean-square transverse magnetic-field amplitudes (bottom row) in simulations Gy2 (lower CR density, left column) and Gy4 (higher CR density, right column). Vertical dashed lines in the bottom row correspond in color to the times at which the CR momenta are displayed in the top row. Solid and dotted semicircles correspond to constant energy surfaces in the wave and laboratory frames, respectively. The vertical black dashed lines in the top row denote the location of $\mu = v_{\rm ph}/v$ for the initial value of the CR velocity $v$. The large-amplitude waves of Gy4 easily isotropize CRs, while those of Gy2 struggle to scatter CRs into $\mu < v_{\rm ph}/v$.}
\label{fig:scat}
\end{figure*}

The wave spectra produced by the ring distribution are quasimonochromatic -- a small number of narrow spectral peaks dominate the dynamics of particles in these simulations. Figure \ref{fig:scat} shows the change over time of the CR distribution functions in the $p_x-p_\perp$ plane for the simulations Gy2 (lower CR density, left column) and Gy4 (higher CR density, right column), along with the associated time dependence of the root-mean-square transverse magnetic-field amplitudes. The primary difference between the displayed simulations lies in the amplitudes to which the fluctuations grow. The CR density of Gy4 is a factor of twenty larger than that of Gy2, and the peak wave amplitudes are correspondingly larger in the former. This discrepancy manifests itself in the motions of CRs. The distribution functions are elongated roughly along the trajectories predicted by Eq.\  \eqref{eqn:constvel} for a parallel-propagating right-handed wave $k_{\rm max}^{\rm PR}$ (solid semicircular lines) as the simulations progress through the phases of instability. Diffusion in total momentum $p$ occurs owing to the deviation of the wave spectra from pure monochromaticity caused by the antiparallel left-handed mode $k_{\rm max}^{\rm AL}$ \citep{1991GeoRL..18.1063M}.

Other than the difference in peak wave amplitudes, the most notable divergence between the evolution of Gy2 and Gy4 is that the CRs of the former are not fully isotropized, and the unstable growth stalls prior to total saturation as we have defined it. The wave modes generated in the linear and nonlinear instability phases of Gy2 scatter the CRs within their respective resonant bands, allowing CRs to cascade to smaller $\mu$. The CRs approach the $\mu = v_{\rm ph}/v$ (vertical black dashed lines, top row of Figure \ref{fig:scat}), but are unable to efficiently cross it. CRs instead remain trapped by the effective potential wells of the largest amplitude waves. The CRs of simulation Gy4 (Figure \ref{fig:scat}, right column) are not constrained to the $\mu \gtrsim v_{\rm ph}/v$ region. The large-amplitude waves generated in the linear instability phase are able to impart forces of sufficient magnitude on CRs such that they are able to cross $\mu = v_{\rm ph}/v$. Positive-helicity waves that facilitate the isotropization process are subsequently generated in the nonlinear phase of instability. Ultimately, CRs populate the entire length of the semicircular momentum-space scattering trajectory.

\begin{figure}[]
\centering\includegraphics[width=\linewidth,clip=true]{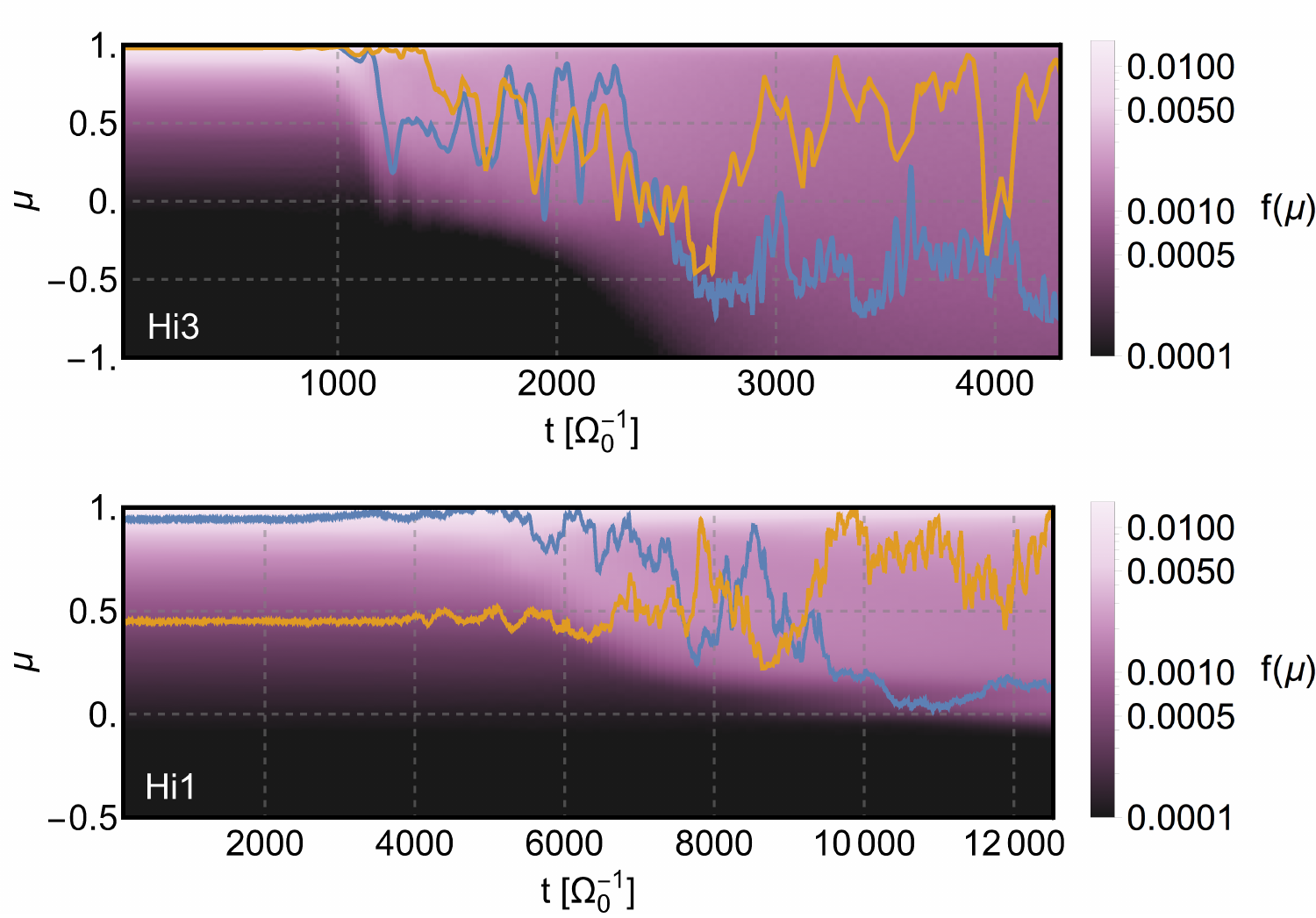}
\caption{Evolution of marginal CR distributions $f(\mu)$ over time in simulations Hi3 (top) and Hi1 (bottom). The large density of CRs in simulation Hi3 rapidly produces large-amplitude right-handed modes that saturate linear growth ($t \sim 1200 \ \Omega_0^{-1}$) and, subsequently, left-handed modes that lead to CR isotropy ($t > 2000 \ \Omega_0^{-1}$). In the lower CR density simulation Hi1, a buildup of CRs forms in the $\mu \sim 0$ region. The right-handed modes generated by linear growth efficiently scatter CRs, but the small-amplitude left-handed waves are unable to continue scattering them into negative $\mu$ on the timescales of the simulation. The pitch-angle cosines $\mu(t)$ are shown for two example CRs in each panel. These trajectories demonstrate the strong scattering of CRs within the range encompassed by moderate- to large-amplitude resonant modes.}
\label{fig:fmut}
\end{figure}

\begin{figure}[t]
\centering\includegraphics[width=\linewidth,clip=true]{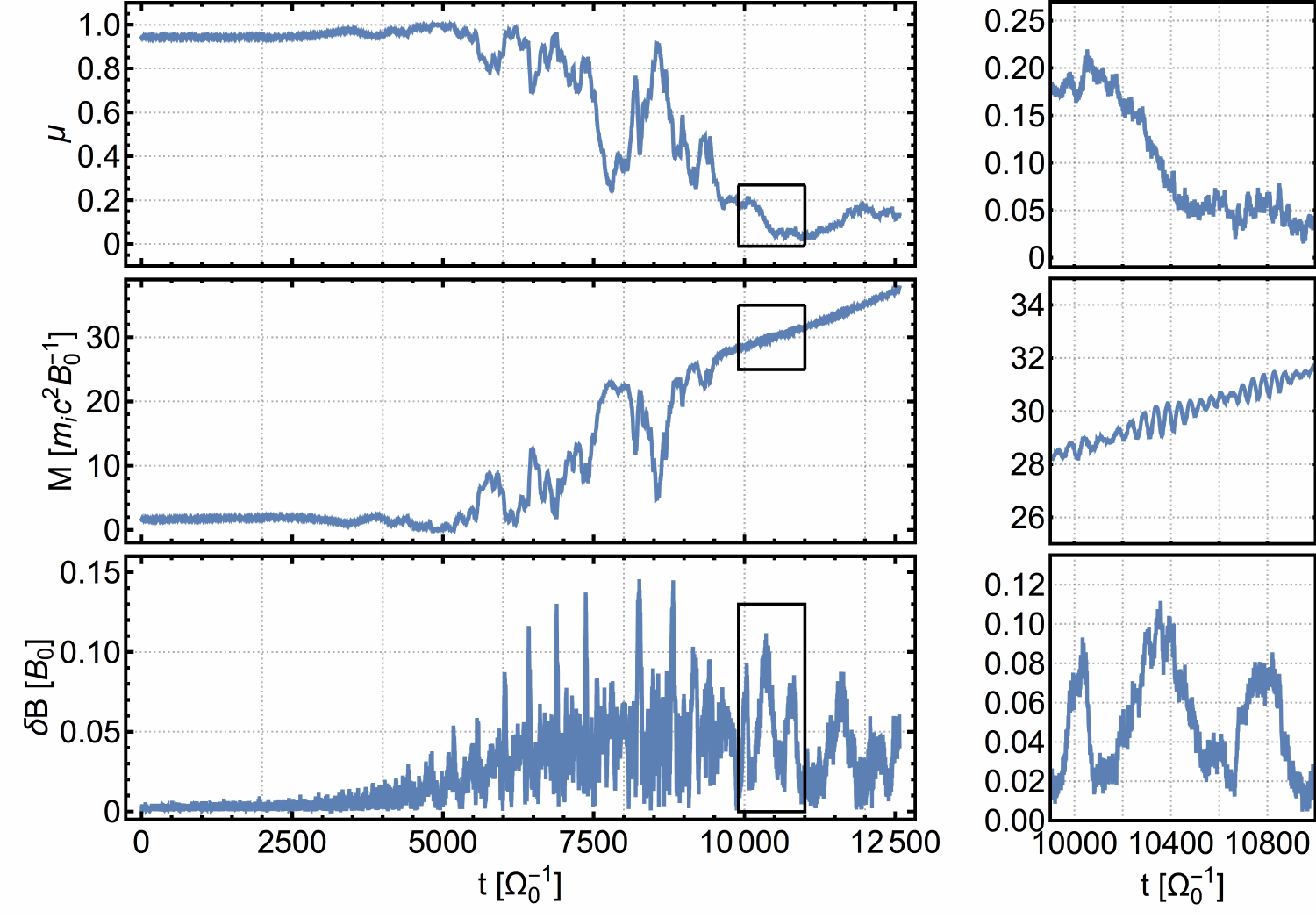}
\caption{Temporal dependence of the pitch-angle cosine $\mu$ (top), magnetic moment $M$ (middle), and transverse magnetic field (bottom) for a particular CR in the low CR density simulation Hi1. The left column shows the trajectory of the particle over the entire simulation. Only small-angle scatterings occur until $t\sim 5000\ \Omega_0^{-1}$, and the particle motion is relatively quiescent. Scattering becomes more severe as the instability progresses to wave amplitudes $\delta B \gtrsim 0.01$, producing large jumps in $\mu$ and $M$. At $t\sim 10100\ \Omega_0^{-1}$ the CR reaches $\mu \approx 0.18$, allowing it to undergo a mirror interaction with the $\delta B \approx 0.1$ peak that it subsequently encounters. The right column shows an expanded view of the mirror process (indicated by inset black boxes in the left column). The magnetic moment increases by $\Sim 5\%$ as the CR reverses its direction of propagation with respect to parallel traveling Alfv\'en waves. This particle is unable to continue into the negative $\mu$ half-plane and remains trapped within the confines of two adjacent $\delta B$ peaks for the remainder of the simulation.}
\label{fig:mirror}
\end{figure}

The time-dependent pitch-angle cosine distributions $f(\mu)$ of power-law simulations for high CR density (Hi3, top panel) and low CR density (Hi1, bottom panel) are displayed in Figure \ref{fig:fmut}. In the initial and linear phases of instability the CRs remain unperturbed. At $t\sim 1200 \ \Omega_0^{-1}$ for Hi3 (top) and $t\sim 5000 \ \Omega_0^{-1}$ for Hi1 (bottom), waves of substantial amplitude are generated and the CRs are moved out of their initial resonant bands, transitioning from the linear to the nonlinear phase of instability. The behaviors of these simulations subsequently diverge. 

The higher CR density of simulation Hi3 enhances the isotropization process in two ways. The first is that the large-amplitude waves produced in the linear phase of Hi3 allow CRs to efficiently scatter beyond $\mu = v_{\rm ph}/v$. The spatial fluctuations of the transverse magnetic field reach peak amplitudes of up to  $\delta B/B_0 \sim 0.5$, and the magnetic mirroring mechanism has a correspondingly broad reach $\mu_M \approx \mu_M' + v_{\rm A}/v \sim 0.45$ which allows CRs to easily move deep into the $\mu < 0$ region. The second way is that the large influx of CRs into the $\mu <0$ region rapidly excites the parallel-propagating left-handed modes that are required to continue scattering towards $\mu \sim -1$. The combined effect of these behaviors leads to the rapid isotropization of CRs with respect to the parallel-propagating Alfv\'en waves in the nonlinear phase of instability. By $t\sim 4000 \ \Omega_0^{-1}$ the system approaches total saturation of instability with a nearly constant CR pitch-angle distribution $f(\mu)$. 

In the less-energetic wave spectrum of Hi1, CRs are unable to efficiently cross the pitch-angle gap into negative $\mu$. A buildup of CRs forms around $\mu\sim 0$ as they cascade down the predominantly parallel-propagating right-handed wave spectrum. There they are met with the parallel-propagating left-handed modes of the spectral noise floor, with amplitudes roughly three orders of magnitude smaller than $k_{\rm max}$ ($\delta B_{k_{\rm max}} \sim 0.01 B_0$). The density associated with these $\mu\sim 0$ CRs does not translate to rapid growth in the left-handed modes, and the instability stalls for some protracted (but likely finite) period of time beyond the duration of the simulation. 

Figure \ref{fig:fmut} also features the pitch-angle trajectories $\mu(t)$ of example CRs (blue and orange lines), allowing us to examine the scattering behaviors of individual particles in regions with and without power in the corresponding resonant waves. For both simulations shown, the CRs experience only small angle deflections prior to the nonlinear phase of instability, while violent scattering events take place following the transition to the nonlinear phase. In the later stages of instability, CRs are free to stochastically explore the entirety of the $\mu$-space regions corresponding to resonance with the dominant wave modes. For simulation Hi3, the region available to CRs extends to the full range of $\mu$ as positive-helicity waves are generated in the nonlinear phase of instability that accommodate resonant interactions at $\mu \sim -1$ (see Figure \ref{fig:d1spec}). The majority of CRs display comparable behavior, i.e., the chosen CR trajectories are not special in this respect.

As may be expected from the preceding discussion, the region available to the majority of CRs via resonant scattering in simulation Hi1 (low CR density) is restricted to $\mu \gtrsim 0.1$. However, a nonzero fraction of CRs are ostensibly able to rotate from $\mu \sim 0.2$ to $\mu \sim 0$ by the mirroring mechanism, including the particle represented by the blue line in the bottom panel of Figure \ref{fig:fmut}. This particle is again displayed in Figure \ref{fig:mirror}, where we show its pitch-angle cosine $\mu(t)$ (top) and magnetic moment $M(t)$ (middle), as well as the transverse magnetic field $\delta B (t)$ that it experiences as it travels. In the period between $t \sim 5000$ to $ \sim 9500 \ \Omega_0^{-1}$ the CR undergoes severe scattering as it resonates with waves of amplitude up to $\delta B/B_0 \sim 0.1$, where the magnetic moment $M$ fluctuates by up to hundreds of percent. Around $t \sim 10500 \ \Omega_0^{-1}$, it undergoes a smooth transition from $\mu \sim 0.2$ to $\mu \sim 0.05$ as it reverses direction relative to a $\delta B/B_0 \sim 0.1$ fluctuation (as indicated by the symmetry of the bottom right panel). During this mirroring event the magnetic moment $M$ increases by only a few percent, which can be accounted for by the secular growth of $M$ that occurs in the final quarter of the simulation. The duration of this interaction is $\Sim 2\times 10^2 \ \Omega_0^{-1}$, in good agreement with the predicted value $t_M$ (section \ref{sec:physsat}). Again, the lack of power in left-handed waves prevents the particle from continuing into negative $\mu$. Instead, it undergoes another mirror reversal before the end of the simulation, apparently trapped between peaks in $\delta B$.

\subsection{Drift Velocities \& Saturation}

In section \ref{sec:physsat} we discussed the saturation of the linear growth phase via the diffusive depletion of free momentum carried by CRs and, alternatively, by the trapping of particles in the effective potential well of a resonant wave. Additionally, progress toward total saturation of instability can be measured via momentum conservation arguments. However, the momentum balance procedure depends on the CR drift velocity as a function of time, which is not known \emph{a priori}. In the quasimonochromatic spectra of the simulations with ring CR distributions, particle dynamics in the saturation stage are dominated by the largest amplitude mode. Thus we expect that the condition for saturation due to particle trapping (Eq.\  \eqref{eqn:beamsat}) to be of particular relevance to the ring distribution. 

\begin{figure}[t]
\centering\includegraphics[width=\linewidth,clip=true]{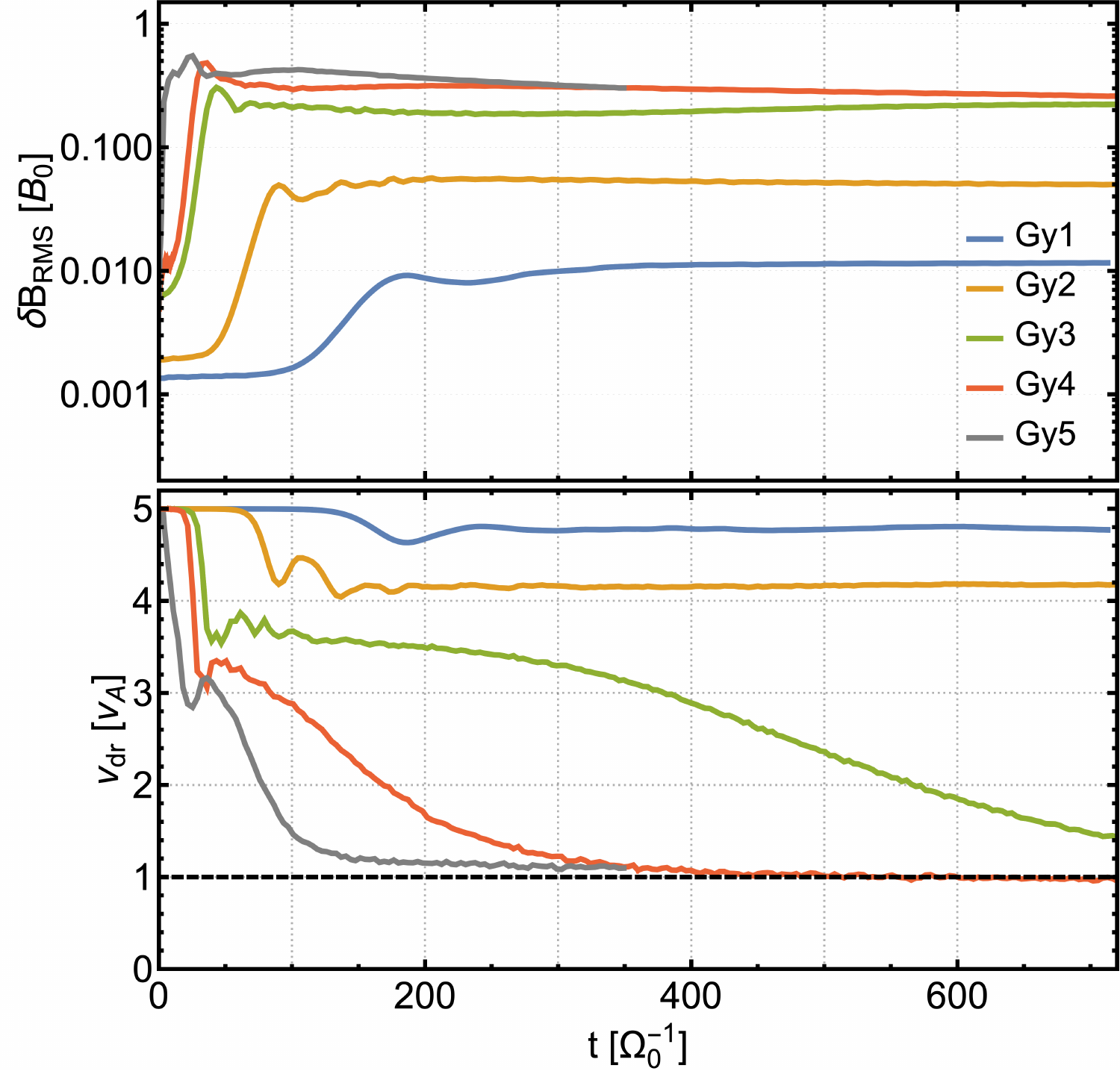}
\caption{Root-mean-square transverse magnetic-field amplitude $\delta B_{\rm RMS}$(top) and CR bulk drift velocity $v_{\rm dr}$ (bottom) over time for ring distributed CRs. Differences in the initial fluctuation amplitudes arise from the variation in the number of simulated particles per cell (Table \ref{tab:prop}). The high-density simulations Gy4 and Gy5 quickly form large-amplitude waves that reduce the bulk drift velocity to $\Sim v_{\rm A}$. The low CR density simulations Gy1, Gy2, and Gy3 enter a state of quasistability prior to total saturation as resonant trapping prevents CRs from scattering to smaller $\mu$.}
\label{fig:gyrodrift}
\end{figure}

In the top panel of Figure \ref{fig:gyrodrift} we show the growth of the RMS transverse magnetic-field amplitudes for the ring distribution runs Gy1-5. In all cases, saturation is driven by the resonant trapping mechanism. Oscillations in the wave amplitudes are seen to occur at the end of the exponential growth phase. The wave amplitudes overshoot their equilibrium levels, resulting in inverted CR gradients and subsequent reabsorption of wave energy by the trapped particles. These oscillations occur on the trapping timescale $\Sim2\pi/\Omega_{\rm tr}$, where $\Omega_{\rm tr}$ is defined in Eq.\  \eqref{eqn:trapfreq}. 

\begin{figure}[t]
\centering\includegraphics[width=\linewidth,clip=true]{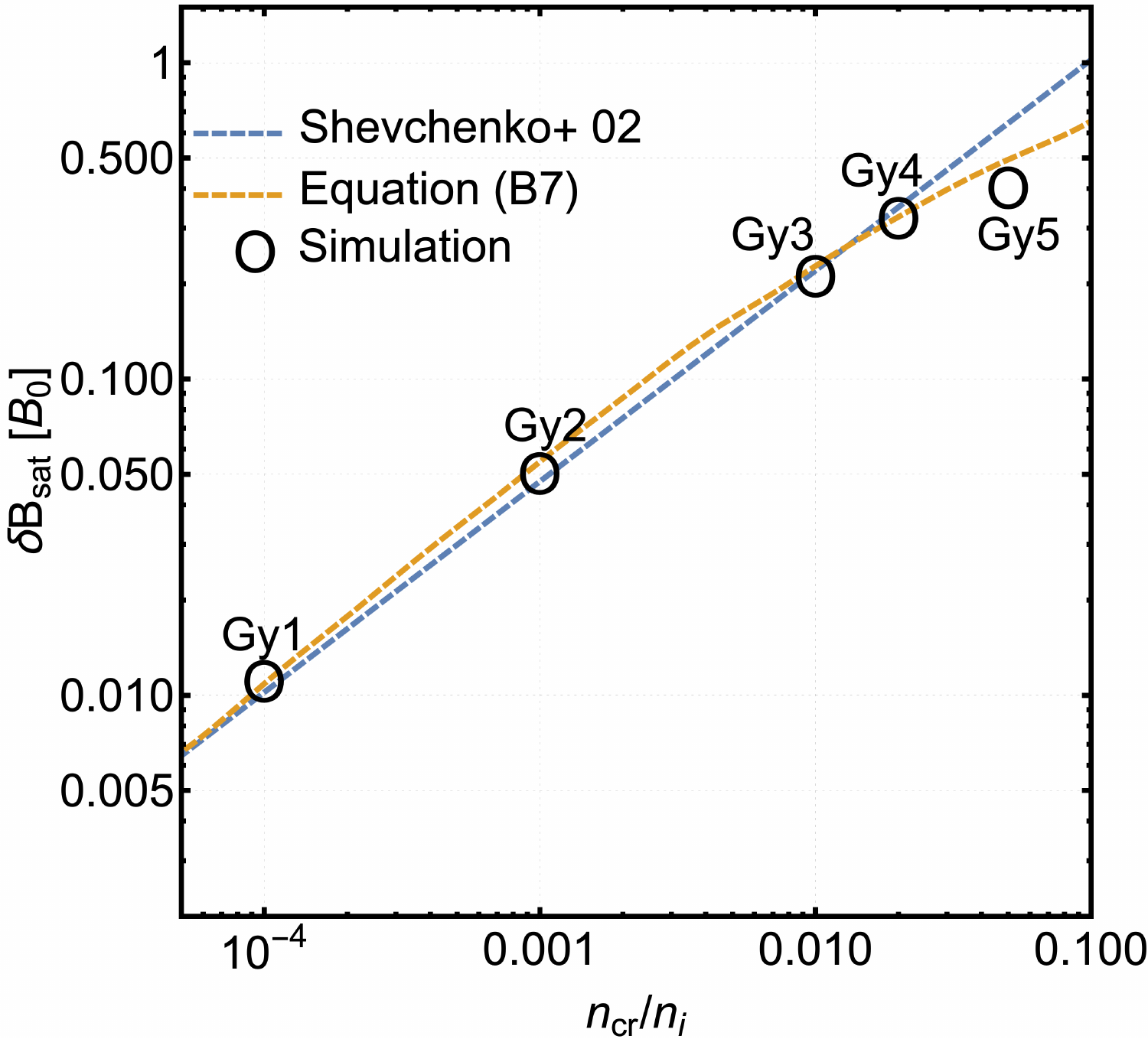}
\caption{Linear-phase saturation amplitudes of ring-distribution simulations. Measurements from the simulations are denoted by ``O" symbols. For comparison, we show two theoretical predictions, both of which derive from the estimate $\Omega_{\rm trap} = 1.95\times \Gamma_{\rm cr}^{\rm PR}$, where the factor 1.95 was chosen to give rough alignment with the simulation data. The blue dashed line corresponds to the approximate growth rate presented in \cite{2002JGRA..107.1367S}, with the saturation amplitude proportional to $(n_{\rm cr}/n_{\rm i})^{2/3}$ (Eq.\  \eqref{eqn:beamsat}). The orange dashed line corresponds to a numerical computation of the maximal growth rate using the full dispersion relation (Eq.\ \eqref{eqn:ringdisp}). Note that these measurements are no more accurate than about the $\Sim 10\%$ level, since the determination of the saturation amplitude is not well defined. }
\label{fig:ringsat}
\end{figure}

In Figure \ref{fig:ringsat} we record the linear-phase saturation amplitudes (``O" symbols) and compare them against the (normalized) scaling relation presented in section \ref{sec:physsat} (Eq.\ \eqref{eqn:beamsat}, blue dashed line). Only in the simulations with the densest CR distributions ($n_{\rm cr} \gtrsim 0.02 n_{\rm i}$) does Eq.\  \eqref{eqn:beamsat} begin to fail to predict the scaling of the saturated wave amplitude, as evidenced by the measurement of Gy5 falling below the blue dashed line. The derivation of Eq.\ \eqref{eqn:beamsat} utilized the approximation of \cite{2002JGRA..107.1367S} for the growth rate of the fastest-growing mode. A detailed numerical solution to the dispersion relation (Appendix \ref{sec:disp}) shows that, in the high CR density regime, the fastest-growing mode shifts away from the gyroresonance prediction of $k_{\rm res}^{\rm PR}$. The shift of the wavenumber causes the trapping frequency to increase, thereby reducing the saturation magnetic-field amplitude -- the formula for the approximate growth rate does not include this effect. Figure \ref{fig:ringsat} also includes predictions for the saturation amplitude by comparing numerically derived growth rates (Eq.\ \eqref{eqn:ringdisp}) to the trapping frequency (orange dashed line). The latter predictions are seemingly able to capture the high CR density reduction away from Eq.\  \eqref{eqn:beamsat} in the saturated amplitude. Note that there is some ambiguity in the measurement of the saturation amplitude.

The phases of instability are again reflected in the change in bulk drift velocity of CRs associated with the growth of transverse waves (bottom panel of Figure \ref{fig:gyrodrift}). A period of quiescence occurs in the initially noisy background fields, with duration dependent on the growth rate of instability. A sharp decline in the drift velocity indicates the cessation of the linear growth phase. The transverse magnetic fields reach an initial peak in coincidence with the abrupt disruption of the CR distribution and simultaneous decrease in the growth rates. As discussed above, oscillations occur around this peak at the trapping frequency (Eq.\  \eqref{eqn:trapfreq}). 

In the nonlinear stages of instability the behavior of CR drift velocity varies -- the low CR density simulations are not able to efficiently reach isotropy. The ability of the unstable system to isotropize the CR distribution depends on its propensity to generate an appropriate spectrum of waves. The effective potential of a resonant wave has an associated velocity width which trapped particles oscillate around, given by
\begin{align}\label{eqn:trapvel}
\Delta v_{\parallel}' &\approx 2^{\frac{3}{2}} \sqrt{\frac{\delta B'}{B_0} \frac{v_\perp}{k} \Omega_{\rm cr}'},
\end{align}
where primed quantities refer to quantities in the wave reference frame \citep{1971JGR....76.4463S}. Dividing through by $v'$ gives a range of pitch angles $\Delta \mu'$ in which a trapped particle can oscillate about the initial pitch-angle cosine $\mu_0'$.  Resonant trapping effects in monochromatic wave packets are unable to isotropize the CR distribution unless the amplitude is exceptionally large ($\delta B/B_0 \sim 1$), so the generation of additional waves is generally required to progress towards total saturation. Unstable systems that produce large-amplitude waves such that the trapping width $\Delta v_\parallel$ allows CRs to cross $\mu \approx v_{\rm ph}/v$ are advantaged because they can quickly gain access to the positive helicity modes required to obtain isotropy.

The simulations that are dominated by resonant trapping retain super-Alfv\'enic drifts indefinitely (Gy1 and Gy2; Figure \ref{fig:gyrodrift}, bottom panel). In contrast, the CR drift velocity is reduced to the local Alfv\'en speed for the unstable systems that are able to isotropize the CRs with respect to the excited waves  (Gy3-5; Figure \ref{fig:gyrodrift}, bottom panel). While the effect of linear-amplitude Alfv\'en modes ($\delta B^2 \sim 0$) on the background medium is negligible, the fields of nonlinear-amplitude modes can drive nontrivial drifts of the background plasma along the axis of wave propagation \citep{2019ApJ...873...57W}. The latter effect can be clearly observed in simulation Gy5 in the bottom panel of Figure \ref{fig:gyrodrift} -- the CR drift velocity ultimately reduces to $v_{\rm A} + v_{\rm i}$, where $v_{\rm A}$ is the (fixed) Alfv\'en speed given in the laboratory frame and $v_{\rm i}$ is the drift imparted to the ions of the background plasma.

\begin{figure}[t]
\centering\includegraphics[width=\linewidth,clip=true]{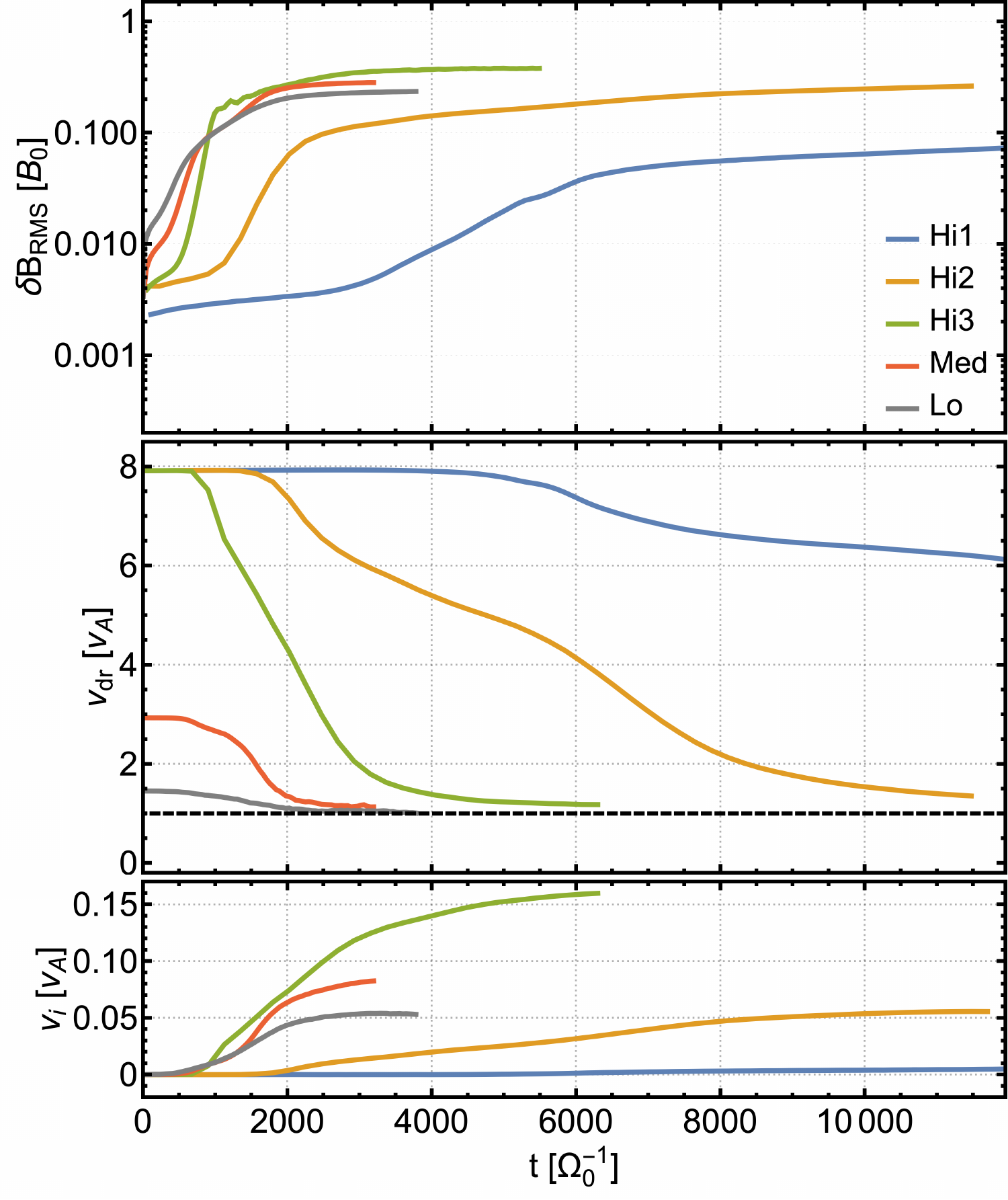}
\caption{Root-mean-square transverse magnetic-field amplitude $\delta B_{\rm RMS}$(top), CR bulk drift velocity $v_{\rm dr}$ (middle), and background ion bulk drift velocity $v_{\rm i}$ (bottom) over time for power-law distributed CRs. The drift of the background plasma causes CRs to reach total saturation at $v_{\rm dr} = v_{\rm A}+v_{\rm i}$ when measured in the stationary laboratory frame.}
\label{fig:growthhi}
\end{figure}

We have seen previously that the range of particle momenta and pitch angles of the power-law distribution produces a broad spectrum of negative-helicity waves and, if the initial anisotropy is sufficiently small, a similarly broad positive-helicity component as well. These features are the basis for the qualitative divergence between the behaviors of the power-law and ring distributed CR systems. In Figure \ref{fig:growthhi} we show the RMS transverse magnetic-field amplitude (top), the bulk CR drift velocity (middle), and the bulk velocity of the background ions (bottom) over time for the power-law simulations. Exponential growth of the fastest-growing mode transitions into an extended nonlinear instability phase where the initial distribution has been disrupted but substantial unstable growth continues on longer time scales. 

\begin{figure}[t]
\centering\includegraphics[width=\linewidth,clip=true]{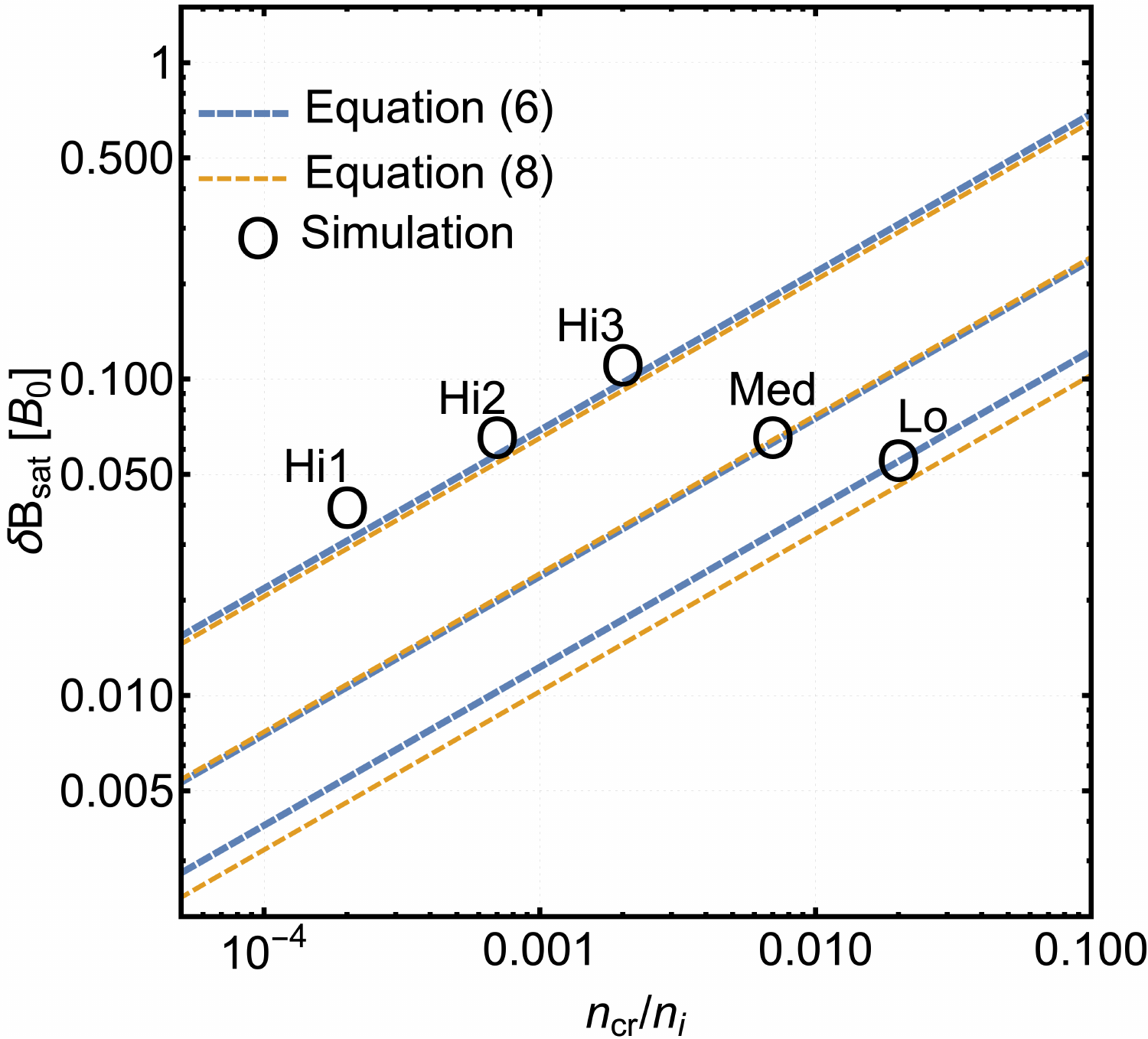}
\caption{Same as Figure \ref{fig:ringsat} but for power-law distributed CRs. The orange dashed lines correspond to saturation estimates, Eq.\  \eqref{eqn:plsat}, when using the approximate growth rate (Eq.\ \eqref{eqn:kuls} with relativistic correction). The blue dashed lines utilize the numerical computation of the growth-rate integral, Eq.\ \eqref{eqn:zwei}, instead.  Both theoretical predictions come from the estimate $\nu_{\rm QLT} = \frac{1}{4}\times\Gamma_{\rm cr}$, where the factor $\frac{1}{4}$ was chosen to give rough alignment with the simulation data. For the strongly right-hand polarized simulations (Hi1-3) we used $\Gamma_{\rm cr} = \Gamma_{\rm cr}^{\rm PR}$, while for the lower anisotropy simulations (Lo and Med) we used $ \Gamma_{\rm cr} = \Gamma_{\rm cr}^{\rm PR} + \Gamma_{\rm cr}^{\rm PL}$. Note that both growth rates utilized here have the same scaling with the CR density, thus the saturation amplitudes scale with $(n_{\rm cr}/n_{\rm i})^{1/2}$ as suggested by Eq.\  \eqref{eqn:plsat}. }
\label{fig:plsat}
\end{figure}

What we have called the ``nonlinear phase of instability," as embodied by the evolution of the transverse magnetic fields, consists of two sequential behaviors of the initially anisotropic CR distributions. First, following the cessation of exponential growth at the linear rate, continued growth of other modes flattens the CR distribution function within the region $\mu \gtrsim v_{\rm ph}/v$. The inefficiency of crossing the 90 degree barrier (due to the absence of left-handed modes) results in a reduced slow-down of the drift velocity during this phase. Unlike the ring distribution-driven instability, the majority of the total wave energy comes from growth in the nonlinear phase of instability, leading to the second behavior. As waves grow, diffusion across the 90 degree barrier and into $\mu <0$ becomes more efficient, resulting in a second and steeper decline in the drift velocity until isotropy is nearly achieved.

The growth rate of simulation Lo (low anisotropy) is comparable to simulations Hi2 and Hi3 (high anisotropy), but the progression of the instability is qualitatively different. Systems with less severe CR anisotropy have smoother transitions between the linear phase disruption, $\mu \gtrsim v_{\rm ph}/v$ gradient flattening, and finally diffusion across the 90 degree barrier. Beyond the trivial explanation that systems with less anisotropy are closer to $v_{\rm dr} = v_{\rm ph}$ by definition, the content of the excited wave spectra plays a role here. In particular, less isotropy translates to a larger fraction of the free momentum going into parallel-propagating left-handed modes. These positive-helicity modes are required to scatter CRs in the post-mirroring region $\mu \lesssim v_{\rm ph}/v - \mu_M'$. The existence of these modes allows simulation Lo to reach total saturation of instability before simulation Hi3, despite the latter having more energy in the transverse magnetic field.

In the bottom panel of Figure \ref{fig:growthhi} we show the response of the background ions $v_{\rm i}$ to the presence of the relatively large-amplitude Alfv\'en waves, where $v_{\rm i}$ is the mean velocity of background ions in the $\hat{x}$ direction. The momentum given up by CRs flows to the background plasma via the $\boldsymbol{E}\times\boldsymbol{B}$ drifts of individual particles. Conservation of momentum implies a bulk flow of the background plasma. Since the Alfv\'en wave frame increases in velocity by an equal amount, total isotropy is obtained when CRs reach a drift velocity $v_{\rm dr}=v_{\rm A} + v_{\rm i}$ (Figure \ref{fig:growthhi}, middle panel), typically with $v_{\rm i} \ll v_{\rm A}$. This is the microphysical basis for CR-driven winds. However, the lack of wave damping combined with the finite CR momentum reservoir in our simulations leads to substantially weaker acceleration of the background plasma compared to the standard CR wind setup (e.g., \citealt{2008ApJ...674..258E}).

Resonant trapping is not the relevant saturation mechanism for growth driven by power-law CRs. The influence of a spectrum of waves provides additional scattering that alleviates trapping effects by preventing a single mode from dominating the dynamics of resonant CRs. Accordingly, the trapping frequency criterion used for the ring distribution underestimates the saturation amplitudes of the power-law simulations by orders of magnitude, in addition to incorrectly predicting the scaling with $n_{\rm cr}/n_{\rm i}$. In figure \ref{fig:plsat} we compare the observed saturated (RMS) field amplitudes against the scaling predicted by Eq.\  \eqref{eqn:plsat}. The estimates provided by $\nu_{\rm QLT} \approx \Gamma_{\rm cr}$ are in moderate agreement with the simulations. Discrepancies likely arise on account of the deviation of measured growth rates from the theoretical values (Table \ref{tab:resultspl}), the uncertainties in measuring the saturation amplitudes, and the assumptions made in deriving these estimates (including the validity of QLT for large-amplitude waves).

\begin{figure}[t]
\centering\includegraphics[width=1.0\linewidth,clip=true]{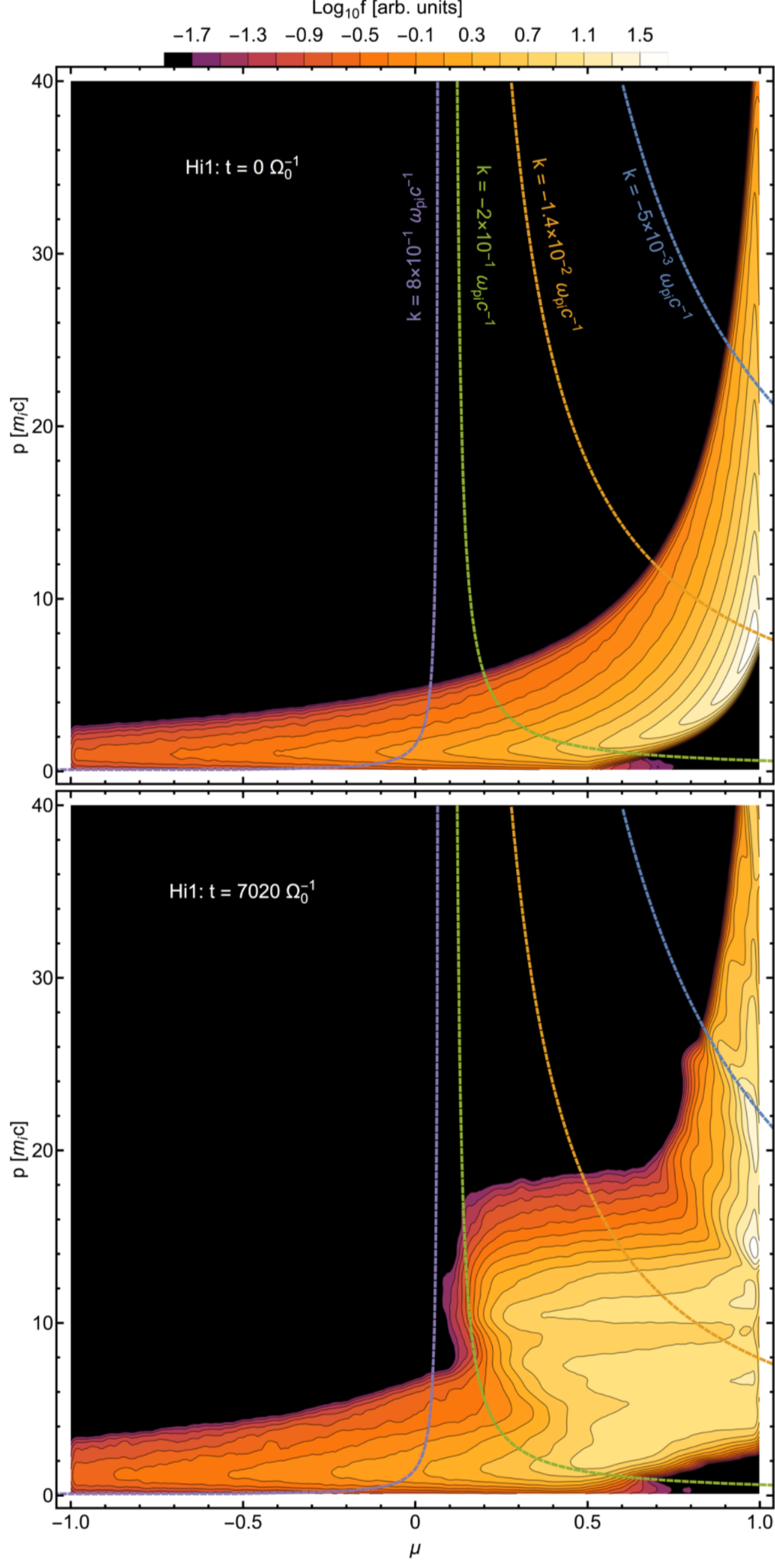}
\caption{Evolution of the CR distribution function in simulation Hi1. Momentum space density contours for CR distribution functions $f(p,\mu)$ are shown at initialization (top) and at the early nonlinear phase (bottom).  Saturation of linear growth occurs when the CR distribution is flattened along the $\mu$ coordinate in the densest regions of the initial distribution. Dashed lines depict the resonance conditions (Eq.\ \eqref{eqn:res}) for particular values of $k$ (see text).}
\label{fig:fpmu}
\end{figure}

In Figure \ref{fig:fpmu} we show the full CR distribution function $f(p,\mu)$ for simulation Hi1  (low CR density) at initialization (top panel, $t= 0 \ \Omega_0^{-1}$) and the end of the linear growth phase (bottom panel,  $t= 7020 \ \Omega_0^{-1}$). Dashed lines depict the resonance conditions of four modes with particular properties (see the following paragraph for details).  The flow of CRs along the resonant scattering trajectories given by Eq.\  \eqref{eqn:constvel} is apparent. Saturation of the fastest-growing modes is observed to occur in coincidence with the flattening of the distribution function in the densest regions of $p$-$\mu$ space, while relatively sparse regions evolve on longer time scales. 

Once again we observe that the resonant cascade is unable to efficiently scatter particles through $\mu \sim v_{\rm ph}/v \sim 0.15$ in Figure \ref{fig:fpmu}. Steep gradients build up as CRs are funneled into this region of momentum space, but the CRs are too sparse to quickly excite high-$k$ waves to scatter on, resulting in the stalled decline of bulk drift velocity observed in Figure \ref{fig:growthhi}. Dashed lines in Figure \ref{fig:fpmu} depict the resonance conditions for the following modes at the time displayed in the bottom panel: the parallel-propagating right-handed waves corresponding to the longest-wavelength mode with amplitude greater than $\delta B/B_0 \sim 0.01$ (blue, $k=5\times10^{-3} \ \omega_{\rm pi}c^{-1}$), the fastest-growing mode (orange, $k=1.4\times10^{-2}\ \omega_{\rm pi}c^{-1}$), roughly the shortest wavelength mode with amplitude greater than the noise floor (green, $k=2\times10^{-1}\ \omega_{\rm pi}c^{-1}$), and the parallel-propagating left-handed mode of largest amplitude (purple, $k=8\times10^{-1} \ \omega_{\rm pi}c^{-1}$). While the negative-helicity waves (e.g., the green, orange, and blue dashed lines) are of sufficient amplitude to eventually bridge the $\mu$ gap, the dearth of positive helicity waves (e.g., the purple dashed line) results in a bottleneck in the isotropization process. The requisite waves are slowly generated, allowing the drift velocity to ultimately decline (e.g.,\ simulation Hi2).

Equation \eqref{eqn:tmu} of section \ref{sec:physsat} provides a rough estimate for the time scale of the nonlinear phase of instability. Caution should be exercised in the application of this calculation. As demonstrated in Appendix \ref{sec:relax}, the time scales as $t_\mu \propto \delta B^{-3}$. It is therefore highly sensitive to $\delta B$, which is itself an estimate. In Table \ref{tab:trel} we compare the observed duration of the nonlinear phase $t_{\rm meas}$, the time elapsed between the end of the linear phase and the time at which $v_{\rm dr} \approx v_{\rm A}$, against predicted values $t_\mu$ (Eq. \eqref{eqn:tmu}). We use the observed saturation values (Figure \ref{fig:plsat}) to obtain more accuracy in the relaxation time $t_\mu$. Despite the large uncertainty of this comparison, a notable trend is visible. The low-anisotropy simulation Lo relaxes more quickly than $t_\mu$ would suggest, while the high-anisotropy simulations Hi2 and Hi3 relax on longer time scales than predicted. Simulation Hi1 does not approach saturation within the duration of the calculation, but appears consistent in behavior relative to Hi2 and Hi3. The effects of large-amplitude waves likely reduce the time for relaxation, since QLT estimates are not valid in this regime. The lack of left-handed waves in the Hi1-3 runs has the opposite effect, lengthening the relaxation time. These effects appear to roughly balance in the intermediate simulation Med, where the observed relaxation time is closer to the predicted value. \\ \\

\begin{table}
\centering
\begin{tabular}{|c|cc|}
\hline
Simulation		& $t_{\rm meas}$ [$\Omega_0^{-1}$]& $t_\mu$ [$\Omega_0^{-1}$] \\
\hline
\hline
Lo			& $\Sim 1500$ 					& $\Sim 3400$ \\
\hline
Med                  & $\Sim 1500$                   & $\Sim 1800$ \\
\hline
Hi1			& $\gg 12000$					& $\Sim 30000$\\
Hi2			& $\gtrsim 9000$				& $\Sim 6700$\\
Hi3			& $\gtrsim 4000$ 				& $\Sim 1200$ \\
\hline
\end{tabular}
\caption{Nonlinear relaxation times for power-law distribution. The predicted duration of the nonlinear phase $t_\mu$ (the mean free time, Eq.\ \eqref{eqn:tmu} and Appendix \ref{sec:relax}) is listed for each simulation. These are compared to $t_{\rm meas}$, the observed duration of the nonlinear phase. }\label{tab:trel}
\end{table}

\section{Discussion}\label{sec:disc}

In section \ref{sec:res} we observed that the evolution of the unstable system depends on the form of the CR distribution. Assuming super-Alfv\'enic drift, the simple ring distribution (Eq.\  \eqref{eqn:ring}) generates a narrow spectrum of parallel-propagating right-handed and antiparallel-propagating left-handed modes (negative helicity only), while the power-law distribution (Eq.\  \eqref{eqn:pow}) produces a broader spectrum of parallel right-handed and left-handed modes (negative \emph{and} positive helicity). The ensuing changes in the initial CR distribution depend on these differing spectral properties. In a broad sense, the distribution functions presented here span a continuum of CR anisotropy, with low drift-velocity power-law, high drift-velocity power-law, and ring distribution simulations from low to high anisotropy, respectively. The linear dispersion relations for highly anisotropic CR distributions predict the emergence of wave spectra that would be unable to fully isotropize the CRs at quasilinear wave amplitudes. The missing modes must therefore be generated in the nonlinear phase of instability if isotropy is to be obtained. The quasimonochromatic spectra evoked by the ring distribution resulted in inefficiencies in scattering due to resonant trapping, which in turn led to an indefinite period of super-Alfv\'enic drift. The spectra from the power-law CR distributions are less susceptible to this effect, particularly in the low drift velocity case where left-handed modes are plentiful. 

Thus far we have not provided a quantitative measure of CR anisotropy, and instead have simply delineated low and high anisotropy as producing predominantly linear and right-hand polarized resonant modes, respectively. For the application at hand, the most desirable measure of anisotropy would be the ratio of maximal right-handed growth rate to the left-handed counterpart $\Delta\Gamma\equiv \Gamma^{\rm PR}_{\rm cr}/\Gamma^{\rm PL}_{\rm cr}$. Unfortunately, this ratio does not have an analytical form in the general case (Eq.\ \eqref{eqn:zwei}), and the assumptions used in deriving Eq.\ \eqref{eqn:kuls} give $\Delta\Gamma = 1$ for all values of the CR drift velocity $v_{\rm dr}$. One alternative measure of anisotropy is the relative fraction of CRs on either side of the $v_x = v_{\rm A}$ divide,
\begin{align}\label{eqn:anis}
a &\equiv \frac{\int_0^\infty \int_{\mu_{\rm A}}^1 \int_0^{2\pi} f(p) p^2 \du p \du\mu \du\phi}{\int_0^\infty \int_{-1}^{\mu_{\rm A}} \int_0^{2\pi} f(p) p^2 \du p \du\mu \du\phi}\\
&= \frac{1-\mu_{\rm A}}{1+\mu_{\rm A}},
\end{align}
where all quantities are measured in the isotropic CR frame. The pitch-angle cosine 
\begin{align}
\mu_{\rm A} &= \frac{v_{\rm A} - v_{\rm dr}}{v_{\rm cr}(1 - \frac{v_{\rm dr}v_{\rm A}}{c^2})},
\end{align} 
with characteristic CR velocity $v_{\rm cr}\sim c $, represents the pitch angle at which a typical CR travels with the Alfv\'en velocity in the CR frame (including the relativistic velocity correction).

\begin{figure}[t]
\centering
\includegraphics[width=1.0\linewidth,clip=true]{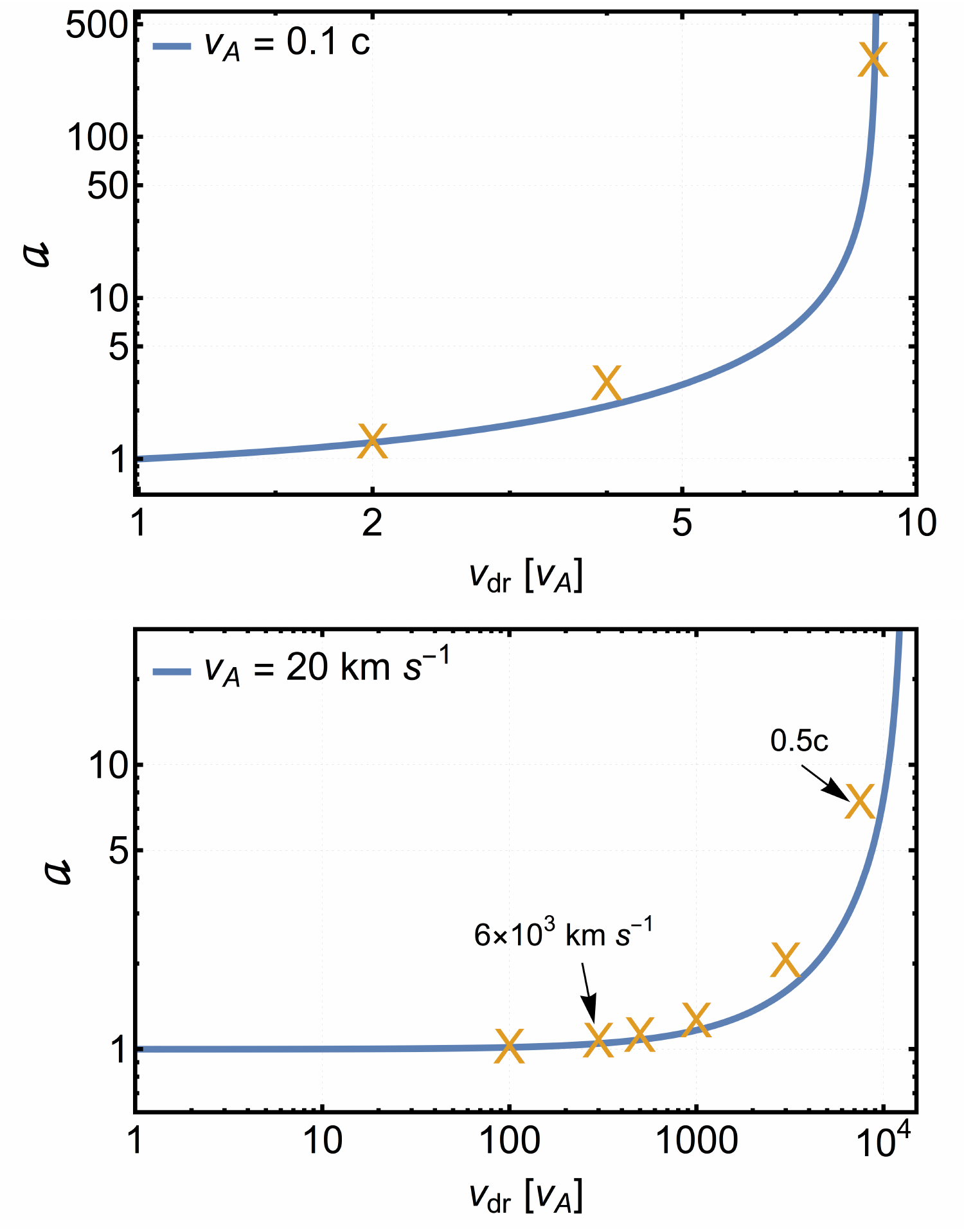}
\caption{The anisotropy parameter $a$ (Eq.\ \eqref{eqn:anis}) as a function of CR drift velocity $v_{\rm dr}$, where $v_{\rm A} = 0.1 c$ (top panel) and $v_{\rm A} = 20$ km s$^{-1}$ (bottom panel). The numerically obtained growth rate ratios $\Delta \Gamma \equiv \Gamma^{\rm PR}_{\rm cr, max}/\Gamma^{\rm PL}_{\rm cr, max}$ are shown for comparison (orange X symbols), where we used Eq.\ \eqref{eqn:zwei} and the power-law CR distribution Eq.\ \eqref{eqn:pow}. For $a$, we use $v_{\rm cr} = 0.87c$, corresponding to $\gamma_{\rm cr} = 2$.}
\label{fig:muA}
\end{figure}

In Figure \ref{fig:muA} we plot the anisotropy parameter $a$ as a function of drift velocity $v_{\rm dr}$. In the top panel we use parameters motivated by our power-law CR simulations ($v_{\rm A}= 0.1$c), while the parameters of the bottom panel are motivated by the conditions around SNR shocks ($v_{\rm A} = 20$ km s$^{-1}$). The anisotropy parameter is only useful in the range $a> 1$, since a value one (or less) would imply stability with the canonical streaming instability setup. It is also only valid up to the pole at $\mu_A  = -1$, after which there are no CRs of the given $v_{\rm cr}$ with $v_x < v_A$. The instability growth rate depends on both the CR density \emph{and} the slope of the CR distribution around $k_{\rm res}$, so $a$ is an imperfect proxy for $\Delta\Gamma$. However, we have that $a \le \Delta\Gamma$ for all valid values of $a$, with $a \approx \Delta\Gamma$ to within $\Sim 10\%$ for $a \lesssim 1.5$. For comparison, we mark numerically calculated values of $\Delta\Gamma$ in Figure \ref{fig:muA} (orange ``X" symbols). 

In the bottom panel of Figure \ref{fig:muA} we use $v_{\rm A} = 20$ km s$^{-1}$, which is a typical value for the ISM surrounding a SNR shock. Near the shock we have $v_{\rm dr} \approx v_{\rm sh}$, where $v_{\rm sh}$ is the velocity of the shock along the background magnetic field. If we take as a characteristic shock velocity $v_{\rm sh}= 6\times 10^{3}$ km s$^{-1} = 300 v_{\rm A}$, then we have $a \approx 1.05$ and $\Delta\Gamma = 1.08$. These conditions imply a modest predominance of right-handed waves over their left-handed counterparts. Far from the SNR shock, CRs will expand anistropically into the ISM, suggesting that $v_{\rm dr}\approx 0.5c = 7500 v_{\rm A}$. This higher value of $v_{\rm dr}$ gives $a=3.7$ and $\Delta\Gamma = 7.4$, indicating the production of strongly right-handed wave spectra. 

The contrivances of periodic simulations do not provide a wholly accurate representation of CR transport in general. Nevertheless, the qualitative transition to the predominance of right-handed resonant modes is applicable to the physical systems that these models are intended to represent. Indeed, the suppression of left-handed modes in an aperiodic CR outflow is likely to be more severe than indicated here, due to the complete absence of CRs with $\mu < v_{\rm A}/v \sim 0$. In this case we have $a$, $\Delta \Gamma \rightarrow \infty$, so that these systems will suffer the inefficiencies in obtaining CR isotropy detailed above.

The simulations performed in this study are limited by the numerical constraints and computational costs incurred by the use of the PIC method.  The electromagnetic noise floor established by small-scale variations in the number of particles per numerical cell necessitates the use of high-current CR distributions to produce satisfactory signal-to-noise ratios. Pushing to smaller growth rates becomes prohibitively expensive owing to the slow scaling of the noise-floor amplitude with particles per cell. Reducing the temperature of the background plasma can similarly reduce the noise floor amplitude. However, the PIC method cannot sustain plasma temperatures such that the Debye length is shorter than a few cells, thus requiring additional spatial resolution as an overhead cost. To suppress the growth of nonresonant (Bell) modes while maintaining a cold plasma ($\beta_{\rm i} \equiv 2v_{\rm th,i}^2/v_{\rm A}^2\ll 1$), we set the Alfv\'en speed to the unnaturally high value of $v_{\rm A} = 0.1 c$,  significantly departing from the standard magnetostatic approximation. These large wave velocities have the effect of expanding the resonance gaps between modes with differing propagation-polarization types. Sufficient signal-to-noise ratios were obtained by pushing waves to large amplitudes $\delta B \sim 0.1$, invalidating the QLT approximation. Additionally, the finite amplitude wave electric fields $\delta E \sim v_{\rm A}\delta B/c$ cause diffusion in momentum on top of the basic pitch-angle scattering (e.g., Figure \ref{fig:scat}). Finally, the temporal and spatial scales over which instability is realized in our power-law CR simulations begin to exceed the limits of what could be considered acceptable usage of the PIC method. In our worst case, for example, the total energy of simulation Hi1 has increased $29\%$ from its initial value by $t=12000 \ \Omega_0^{-1}$ ($\Sim 10^7$ time steps) owing to interpolation errors in calculating the the Lorentz force on particles from the discrete electromagnetic fields \citep{1991ppcs.book.....B,2013A&A...558A.133M}. While the short time scales of the ring distribution simulations do not noticeably suffer in this respect, the nonconservation of energy in our power-law distribution simulations begins to cast doubt on the long-term results beyond $t \gtrsim 10^4 \ \Omega_0^{-1}$, preventing the execution of simulations on longer time scales.

The primary advantage of employing the PIC method is the resolution of physics down to electron-kinetic scales. When the wave spectrum is in the quasilinear regime, pitch-angle scattering occurs via small deflections. The resolution of high-$k$ modes becomes very important for the cascade of CRs to sufficiently small $\mu$ such that magnetic mirroring (or other post-quasilinear effects) can bridge the resonance gap. Numerical methods invoking magnetohydrodynamical approximations may fail to sustain the short wavelength modes necessary to achieve Alfv\'enic CR drift. Although we did not discuss the behavior of the CR electrons (CRe) in detail in this work, this fine spatial resolution permitted us to observe the gyroresonant streaming instability for electrons under certain conditions. The higher frequency of electron gyration results in shorter wavelengths of resonant waves, while their negative charge reverses the polarization relationship compared to CR ions. These high $k$, left-handed CRe resonant waves are quickly damped away by ion-cyclotron resonance in the background plasma.

One goal of this study was to observe the mechanisms of saturation that are intrinsic to the gyroresonant streaming instability, i.e.\ the cessation of unstable growth owing only to wave-CR interactions, such as gradient flattening and resonant trapping. However, wave damping can contribute to saturation in general. The most commonly discussed types include ion-neutral friction \citep{1969ApJ...156..445K,1971ApL.....8..189K,1982ApJ...259..859Z}, nonlinear Landau resonance \citep{1971PhRvL..27.1349H,1973Ap&SS..24...31L,Volk:1982fk}, and turbulent cascade \citep{2004ApJ...604..671F, 2002PhRvL..89B1102Y,2004ApJ...614..757Y}. These extrinsic wave damping channels are expected to regulate unstable growth and make important contributions to the steady-state transport of CRs in the interstellar and intracluster media (e.g., \citealt{2001ApJ...553..198F,2013MNRAS.434.2209W}). Therefore no study on the streaming instability can be widely applicable to CR transport in nature without taking the latter mechanisms into consideration. The simulations herein represent the undamped limit of unstable CR behavior. They provide an upper bound on the strength of CR scattering in the presence of self-generated turbulence, since the wave spectra will generally be smaller in amplitude when damping is present. In principle, nonlinear Landau damping could be captured by PIC simulations without any additional physical modeling; however, this mechanism does not enter into the cold simulations presented here because it becomes significant only at moderate plasma beta $\beta_{\rm i}\gtrsim 1$.

The use of periodic simulations simplifies the interpretation of the physics, but the finite reservoir of free momentum available in the initial CR distribution artificially limits the amplitude that unstable waves can reach (Eq.\  \eqref{eqn:sat}). In this sense, the periodicity of the computational domain imitates the effects of extrinsic damping. In a scenario where CR current is continuously injected, as in the outskirts of supernova remnants or galactic halos, waves should easily reach $\delta B/B_0 \sim 1$ in the absence of extrinsic damping owing to the effectively unlimited supply of free momentum. CRs that subsequently propagate in this large-amplitude turbulent field would have their anisotropy rapidly reduced, establishing  self-confinement. On the other hand, strongly damped media would permit CRs to propagate unhindered by wave interactions \citep{2001ApJ...553..198F}.

The ISM is not a monolithic substrate throughout the extent of a galaxy -- the propagation of CRs within the various phases of the ISM adds an additional layer of complexity to the problem. The behavior of instability and subsequent CR transport can vary drastically depending on the properties of the local medium. For present purposes, the most important phases of ISM are the warm neutral medium (WNM), warm ionized medium (WIM), and hot ionized medium/coronal gas (HIM), which together take up the vast majority of the volume in the Milky Way galactic disk, and also the CR halo which surrounds the latter out to a few kpc.

The WNM and WIM are characterized by $\beta_{\rm i}\sim 0.1$ plasmas with small and large ionization fractions, respectively. In these phases, neutral atoms couple to the ionized component via collisions, allowing the transfer of wave energy in the electromagnetic fields to thermal energy in the neutrals. The ion-neutral damping rate $\Gamma_{\rm IN}$ is a nearly constant function of wavenumber $k$ for sufficiently high frequency waves $\omega(k) > \nu_{\rm IN}$, where $\nu_{\rm IN}$ is the ion-neutral scattering rate \citep{1982ApJ...259..859Z,2016MNRAS.461.3552N}. Resonant waves in the WNM are expected to be completely damped, allowing CRs to freely stream through these regions \citep{2001ApJ...553..198F}. 

\begin{figure}[t]
\centering\includegraphics[width=\linewidth,clip=true]{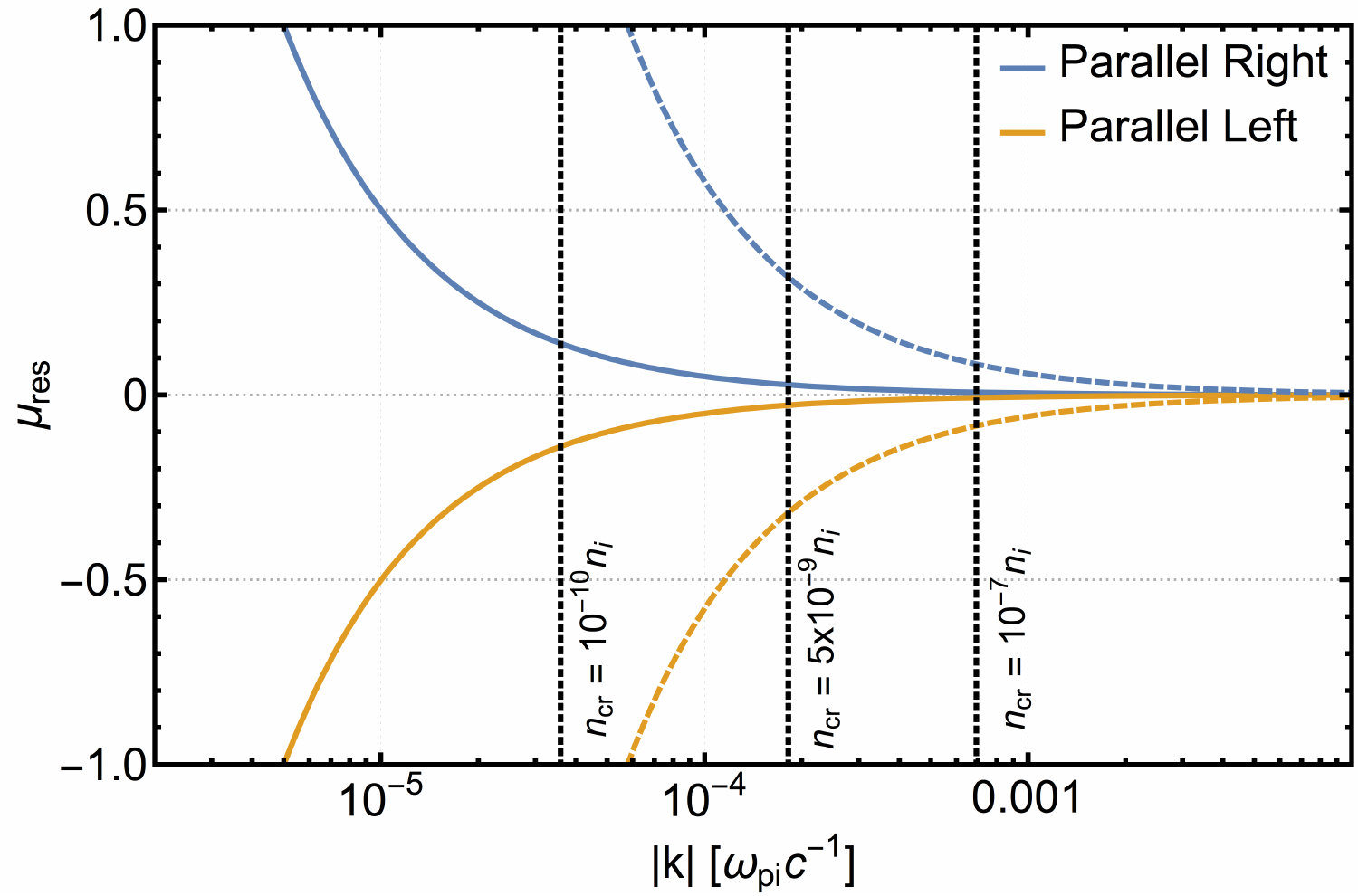}
\caption{Expansion of resonance gaps via ion-neutral damping. Gyroresonance curves (Eq.\  \eqref{eqn:mures}) are shown for parallel right-handed (blue) and parallel left-handed (orange) modes interacting with CRs of Lorentz factors $\gamma = 2$ (dashed) and $\gamma = 20$ (solid). Vertical black dotted lines denote the critical wavenumber $k_{\rm IN}$ at which ion-neutral damping overtakes unstable growth from CR streaming instability at CR densities of $n_{\rm cr}/n_{\rm i} = 10^{-10}$, $5\times 10^{-9}$, and $10^{-7}$ and typical parameters for the warm ionized medium (see text). At a given CR density, modes with $k \ge k_{\rm IN}$, where $k_{\rm IN} = 3.6\times 10^{-5}$, $1.8\times10^{-4}$, and $6.9\times 10^{-4} \ \omega_{\rm pi} c^{-1}$,  are unable to grow, resulting in substantially expanded resonance gaps. Low energy CRs are particularly susceptible to this effect.}
\label{fig:resgapIN}
\end{figure}

Even in the WIM, where streaming instability growth rates can plausibly exceed the ion-neutral damping rates if CR densities are large \citep{2013ApJ...767...87W}, the spectra of waves will be significantly impacted. Obtaining isotropy in streaming CRs requires efficiently mirroring particles across the $\mu \sim 0$ resonance gap, which in turn depends on the existence of short-wavelength modes to resonantly scatter CRs into the mirroring region $|\mu' | \le \mu_M' $. These high-$k$ modes are particularly susceptible to damping because of the $\Gamma_{\rm cr} \propto k^{3-\alpha}$ scaling of Eq.\  \eqref{eqn:kuls}, and the effective resonance gap will widen should damping dominate over growth for these modes.  Assuming typical WIM parameters along with a nearly isotropic power-law CR distribution with drift $v_{\rm dr} = 2 v_{\rm A}$, we can calculate the wavenumber $k_{\rm IN}$ at which ion-neutral damping overtakes gyroresonant growth using Eq.\  \eqref{eqn:kuls} and Eq.\  (A4) of \cite{1982ApJ...259..859Z}. We illustrate this behavior in Figure \ref{fig:resgapIN}, which shows gyroresonance curves (Eq.\  \eqref{eqn:mures}) for parallel right-handed (blue) and parallel left-handed (orange) modes interacting with CRs of Lorentz factors $\gamma = 2$ (dashed) and $\gamma = 20$ (solid). Here we have adopted the parameters $T = 8\times 10^3$ K, $v_{\rm A} = 10^{-4} c$, $n_{\rm i} = 0.315$ cm$^{-3}$, and $n_n = 0.035$ cm$^{-3}$, where $n_n$ is the density of neutral hydrogen atoms, giving an ion-neutral scattering rate of $\nu_{\rm IN} \approx 5.4\times 10^{-10}$ s$^{-1}$ \citep{1971ApL.....8..189K}. We choose a power-law index $\alpha=4$ and CR densities across the range $n_{\rm cr}/n_{\rm i} = 10^{-10}$, $5\times 10^{-9}$, and $10^{-7}$ to capture reasonable lower and upper bounds on the instability growth rates. 

The severity of resonance-gap broadening depends on both the growth rates of instability and the energies of CRs under consideration. CRs with higher energy are less impacted by this effect because their reduced gyrofrequencies push the corresponding gyroresonance to longer wavelengths (\cite{1981A&A....98..161A} reached a similar conclusion for the influence of ion-cyclotron damping in high $\beta$ plasmas). Examining the $k_{\rm IN}$ line corresponding to  $n_{\rm cr}/n_{\rm i} = 10^{-10}$ in Figure \ref{fig:resgapIN}, we see that the entire range of $k$ that $\gamma = 2$ CRs are capable of interacting with will be damped away, allowing them to ballistically stream through the WIM. Even for faster growth rates, self-confinement for low energy CRs can still fail. With $n_{\rm cr}/n_{\rm i} = 10^{-7}$, the resonance gap for the $\gamma = 2$ CRs is still of the order $\delta \mu \sim 0.1$, requiring $|\delta B/B_0|$ fluctuations of comparable amplitude for mirroring to operate. Highly anisotropic CR distributions are even more vulnerable to ion-neutral damping, owing to the suppressed growth rates for left-handed modes that would be easily dominated by damping.

Anisotropy may also play an important role for CRs in the HIM, galactic halo, and ICM, where plasma temperatures are too high to facilitate ion-neutral damping. In these moderate to high-beta environments ($\beta_{\rm i}\gtrsim 1$), the nonlinear Landau damping (NLLD) and turbulent damping mechanisms are considered the largest threats to CR self-confinement. A commonly overlooked property of NLLD is that the flow of energy is qualitatively different in linearly and circularly polarized spectra. In the linearly polarized case, waves are purely damped while heating background particles. The dependence of the damping rate on the wave amplitude $\Gamma_{\rm NLLD}\propto (\delta B/B_0)^2$ allows the derivation of simple estimates for critical wave amplitudes such that unstable growth balances damping \citep{2005ppfa.book.....K,2013MNRAS.434.2209W}. However, in the case of NLLD of a circularly polarized spectrum, high-frequency waves damp while low-frequency waves grow, resulting in an inverse cascade of wave energy \citep{1973Ap&SS..24...31L}. As the low frequency waves grow, the high frequencies are damped at increasing rates, invalidating the assumption of balance between growth and damping. For this reason, we cannot offer simple estimates of steady-state damping rates and wave amplitudes.  

To fully understand the behavior of the CR instability under the influence of NLLD would require numerical simulations of the nonlinear physics. However, self-consistently simulating CR-driven growth and NLLD with the PIC method is prohibitively difficult because of the computational cost of reducing the electromagnetic noise floor, particularly if wave amplitudes of $\delta B/B_0 \sim 10^{-3}$ (as estimated by, e.g., \cite{2013MNRAS.434.2209W}) are desired. Such simulations might be achieved with hybrid-kinetic codes that treat the background ions as individual particles, assuming that the noise floor is substantially reduced compared to PIC simulation at comparable computational expense. An additional difficulty arises due to the possibility for the NLLD to saturate via resonant trapping of background ions \citep{Volk:1982fk} by a comparable mechanism to the saturation of the ring distribution-driven instability for CRs discussed above. Should the NLLD mechanism reach saturation, unstable growth would continue unhindered and the long-term evolution of the system would be only trivially altered from the undamped case. \cite{2001ApJ...553..198F} have argued that collisions in the background plasma would prevent saturation of NLLD, at least in the ISM. Thus, a full study of the growth and damping problem would require the inclusion of a model for ion-collision physics. An adequately detailed quantitative treatment of the CR-NLLD growth and damping problem is beyond the scope of this article; we therefore defer further to forthcoming publications for further discussion.

Although there is ample evidence that CRs are strongly isotropized by scattering in the ISM, the origin of CRs from point sources (e.g., supernova remnants) suggests that their initial distributions are anisotropic. While the large Lorentz boosts of our high-anisotropy power-law simulations are unrealistic for the most likely CR sources, these distributions qualitatively capture the realistic behavior of anisotropic CR distributions as they propagate away from their sources. The time-dependent evolution of this natal anisotropy has long been ignored in favor of the simple analytical properties of steady-state solutions using nearly isotropic CR distribution functions. Such approximations have yet to receive validation from nonlinear numerical computations. 

The simulations presented herein are intended to provide an initial step towards the time-dependent solution of CR evolution in natural environments. With the caveats addressed above, these results suggest that there are previously under-appreciated difficulties in obtaining self-confinement when the CRs are strongly skewed towards a single propagation direction. If this is indeed the case for a more realistic physical setup, then the self-confinement paradigm may be insufficient to explain the observed CR properties. The most likely mechanism for CR confinement would then be scattering on the turbulent field of compressible MHD modes, which \cite{2008ApJ...673..942Y} have demonstrated can account for most, if not all, CR scattering in the ISM. Scattering on the fast wave cascade is particularly appealing as a confinement mechanism because it is, in principle, insensitive to the initial CR anisotropy -- the full range of waves needed to isotropize CRs exists independently of the streaming instability.

\section{Conclusions}\label{sec:conc}
We have demonstrated the excitation of cosmic ray self-generated turbulence via the gyroresonant streaming instability. Using the particle-in-cell numerical method in one spatial dimension, we showed that the instability growth rates are in satisfactory agreement with the predictions of the linear dispersion relation from plasma kinetic theory. Diverging from the standard assumptions of small CR drifts with $v_{\rm dr} \gtrsim v_{\rm A}$, we gave special attention to the behavior of instability when CRs are highly anisotropic. This anisotropic setup idealizes the case where CRs flow outward from a relatively small injection region. In this regime, with power-law CR distributions, the growth rates of parallel-propagating right-handed modes become enhanced and shift to longer wavelength compared to the standard approximate growth rate formula (Eq.\  \eqref{eqn:kuls}). Parallel-propagating left-handed modes are suppressed owing to the reduction in CRs that satisfy the relevant resonance conditions.

The properties of the emergent wave spectra played an important role in the subsequent backreaction on the CR distribution, particularly during the nonlinear phase of instability. We saw that CRs were not able to efficiently go beyond the small-$\mu$ barrier when positive-helicity waves (i.e., parallel-propagating left-handed modes) did not contain substantial energy density, even with relatively large-amplitude waves ($\delta B/B_0 \sim 0.1$). While large-amplitude effects (e.g.,\ magnetic mirroring) did occur as anticipated, these mechanisms are only important around $\mu \sim v_{\rm ph}/v$. In order to reach full isotropy, it was necessary to excite additional positive helicity modes in the nonlinear phase of instability that allow CRs to span the entirety of $-1 \le \mu \le 1$. These additional waves necessarily grow at slower rates than the fastest-growing mode predicted by linear theory, owing to the reduced number densities of CRs that achieve resonance with these waves, and therefore are more susceptible to domination by extrinsic damping. In the latter case, it is conceivable that highly anisotropic CR distributions can retain large drift velocites $v_{\rm dr}\sim c$ even in lightly damped media.

As the initial CR anisotropy is reduced, the canonical behavior of the streaming instability emerges. When $v_{\rm dr} \gtrsim v_{\rm A}$, the growth rates of left- and right-handed modes become degenerate, resulting in linearly polarized Alfv\'en waves. Here, the generation of additional modes is not necessary to achieve isotropy so long as magnetic mirroring is effective in moving CRs through the resonance gaps.

Ultimately, the saturation of resonant instability occurs via the flattening of CR distribution gradients. The action of stochastic wave-particle interactions compounds into nonlinearities that make the prediction of exact amplitudes of individual modes intractable, except in highly simplified scenarios. The saturation of instability in a periodic setting is constrained by the finite quantity of free momentum contained by the initial CR anisotropy. This allowed us to predict an upper bound on the amplitude of the self-generated turbulence excited by an arbitrary CR distribution (Eq.\  \eqref{eqn:sat}). This upper bound is achieved when CRs are able to reduce their drift velocity to that of the resonant waves, $v_{\rm dr} = v_{\rm A}$, via pitch-angle scattering. The resonant trapping effect provided a method for explicitly calculating the saturated wave amplitude of quasimonochromatic waves produced by the ring distribution (Eq.\  \eqref{eqn:beamsat}), while the quasilinear resonant scattering rate facilitated the same for the broad spectra of power-law distributed CRs (Eq.\  \eqref{eqn:plsat}). Finally, the mean free time for quasilinear scattering (Eq.\  \eqref{eqn:tmu}) allowed us to predict the duration over which a CR distribution with small initial anisotropy would relax. We emphasize that caution should be maintained when applying these estimates to aperiodic domains.

Although the model systems studied in this work are contrived to simplify the complexities of wave-particle interactions, we have distilled the problem of the gyroresonant streaming instability down to a foundation of physics from which future studies can proceed. In the ``undamped" limit we have considered here, sufficiently dense CR distributions can always produce large-amplitude waves that in turn strongly scatter the CRs.  Conversely, CR distributions of small density and/or initially large anisotropy may have difficulty in reaching Alfv\'enic drift velocities. The inclusion of extrinsic wave damping channels would exacerbate any of the inefficiencies in developing isotropy observed in this work. Given the expected importance of the compressible MHD cascade in scattering CRs, in addition to its role in dissipating the energy of resonant modes, particular attention in future studies should be given to the behavior of instability in the presence of an empirically motivated turbulent spectrum. Finally, galactic winds are expected to manifest via instability driven by large-scale spatial inhomogeneities of CRs. A more realistic simulation setup would remove the constraints of periodicity to examine the formation of spatially dependent quasi-steady state structures in an expanding bubble of CRs.

\acknowledgements
We appreciate the enlightening discussions held with X.\ Bai, D. Caprioli, R.\ Kumar, S.\ A.\ Mao, S.~P.\ Oh, E.\ Ostriker, I. Plotnikov, and E.\ Zweibel on the topics presented herein. We also acknowledge the comments and suggestions of S.\ Komarov, M.\ Shalaby, and the anonymous referee in revising the pre-print version of this article. C.~H. gratefully recognizes support from the Department of Energy National Nuclear Security Administration Stewardship Science Graduate Fellowship under grant DE-NA0002135, and from Krell Institute. A.~S.\ acknowledges the support of NSF grant AST-1814708 and the Simons Foundation (grant 267233).
The simulations presented in this article used computational resources supported by the PICSciE-OIT High Performance Computing Center and Visualization Laboratory at Princeton University.

\appendix
\section{On Polarization, Helicity, \& Gyroresonance}\label{sec:pol}
The canonical Alfv\'en wave is a periodic transverse fluctuation with circular polarization. Although there are many conventions for defining the wave quantities, we adopt the one that allows us to self-consistently treat waves with all combinations of parallel/antiparallel-propagation (with respect to $\boldsymbol{B}_0 = B_0\hat{x}$) and right/left-handed circular polarization. The transverse magnetic-field amplitude is of the form $\delta B = |\delta B|\exp{(ikx - i\omega t)}$, where $B_y = \operatorname{Re} \delta B$ and $B_z = \operatorname{Im} \delta B$, and similarly for other wave quantities.  

We distinguish between the \emph{polarization} and the \emph{helicity} of a transverse wave. The polarization is defined by the sense in which a wave field vector rotates as a function of time when viewed at fixed position (e.g., $x = 0$), and is generally a frame-dependent property. When using the plasma rest frame as a basis for the definition of polarization, left-handed waves are capable of resonating with stationary ($v_x = 0$) ions, while right-handed waves can resonate with stationary electrons. The helicity is defined by the sense in which a wave field vector rotates as a function of position at fixed time (e.g., $t=0$), and is not a frame-dependent property. We have adopted a sign convention that allows the wave frequency $\omega$ and wavenumber $k$ to take on both positive and negative values. Thus we can describe the parallel left-handed and antiparallel right-handed waves as having positive helicity ($k>0$) and the parallel right-handed and antiparallel left-handed waves waves as having negative helicity ($k<0$), where parallel and antiparallel refer to positive and negative phase velocity $v_{\rm ph} = \omega/k$. To avoid further confusion we will refer to left/right-handedness in reference to polarization only and to positive/negative in reference to helicity only.

According to Eq.\ \eqref{eqn:res}, gyroresonance occurs for wave-particle pairs that rotate together. The gyroresonance condition is thus interpreted as a Doppler shift due to the relative velocity between wave and particle -- the wave frequency is equal to the gyrofrequency in the reference frame traveling with the particle parallel velocity $v_x$. The resonant particle then exchanges energy and momentum with a constant amplitude electromagnetic field. The same interaction can be considered in the frame traveling with the wave phase velocity, $v_{\rm ph} = \omega/k$, where the wave electric field goes to zero. In the frame of a wave with amplitude $\delta B$, resonant particles experience a constant $\boldsymbol{v}_\perp \times \delta\boldsymbol{B}$ Lorentz force that perturbs the gyromotion about the background magnetic field, resulting in pure pitch-angle scattering \citep{2005ppfa.book.....K}.

A particle of given charge (and associated gyromotion handedness) can resonate with right and/or left-handed waves, depending on the relative velocity between particle and wave, because of the frame dependence of polarization. For example, a positively charged ion (a ``left-handed" particle) with $\mu > v_{\rm ph}/v$ resonates with parallel right-handed (the so-called anomalous gyroresonance) and antiparallel left-handed waves because these waves both appear to rotate in the left-handed sense in the particle frame traveling with $v_x$. Similarly, a positively charged ion with $-v_{\rm ph}/v < \mu < v_{\rm ph}/v$ resonates with parallel and antiparallel-propagating left-handed waves. The relation between particle pitch-angle cosine and gyroresonance with wavenumber $k$ is exemplified in Figure \ref{fig:mures}.

\section{Dispersion Relations}\label{sec:disp}

In this section we examine the dispersion relations that are relevant to the simulations performed, looking both at the branches of waves that appear and their respective growth rates. One can derive the (linear) dispersion relation for parallel-propagating transverse waves in the presence of arbitrary particle species (denoted with index $s$) by linearizing the Vlasov equation \citep{Stix:1992td}:
\begin{align}\label{eqn:disp}
\frac{k^2 c^2}{\omega^2} &= 1   +\sum_{s} \chi_s ,
\end{align}
\begin{align}\label{eqn:chi}
\chi_s &\equiv \frac{2\pi^2 q_s^2}{\omega} \int_0^\infty dp \int_{-1}^{1} d\mu \frac{vp^2(1-\mu^2)}{\omega - k v \mu- \Omega_s} A[f_s]
\end{align}
\begin{align}\label{eqn:appa}
A[f] &\equiv \frac{\partial f}{\partial p} + \bigg(\frac{kv(p)}{\omega}- \mu\bigg)\frac{1}{p}\frac{\partial f}{\partial \mu},
\end{align}
where $q_s$, $\Omega_s$, and $f_s$ are, respectively, the charge, relativistic gyrofrequency, and distribution function of species $s$. Although our simulations will utilize small, but finite, temperatures for the background plasma, we will assume here a cold isotropic background for analytical simplicity. The background consists of ions and electrons with the distributions
\begin{align}
f_{\rm i}(p) &= \frac{n_{\rm i}}{4\pi p^2}\delta(p) \\
f_{\rm e}(p) &= \frac{n_{\rm i}}{4\pi p^2}\delta(p),
\end{align}
where $n_{\rm i}$ is the number density of background ions (and electrons, for charge neutrality). We now specify the CR distributions and discuss their properties separately.

\subsection{Ring Distribution}
Here we discuss ring distributed CRs (Eq.\  \eqref{eqn:ring}). Charge and current neutrality is enforced by making use of a cold beam of ``CR electrons" 
\begin{align}
f_{\rm cre}(p) &= \frac{n_{\rm cr}}{2\pi p^2}\delta(p - p_{\rm e}) \delta(\mu - 1),
\end{align}
where $n_{\rm cr}$ is the density of CRs (and CR electrons), $p_{\rm e}$ satisfies $v(p_{\rm e}) = v_{\rm dr}$, and $v_{\rm dr}$ is the bulk velocity of the CR distribution. For these distributions the integrals in Eq.\  \eqref{eqn:disp} are readily computed analytically to yield
\begin{align}
\begin{split}\label{eqn:ringdisp}
\frac{k^2 c^2}{\omega^2} &= 1 - \frac{\omega_{\rm pe}}{\omega}\frac{\omega_{\rm pe}}{(\omega - \Omega_{\rm e})} - \frac{\omega_{\rm pi}}{\omega}\frac{\omega_{\rm pi}}{(\omega - \Omega_0)}  \\
&- \frac{n_{\rm cr}}{n_{\rm i}} \frac{\omega_{\rm pi}^2}{\omega^2}\frac{(\omega - k v_{\rm dr})}{\gamma_{\rm cr}(\omega - kv_{\rm dr} - \Omega_{\rm cr})} - \frac{1}{2}\frac{n_{\rm cr}}{n_{\rm i}} \frac{\omega_{\rm pi}^2}{\omega^2}\frac{(k^2c^2 -\omega^2) v_\perp^2/c^2}{\gamma_{\rm cr}(\omega - kv_{\rm dr}- \Omega_{\rm cr})^2} \\
&- \frac{n_{\rm cr}}{n_{\rm i}} \frac{\omega_{\rm pe}^2}{\omega^2}\frac{(\omega - k v_{\rm dr})}{\gamma_{\rm cre}(\omega - kv_{\rm dr} - \Omega_{\rm cre})},
\end{split}
\end{align}
where $\omega_{\rm pe}$ and $\omega_{\rm pi}$ are the electron and ion plasma frequencies, $\Omega_0$, $\Omega_{\rm e}$, $\Omega_{\rm cr}$, and $\Omega_{\rm cre}$ are the gyrofrequencies of background ions, background electrons, CRs, and CR electrons, respectively, and $\gamma_{\rm cr}$ and $\gamma_{\rm cre} = (1-v_{\rm dr}^2/c^2)^{-1/2}$ are the CR and CR electron Lorentz factors, respectively \citep{1972JGR....77.5399W}. The solution to the dispersion relation is reduced to finding the roots $\omega(k)$. Fixing $k$ to real values, instability is obtained when $\omega(k)$ becomes complex valued. 

\begin{figure*}[t]
\centering\includegraphics[width=\linewidth,clip=true]{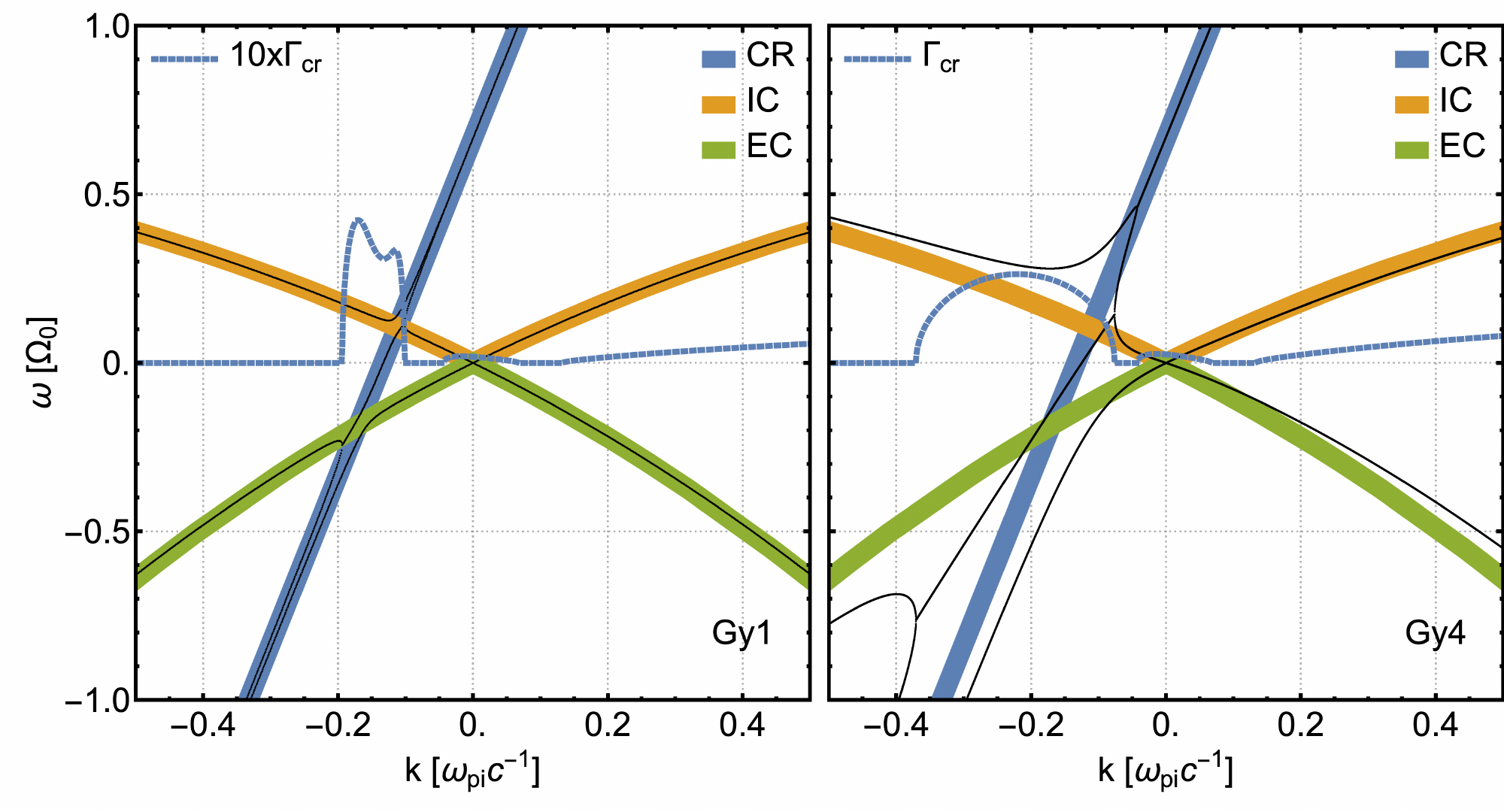}
\caption{Dispersion relations for simulations Gy1 (low CR density, left panel) and Gy4 (high CR density, right panel). We show the real (black lines) and imaginary (dashed blue lines) parts of the dispersion relation Eq.\ \eqref{eqn:ringdisp}. For comparison, we use thick solid lines to highlight the standard branches of the cold dispersion relation for the left-handed ion-cyclotron branch (IC, orange), the right-handed whistler/electron-cyclotron branch (EC, green), and a CR resonant mode obeying the dispersion relation $\omega = k v_{\rm dr} + \Omega_{\rm cr}$ (CR, blue). For simulation Gy1, parallel right-handed (lower left quadrant) and antiparallel left-handed (upper left quadrant) modes dominate the linear growth spectra. As the CR density increases (Gy4), the normal modes of the background plasma are significantly disrupted and the fastest-growing mode shifts to shorter wavelengths.}
\label{fig:beamdisp}
\end{figure*}

In Figure \ref{fig:beamdisp} we show the low-frequency dispersion relation for the parameters of simulation Gy1 (left) and Gy4 (right). A numerical solution of the real part of the full dispersion relation is shown with black lines, while solid colored lines indicate the standard cold dispersion relations for the left-handed ion-cyclotron branch (IC, orange), the right-handed whistler/electron-cyclotron branch (EC, green), and a CR mode obeying the dispersion relation $\omega = k v_{\rm dr} + \Omega_{\rm cr}$ (CR, blue). The numerical solution of the imaginary part of the dispersion relation $\Gamma$ is shown with a blue dashed line, corresponding to growth in the resonant mode. For the low CR density of Gy1, the peaks of the growth rate occur where the CR branch couples to the standard EC and IC branches, producing quasi-Alfv\'enic modes. Both parallel right-handed (lower-left quadrant) and antiparallel left-handed (upper-left quadrant) quasi-Alfv\'en modes are predicted to grow, with the former dominating the wave spectra during the linear phase. At the higher CR density of Gy4, the normal modes of the background plasma are significantly disrupted, while the maximum growth rate of the CR mode shifts to a shorter wavelength.

\subsection{Power-Law Distribution}

\begin{figure}[t]
\centering\includegraphics[width=\linewidth,clip=true]{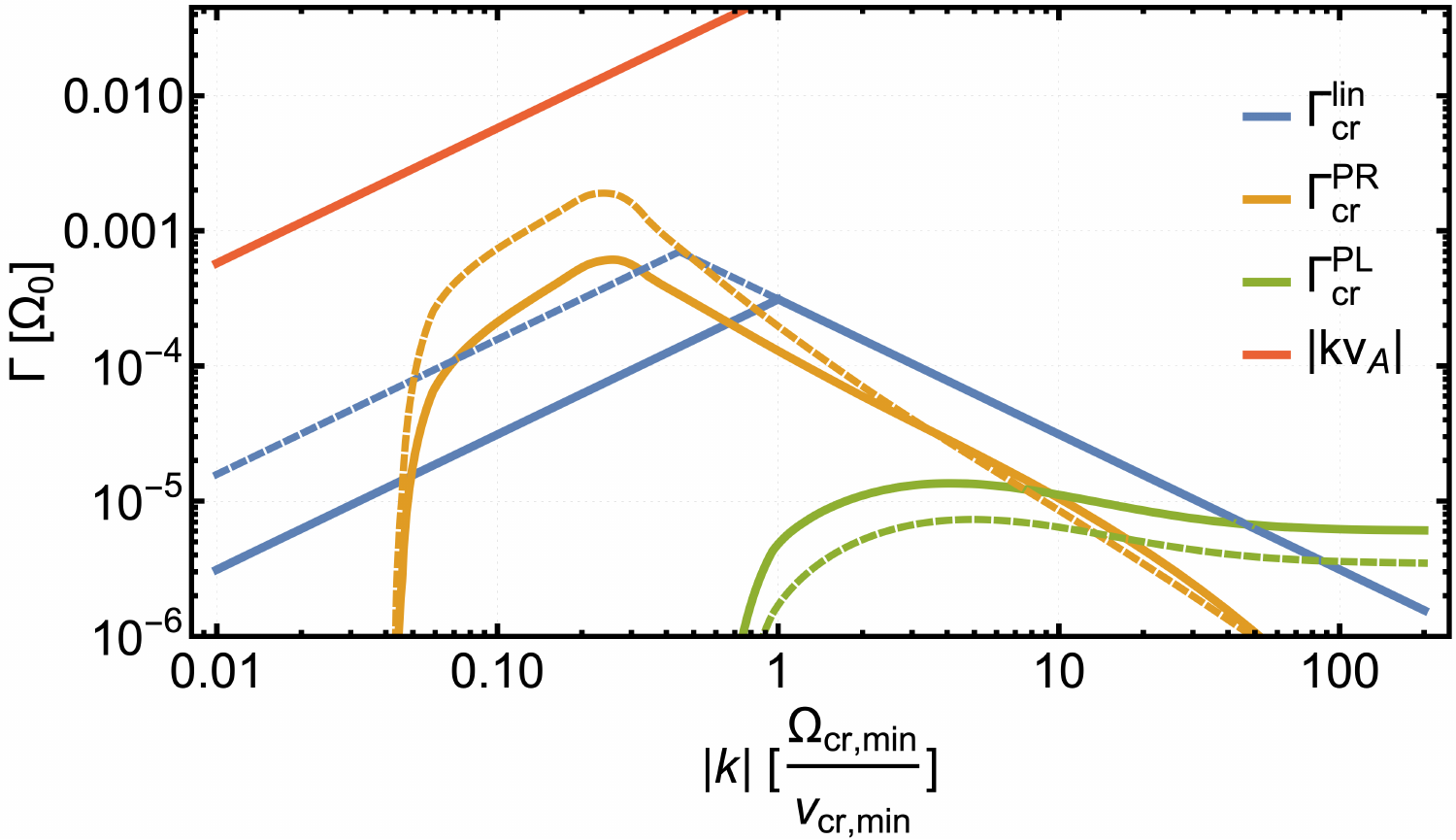}
\caption{Comparison between growth rates from varying distributions using the parameters of simulation Hi1. The orange and green lines depict right- and left-handed modes, respectively, with $f_{\rm PL}''\gamma''/\gamma$ (solid, particle boosted) and $f_{\rm PL}''$ (dashed, invariant). For these growth rates we have used the real part of the Hall-MHD Alfv\'en dispersion relation. The blue lines show the approximate growth rate (solid, Eq.\  \eqref{eqn:kuls}) and the relativistic correction (dashed). The real part of the low-frequency Alfv\'en dispersion relation is marked by the red line.}
\label{fig:appdisp}
\end{figure}

In section \ref{sec:num} we noted that the procedure of Lorentz boosting individual particles does not respect the Lorentz invariance of the distribution function. The reason is that the relativistic contraction of the position-space part of the phase-space volume (i.e., $\du^3x$) is neglected by this transformation. As a result, a factor of $\gamma''/\gamma$ is appended to the CR frame distribution function
\begin{align}
f_{\rm PL}'' \rightarrow f_{\rm PL}'' \frac{\gamma''}{\gamma},
\end{align}
with the latter function entering into Eq.\  \eqref{eqn:zwei} to calculate the growth rate in our simulations. In Figure \ref{fig:appdisp} we examine the effect this transformation has on the growth rates by comparing against the rates obtained with the Lorentz invariant distribution. Boosting individual particles reduces the peak growth rate while creating a shallower decline at $k>k_{\rm max}$. This decline is comparable to the $\propto k^{-1}$ scaling of the approximate formula Eq.\  \eqref{eqn:kuls}, whereas the invariant distribution produces a steeper decrease with $k$. The left-handed modes' growth rates are enhanced, but maintain a similar shape. The decrease/increase of the right/left-handed growth rates in our simulations means that the effects of anisotropy are suppressed relative to an invariant power-law distribution with the same parameters. Consequently, the difficulties of achieving CR isotropy are somewhat understated by our simulations.

We also show in Figure \ref{fig:appdisp} the approximate growth rate (solid blue; Eq.\  \eqref{eqn:kuls}) alongside the relativistic correction (dashed blue) suggested by \cite{1971ApL.....8..189K}, 
\begin{align}
\Gamma_{\rm cr,rel}^{\rm lin}(k) = \gamma_{\rm dr} \Gamma_{\rm cr}^{\rm lin}(\gamma_{\rm dr} k).
\end{align}
The relativistic correction increases the maximum growth rate, but, as discussed in the main text, fails to account for the disparity between right- and left-handed modes. The increased maximum growth rate is not comparable to the true growth rate with the invariant distribution. Instead, it (coincidentally) brings it to the level of the particle boosted distribution, which explains why our saturated amplitude predictions (Eq.\  \eqref{eqn:plsat}) fit the simulation data so closely (Figure \ref{fig:plsat}).

\section{Estimating the resonant relaxation time}\label{sec:relax}

Following chapters 2 and 3 of \cite{Shalchi:2009vg}, we derive the mean free time for scattering in the saturated self-generated turbulence. This will be the time scale for a typical CR to scatter from a characteristic pitch angle $\mu > v_{\rm ph}/v$ down to the mirroring pitch angle $\mu_M$. We must first obtain the QLT diffusion coefficient \citep{1966ApJ...146..480J,1968ApJ...152..997J,1969ApJ...156..445K} in slab geometry
\begin{align}
D_{\mu\mu}(k) &= \frac{(1-\mu^2)}{2} 4\pi^2 \Omega \frac{k g(k)}{B_0^2}, 
\end{align}
where $g(k)$ is the differential wave power spectrum, normalized to the total transverse field energy density
\begin{align}
\frac{\delta B^2}{8\pi} &= \int_{k_0}^\infty  g(k) \du k,
\end{align}
and $k_0$ is a reference wavenumber which we take to be the fastest-growing wavenumber $k_{\rm max}$. Motivated by our simulations, we model the spectrum as a power law that extends to infinity from the fastest-growing mode $k_{\rm max}$ with spectral index $\nu$
\begin{align}
g_\nu(k) &= \frac{\nu }{8\pi} \frac{\delta B^2}{k_0} \bigg(\frac{k}{k_0}\bigg)^{-\nu - 1},
\end{align}
where $\nu \neq 0,1$. For simplicity, we have assumed that the power law is unbroken. The pitch-angle diffusion coefficient is then
\begin{align}
D_{\mu\mu}(k) &= \frac{(1-\mu^2)}{2} \frac{\pi \nu }{2} \bigg(\frac{\delta B}{B_0}\bigg)^2 \bigg(\frac{k}{k_{\rm max}} \bigg)^{-\nu} \Omega. 
\end{align}

To estimate the relaxation time of the anisotropic CR distribution, we calculate the typical time scale for a CR of average energy $\gamma_0$ and pitch angle $\mu_0$ within the resonant band of the fastest-growing mode $k_{\rm max}$ to scatter down to the mirroring range $\mu_M$. This time scale is the mean free time \citep{1966ApJ...146..480J,1969ApJ...156..445K,Shalchi:2009vg}, 
\begin{align}
t_{\mu} &= \frac{3}{8}\int_{\mu_{\rm M}}^{\mu_0}\du\mu \frac{(1-\mu^2)^2}{D_{\rm \mu \mu}} \\
		  &\approx \frac{3}{4\pi} \bigg(\frac{\delta B}{B_0} \bigg)^{-2} \Omega^{-1} C_\mu, \label{eqn:tmuapp} \\
C_\mu  &\equiv \int_{\mu_{\rm M}}^{\mu_0} \du \mu (1-\mu^2)\bigg(\frac{\mu_0 v - v_{\rm A}}{\mu v - v_{\rm A}}\bigg)^{\nu },	  
\end{align}
where we have used the resonant condition (Eq.\  \eqref{eqn:res}) and $v = c\sqrt{1-\gamma_0^{-2}}$. Motivated by the spectra observed in section \ref{sec:res}, we take $\nu = 2$, which is a rough estimate of the power spectrum slope in the early nonlinear phase of instability. We then obtain
\begin{align}\label{eqn:cmu}
C_\mu	  &= \frac{(\mu_0 v- v_{\rm A})}{v^3}\bigg[(v^2 - v_{\rm A}^2)(x_M - 1)  + (\mu_0 v - v_{\rm A})^2 \bigg(\frac{1}{x_M} - 1\bigg) -2 v_{\rm A}(\mu_0 v - v_{\rm A})\ln x_M  \bigg], \\
x_M &\equiv \frac{(\mu_0 v- v_{\rm A})}{(\mu_M v- v_{\rm A})} \\
	&\approx \frac{(\mu_0 v- v_{\rm A})}{\mu_M' v}.
\end{align}
The scaling of $C_\mu$ with $\delta B/B_0$ is predominantly controlled by the $(x_M - 1)$ term, leading to $C_\mu \propto (\mu '_M)^{-1}\propto (\delta B/B_0)^{-1}$ until $\delta B/B_0 \gtrsim 0.1$. At larger wave amplitude, the mirroring pitch angle $\mu_M$ approaches the typical pitch angle of resonant particles $\mu_0$ and the formalism developed here begins to break down. Under typical quasilinear conditions, and with the spectral shape adopted here, we have $t_\mu \propto (\delta B/B_0)^{-3}$.


\begin{thebibliography}{}
\expandafter\ifx\csname natexlab\endcsname\relax\def\natexlab#1{#1}\fi

\bibitem[{Achterberg(1981)}]{1981A&A....98..161A}
Achterberg, A. 1981, Astronomy and Astrophysics, 98, 161

\bibitem[{Amato \& Blasi(2009)}]{2009MNRAS.392.1591A}
Amato, E., \& Blasi, P. 2009, Monthly Notices of the Royal Astronomical
  Society, 392, 1591

\bibitem[{Bai {et~al.}(2015)Bai, Caprioli, Sironi, \&
  Spitkovsky}]{2015ApJ...809...55B}
Bai, X.-N., Caprioli, D., Sironi, L., \& Spitkovsky, A. 2015, The Astrophysical
  Journal, 809, 55

\bibitem[{Bell(2004)}]{2004MNRAS.353..550B}
Bell, A.~R. 2004, Monthly Notices of the Royal Astronomical Society, 353, 550

\bibitem[{Bell(2005)}]{2005MNRAS.358..181B}
---. 2005, Monthly Notices of the Royal Astronomical Society, 358, 181

\bibitem[{Birdsall \& Langdon(1991)}]{1991ppcs.book.....B}
Birdsall, C.~K., \& Langdon, A.~B. 1991, Plasma Physics via Computer Simulation

\bibitem[{Breitschwerdt {et~al.}(1991)Breitschwerdt, McKenzie, \&
  Voelk}]{1991A&A...245...79B}
Breitschwerdt, D., McKenzie, J.~F., \& Voelk, H.~J. 1991, Astronomy and
  Astrophysics (ISSN 0004-6361), 245, 79

\bibitem[{Draine(2011)}]{2011piim.book.....D}
Draine, B.~T. 2011, Physics of the Interstellar and Intergalactic Medium by
  Bruce T. Draine. Princeton University Press

\bibitem[{Dupree(1966)}]{1966PhFl....9.1773D}
Dupree, T.~H. 1966, Physics of Fluids, 9, 1773

\bibitem[{Everett {et~al.}(2008)Everett, Zweibel, Benjamin, McCammon, Rocks, \&
  Gallagher}]{2008ApJ...674..258E}
Everett, J.~E., Zweibel, E.~G., Benjamin, R.~A., {et~al.} 2008, The
  Astrophysical Journal, 674, 258

\bibitem[{Farmer \& Goldreich(2004)}]{2004ApJ...604..671F}
Farmer, A.~J., \& Goldreich, P. 2004, The Astrophysical Journal, 604, 671

\bibitem[{Felice \& Kulsrud(2001)}]{2001ApJ...553..198F}
Felice, G.~M., \& Kulsrud, R.~M. 2001, The Astrophysical Journal, 553, 198

\bibitem[{Galinsky {et~al.}(1997)Galinsky, Shevchenko, Ride, \&
  Baine}]{1997JGR...10222365G}
Galinsky, V.~L., Shevchenko, V.~I., Ride, S.~K., \& Baine, M. 1997, Journal of
  Geophysical Research, 102, 22365

\bibitem[{Girichidis {et~al.}(2018)Girichidis, Naab, Hanasz, \&
  Walch}]{2018MNRAS.tmp.1588G}
Girichidis, P., Naab, T., Hanasz, M., \& Walch, S. 2018, Monthly Notices of the
  Royal Astronomical Society

\bibitem[{Girichidis {et~al.}(2016)Girichidis, Naab, Walch, Hanasz, Mac~Low,
  Ostriker, Gatto, Peters, W{\"u}nsch, Glover, Klessen, Clark, \&
  Baczynski}]{2016ApJ...816L..19G}
Girichidis, P., Naab, T., Walch, S., {et~al.} 2016, The Astrophysical Journal
  Letters, 816, L19

\bibitem[{Guo \& Oh(2008)}]{2008MNRAS.384..251G}
Guo, F., \& Oh, S.~P. 2008, Monthly Notices of the Royal Astronomical Society,
  384, 251

\bibitem[{Hollweg(1971)}]{1971PhRvL..27.1349H}
Hollweg, J.~V. 1971, Physical Review Letters, 27, 1349

\bibitem[{Ipavich(1975)}]{1975ApJ...196..107I}
Ipavich, F.~M. 1975, Astrophysical Journal, 196, 107

\bibitem[{Jiang \& Oh(2018)}]{2018ApJ...854....5J}
Jiang, Y.-F., \& Oh, S.~P. 2018, The Astrophysical Journal, 854, 5

\bibitem[{Jokipii(1966)}]{1966ApJ...146..480J}
Jokipii, J.~R. 1966, Astrophysical Journal, 146, 480

\bibitem[{Jokipii(1968)}]{1968ApJ...152..997J}
---. 1968, Astrophysical Journal, 152, 997

\bibitem[{Jones {et~al.}(1978)Jones, Birmingham, \&
  Kaiser}]{1978PhFl...21..347J}
Jones, F.~C., Birmingham, T.~J., \& Kaiser, T.~B. 1978, Physics of Fluids, 21,
  347

\bibitem[{Kulsrud \& Pearce(1969)}]{1969ApJ...156..445K}
Kulsrud, R., \& Pearce, W.~P. 1969, Astrophysical Journal, 156, 445

\bibitem[{Kulsrud(2005)}]{2005ppfa.book.....K}
Kulsrud, R.~M. 2005, Plasma physics for astrophysics / Russell M. Kulsrud.
  Princeton

\bibitem[{Kulsrud \& Cesarsky(1971)}]{1971ApL.....8..189K}
Kulsrud, R.~M., \& Cesarsky, C.~J. 1971, Astrophysical Letters, 8, 189

\bibitem[{Lebiga {et~al.}(2018)Lebiga, Santos-Lima, \&
  Yan}]{2018MNRAS.476.2779L}
Lebiga, O., Santos-Lima, R., \& Yan, H. 2018, Monthly Notices of the Royal
  Astronomical Society, 476, 2779

\bibitem[{Lee \& V{\"o}lk(1973)}]{1973Ap&SS..24...31L}
Lee, M.~A., \& V{\"o}lk, H.~J. 1973, Astrophys Space Sci, 24, 31

\bibitem[{Lerche(1967)}]{1967ApJ...147..689L}
Lerche, I. 1967, Astrophysical Journal, 147, 689

\bibitem[{Loewenstein {et~al.}(1991)Loewenstein, Zweibel, \&
  Begelman}]{1991ApJ...377..392L}
Loewenstein, M., Zweibel, E.~G., \& Begelman, M.~C. 1991, Astrophysical
  Journal, 377, 392

\bibitem[{Mao \& Ostriker(2018)}]{2018ApJ...854...89M}
Mao, S.~A., \& Ostriker, E.~C. 2018, The Astrophysical Journal, 854, 89

\bibitem[{Melzani {et~al.}(2013)Melzani, Winisdoerffer, Walder, Folini, Favre,
  Krastanov, \& Messmer}]{2013A&A...558A.133M}
Melzani, M., Winisdoerffer, C., Walder, R., {et~al.} 2013, Astronomy and
  Astrophysics, 558, A133

\bibitem[{Miller {et~al.}(1991)Miller, Gary, Winske, \&
  Gombosi}]{1991GeoRL..18.1063M}
Miller, R.~H., Gary, S.~P., Winske, D., \& Gombosi, T.~I. 1991, Geophysical
  Research Letters (ISSN 0094-8276), 18, 1063

\bibitem[{Nava {et~al.}(2016)Nava, Gabici, Marcowith, Morlino, \&
  Ptuskin}]{2016MNRAS.461.3552N}
Nava, L., Gabici, S., Marcowith, A., Morlino, G., \& Ptuskin, V.~S. 2016,
  Monthly Notices of the Royal Astronomical Society, 461, 3552

\bibitem[{Niemiec {et~al.}(2008)Niemiec, Pohl, Stroman, \&
  Nishikawa}]{2008ApJ...684.1174N}
Niemiec, J., Pohl, M., Stroman, T., \& Nishikawa, K.-I. 2008, The Astrophysical
  Journal, 684, 1174

\bibitem[{Park {et~al.}(2015)Park, Caprioli, \&
  Spitkovsky}]{2015PhRvL.114h5003P}
Park, J., Caprioli, D., \& Spitkovsky, A. 2015, Physical Review Letters, 114,
  085003

\bibitem[{Philippov {et~al.}(2015)Philippov, Spitkovsky, \&
  Cerutti}]{2015ApJ...801L..19P}
Philippov, A.~A., Spitkovsky, A., \& Cerutti, B. 2015, The Astrophysical
  Journal Letters, 801, L19

\bibitem[{Recchia {et~al.}(2016)Recchia, Blasi, \&
  Morlino}]{2016MNRAS.462.4227R}
Recchia, S., Blasi, P., \& Morlino, G. 2016, Monthly Notices of the Royal
  Astronomical Society, 462, 4227

\bibitem[{Riquelme \& Spitkovsky(2009)}]{2009ApJ...694..626R}
Riquelme, M.~A., \& Spitkovsky, A. 2009, The Astrophysical Journal, 694, 626

\bibitem[{Riquelme \& Spitkovsky(2010)}]{2010ApJ...717.1054R}
---. 2010, The Astrophysical Journal, 717, 1054

\bibitem[{Ruszkowski {et~al.}(2017{\natexlab{a}})Ruszkowski, Yang, \&
  Reynolds}]{2017ApJ...844...13R}
Ruszkowski, M., Yang, H. Y.~K., \& Reynolds, C.~S. 2017{\natexlab{a}}, The
  Astrophysical Journal, 844, 13

\bibitem[{Ruszkowski {et~al.}(2017{\natexlab{b}})Ruszkowski, Yang, \&
  Zweibel}]{2017ApJ...834..208R}
Ruszkowski, M., Yang, H. Y.~K., \& Zweibel, E. 2017{\natexlab{b}}, The
  Astrophysical Journal, 834, 208

\bibitem[{Schlickeiser(1989)}]{1989ApJ...336..243S}
Schlickeiser, R. 1989, Astrophysical Journal, 336, 243

\bibitem[{Schlickeiser \& Yoon(2015)}]{2015PhPl...22g2108S}
Schlickeiser, R., \& Yoon, P.~H. 2015, Physics of Plasmas, 22, 072108

\bibitem[{Schreiner {et~al.}(2017)Schreiner, Kilian, \&
  Spanier}]{2017ApJ...834..161S}
Schreiner, C., Kilian, P., \& Spanier, F. 2017, The Astrophysical Journal, 834,
  161

\bibitem[{Shalchi(2009)}]{Shalchi:2009vg}
Shalchi, A. 2009, {Nonlinear cosmic ray diffusion theories}

\bibitem[{Shevchenko {et~al.}(2002)Shevchenko, Galinsky, \&
  Ride}]{2002JGRA..107.1367S}
Shevchenko, V.~I., Galinsky, V.~L., \& Ride, S.~K. 2002, Journal of Geophysical
  Research: Space Physics, 107, 1367

\bibitem[{Sironi \& Spitkovsky(2009{\natexlab{a}})}]{2009ApJ...698.1523S}
Sironi, L., \& Spitkovsky, A. 2009{\natexlab{a}}, The Astrophysical Journal,
  698, 1523

\bibitem[{Sironi \& Spitkovsky(2009{\natexlab{b}})}]{2009ApJ...707L..92S}
---. 2009{\natexlab{b}}, The Astrophysical Journal Letters, 707, L92

\bibitem[{Skilling(1971)}]{1971ApJ...170..265S}
Skilling, J. 1971, Astrophysical Journal, 170, 265

\bibitem[{Skilling(1975)}]{1975MNRAS.172..557S}
---. 1975, Monthly Notices of the Royal Astronomical Society, 172, 557

\bibitem[{Socrates {et~al.}(2008)Socrates, Davis, \&
  Ramirez-Ruiz}]{2008ApJ...687..202S}
Socrates, A., Davis, S.~W., \& Ramirez-Ruiz, E. 2008, The Astrophysical
  Journal, 687, 202

\bibitem[{Spitkovsky(2005)}]{2005AIPC..801..345S}
Spitkovsky, A. 2005, in ASTROPHYSICAL SOURCES OF HIGH ENERGY PARTICLES AND
  RADIATION. AIP Conference Proceedings, Kavli Institute for Particle
  Astrophysics and Cosmology, Stanford University, PO Box 20450, MS 29,
  Stanford, CA 94309, 345--350

\bibitem[{Spitkovsky(2008)}]{2008ApJ...673L..39S}
Spitkovsky, A. 2008, The Astrophysical Journal Letters, 673, L39

\bibitem[{Stix(1992)}]{Stix:1992td}
Stix, T.~H. 1992, {Waves in Plasmas} (Springer Science {\&} Business Media)

\bibitem[{Stroman {et~al.}(2009)Stroman, Pohl, \&
  Niemiec}]{2009ApJ...706...38S}
Stroman, T., Pohl, M., \& Niemiec, J. 2009, The Astrophysical Journal, 706, 38

\bibitem[{Sudan \& Ott(1971)}]{1971JGR....76.4463S}
Sudan, R.~N., \& Ott, E. 1971, Journal of Geophysical Research, 76, 4463

\bibitem[{Thomas \& Pfrommer(2019)}]{2019MNRAS.tmp..260T}
Thomas, T., \& Pfrommer, C. 2019, Monthly Notices of the Royal Astronomical
  Society

\bibitem[{V{\"o}lk \& Cesarsky(1982)}]{Volk:1982fk}
V{\"o}lk, H.~J., \& Cesarsky, C.~J. 1982, Zeitschrift f{\"u}r Naturforschung A,
  37, 809

\bibitem[{Weidl {et~al.}(2019{\natexlab{a}})Weidl, Winske, \&
  Niemann}]{2019ApJ...872...48W}
Weidl, M.~S., Winske, D., \& Niemann, C. 2019{\natexlab{a}}, The Astrophysical
  Journal, 872, 48

\bibitem[{Weidl {et~al.}(2019{\natexlab{b}})Weidl, Winske, \&
  Niemann}]{2019ApJ...873...57W}
---. 2019{\natexlab{b}}, The Astrophysical Journal, 873, 57

\bibitem[{Wentzel(1969)}]{1969ApJ...156..303W}
Wentzel, D.~G. 1969, Astrophysical Journal, 156, 303

\bibitem[{Wiener {et~al.}(2013{\natexlab{a}})Wiener, Oh, \&
  Guo}]{2013MNRAS.434.2209W}
Wiener, J., Oh, S.~P., \& Guo, F. 2013{\natexlab{a}}, Monthly Notices of the
  Royal Astronomical Society, 434, 2209

\bibitem[{Wiener {et~al.}(2017)Wiener, Pfrommer, \&
  Peng~Oh}]{2017MNRAS.467..906W}
Wiener, J., Pfrommer, C., \& Peng~Oh, S. 2017, Monthly Notices of the Royal
  Astronomical Society, 467, 906

\bibitem[{Wiener {et~al.}(2013{\natexlab{b}})Wiener, Zweibel, \&
  Oh}]{2013ApJ...767...87W}
Wiener, J., Zweibel, E.~G., \& Oh, S.~P. 2013{\natexlab{b}}, The Astrophysical
  Journal, 767, 87

\bibitem[{Wiener {et~al.}(2018)Wiener, Zweibel, \& Oh}]{2018MNRAS.473.3095W}
---. 2018, Monthly Notices of the Royal Astronomical Society, 473, 3095

\bibitem[{Wu \& Davidson(1972)}]{1972JGR....77.5399W}
Wu, C.~S., \& Davidson, R.~C. 1972, Journal of Geophysical Research, 77, 5399

\bibitem[{Yan \& Lazarian(2002)}]{2002PhRvL..89B1102Y}
Yan, H., \& Lazarian, A. 2002, Physical Review Letters, 89, 281102

\bibitem[{Yan \& Lazarian(2004)}]{2004ApJ...614..757Y}
---. 2004, The Astrophysical Journal, 614, 757

\bibitem[{Yan \& Lazarian(2008)}]{2008ApJ...673..942Y}
---. 2008, The Astrophysical Journal, 673, 942

\bibitem[{Yoon {et~al.}(2014)Yoon, Schlickeiser, \&
  Kolberg}]{2014PhPl...21c2109Y}
Yoon, P.~H., Schlickeiser, R., \& Kolberg, U. 2014, Physics of Plasmas, 21,
  032109

\bibitem[{Zenitani(2015)}]{2015PhPl...22d2116Z}
Zenitani, S. 2015, Physics of Plasmas, 22, 042116

\bibitem[{Zweibel(2013)}]{2013PhPl...20e5501Z}
Zweibel, E.~G. 2013, Physics of Plasmas, 20, 5501

\bibitem[{Zweibel(2017)}]{2017PhPl...24e5402Z}
---. 2017, Physics of Plasmas, 24, 055402

\bibitem[{Zweibel \& Everett(2010)}]{2010ApJ...709.1412Z}
Zweibel, E.~G., \& Everett, J.~E. 2010, The Astrophysical Journal, 709, 1412

\bibitem[{Zweibel \& Shull(1982)}]{1982ApJ...259..859Z}
Zweibel, E.~G., \& Shull, J.~M. 1982, Astrophysical Journal, 259, 859

\end{thebibliography}
\end{document}